\documentclass[fleqn,usenatbib]{mnras}

\usepackage[T1]{fontenc}

\DeclareRobustCommand{\VAN}[3]{#2}
\let\VANthebibliography\thebibliography
\def\thebibliography{\DeclareRobustCommand{\VAN}[3]{##3}\VANthebibliography}

\usepackage{graphicx}	
\usepackage{amsmath}	
\usepackage{amssymb}	
\usepackage{scrextend} 
\usepackage{lipsum}
\usepackage{pdflscape}  
\usepackage[dvipsnames]{color}
\usepackage[percent]{overpic} 
\usepackage{tikz}
\usepackage{float}
\usetikzlibrary{shapes.geometric, arrows}
\usepackage{newtxtext,newtxmath}

\newcommand{\mrm}[1]{\mathrm{#1}}	
\newcommand{\vect}[1]{\boldsymbol{#1}}
\newcommand{\ellsum}[1]{\ell_{\Sigma}} 
\newcommand{\kms}{\mathrm{km\,s^{-1}}} 

\title[PCA analysis of Stokes~$V$ LSD profiles]{Diagnosing large-scale stellar magnetic fields using PCA on spectropolarimetric data}

\author[L. T. Lehmann and J.-F. Donati]{
L. T. Lehmann,\thanks{E-mail: lisa.lehmann@irap.omp.eu}
J.-F. Donati
\\
IRAP, Universit\'e de Toulouse, CNRS / UMR 5277, UPS-OMP, 14 Avenue E. Belin, Toulouse F-31400, France\\
}

\date{Accepted XXX. Received YYY; in original form ZZZ}

\pubyear{2021}

\begin{document}
\label{firstpage}
\pagerange{\pageref{firstpage}--\pageref{lastpage}}
\maketitle

\begin{abstract}
Insights on stellar surface large-scale magnetic field topologies are usually drawn by applying Zeeman-Doppler-Imaging (ZDI) to the observed spectropolarimetric time series. However, ZDI requires experience for reliable results to be reached and is based on a number of prior assumptions that may not be valid, e.g., when the magnetic topology is evolving on timescales comparable to or shorter than the time span over which observations are collected. In this paper, we present a method based on Principal Component Analysis (PCA) applied to circularly polarised (Stokes~$V$) line profiles of magnetic stars to retrieve the main characteristics of the parent large-scale magnetic topologies, like for instance, the relative strength of the poloidal and toroidal components, and the degree of axisymmetry of the dominant field component and its complexity (dipolar or more complex). We show that this method can also be used to diagnose the temporal variability of the large-scale magnetic field. Performing best for stars with moderate projected equatorial velocities hosting relatively simple magnetic field topologies, this new method is simpler than ZDI, making it convenient to rapidly diagnose the main characteristics of the large-scale fields of non-degenerate stars and to provide insights into the temporal evolution of the field topology.
\end{abstract}




\section{Introduction}
\label{Sec:Intro}

For investigating stellar large-scale magnetic field topologies, the method of choice is Zeeman-Doppler-Imaging \citep[ZDI,][]{DonatiBrown1997, Kochukhov2004, Donati2006, Carroll2012, Folsom2016, Hussain2016}. ZDI analyses phase-resolved sets of spectropolarimetric data and recovers the vector magnetic field map at the stellar surface that best fits the observed spectra. ZDI exploits the rotational modulation of the spectropolarimetric spectra accounting for the Doppler and Zeeman effects to retrieve information on the location of surface magnetic features, as well as on the orientation of field lines within these regions. 

ZDI is only sensitive to large-scale fields as the polarisation in line profiles responds to the vector properties of the magnetic field; as a result, tangled fields on small scales cancel out their average contribution to polarized Zeeman signatures. Analysing unpolarised (Stokes~$I$) spectra and the broadening that the Zeeman effect induces on line profiles can reveal the overall (small- and large-scale) magnetic flux emerging from the stellar surface but with little to no information about the field topology \citep{Robinson1980, Saar1988, Reiners2006, Lehmann2015}. Until now, ZDI was the only option to retrieve information on the main characteristics of the parent large-scale magnetic topologies from the Stokes~$V$ signatures.

In this paper, we present a new method to diagnose the large-scale field directly from the circularly polarised Stokes~$V$ profiles using  Principal Component Analysis (PCA). PCA is a data-driven method, that can decompose a given set of data (e.g.\ Stokes~$V$ profiles) into an orthogonal basis of reference profiles (called the principal components, eigenprofiles or eigenvectors). The eigenvectors are ordered by their decreasing importance on the variance and can be used to express the original data set (i.e.\ Stokes~$V$ profiles) as a linear combination of all PCA eigenvectors, see e.g., \citet{Murtagh1987,Rees2000}.

PCA is widely used for noise and dimension reduction and is often applied in the astrophysical context, e.g., for the automatic classification of galactic redshifts \citep{Glazebrook1998}, the direct imaging of exoplanets \citep{Kiefer2021}, for noise-reduction in spectropolarimetric lines \citep{Carroll2007,Casini2019} and many other natural or computer science areas, e.g., for the detection of microplastic \citep{Fang2022}, or facial recognition \citep{Chan2015}.

\cite{Rees2000} applied the PCA technique to the unpolarised and polarised Stokes profiles of sunspot observations and showed its value as a fast method for spectral line inversion. \cite{Eydenberg2005} recovered solar magnetic field vector maps of resolved sunspots using PCA whereas \cite{Lehmann2015} studied the cyclic magnetic field evolution of the young solar analogue $\varepsilon~\mrm{Eri}$ from Stokes~$I$ profiles across a time range of several years. \cite{Skumanich2002} presented the physical background for the PCA analysis of Stokes profiles of spatially-resolved solar observations. In particular, they showed, that the eigenprofiles found by PCA can be related to certain physical quantities, e.g., to the line-of-sight velocity and the magnetic Zeeman splitting for the unpolarised Stokes~$I$ profiles, and to the longitudinal and transverse field components for the polarised Stokes~$Q$, $U$ and $V$ profiles. 

Inspired by these results, we propose to test, whether one can retrieve the main characteristics of the large-scale fields of magnetic stars from sets of observed circularly polarised Stokes~$V$ profiles.

In our study, we apply PCA to circularly polarised Stokes~$V$ profiles extracted from thousands of spectral lines with Least-Square Deconvolution \citep[LSD,][]{Donati1997b, Kochukhov2010}. In principle, PCA can be applied to individual spectral lines; however, working with high-SNR LSD Stokes~$V$ profiles is mandatory here, in order to extract as precisely as possible the subtle shape changes that probe the characteristics of the large-scale field. In contrast to \cite{Skumanich2002} the polarimetric observations are not spatially resolved at a given epoch, but rather spatially unresolved (i.e., averaged over the visible stellar hemisphere) and collected at a number of different observations. 

We aim at deriving simple conclusions about the respective amounts of poloidal and toroidal components in the large-scale field as well as about the degree of axisymmetry  and complexity of the large-scale field by directly applying PCA on the time series of observed Stokes~$V$ profiles. 

In this paper, we quickly review the PCA method and introduce our simulation set-up in Section~\ref{Sec:Methods}. We present the diagnostics of axisymmetric large-scale fields directly from Stokes~$V$ profiles in Section~\ref{Sec:DiagField} and introduce the PCA method for the special case of a dipole magnetic field in Section~\ref{Sec:Dipole} before moving on to the more general case of complex magnetic field topologies in Section~\ref{Sec:Complex}. In Section~\ref{Sec:EvoMaps}, we show our method's value in tracing evolving large-scale fields and in Section~\ref{Sec:RealObs} we demonstrate its applicability to real archival observations of two M~dwarfs. We finally end with a summary in Section~\ref{Sec:Conclusions}.

\section{Method}
\label{Sec:Methods}

\subsection{Principles of PCA}
\label{SubSec:PCA}

PCA \citep{Murtagh1987} expresses the input data in a new orthogonal basis consisting of the eigenvectors of the covariance matrix of the input data. The eigenvectors are ordered by decreasing unsigned eigenvalue, so that the eigenvectors with the highest contribution to the variance, i.e., those with the largest unsigned eigenvalue (the most principal components) are stored first.  
Further, PCA determines the contribution of each eigenvector to the input data (in our case to the Stokes~$V$ profiles) and how this contribution varies across the data set, e.g., with time. 
The more Stokes~$V$ profiles in the data set, the better PCA can distinguish how these profiles vary with time and disentangle real variations from noise.  

In our study, the data matrix ($V(t,v)$) consists of the Stokes~$V$ LSD profiles as a function of time ($t$) and velocity ($v$), forming a $m \times n$ matrix, where $m$~is the number of spectral points per Stokes~$V$ profile and $n$~the total number of Stokes~$V$ profiles. For our Stokes~$V$ time series, we have $n < m$, so that PCA returns the $n$~eigenvectors $f(v)$ of the covariance matrix of $V(t,v)$ and the corresponding coefficients $c(t)$ so that the dot product of $c(t)$ and $f(v)$ returns $V(t,v)$ again.
In the framework of our study, we apply PCA to the mean-subtracted Stokes~$V$ profiles directly, where the mean Stokes~$V$ profile of the given time series is subtracted from each Stokes~$V$ profile before applying PCA. More specifically, the mean Stokes~$V$ profile is first used to diagnose the axisymmetric component of the large-scale field, whereas the PCA analysis gives us information on the non-axisymmetric component of the field.  

We expect the first principal components (i.e.\ the eigenvectors with the largest absolute eigenvalues) of the circularly polarised Stokes~$V$ profiles to show shapes similar to the derivatives of the Stokes~$I$ profiles following the results of \cite{Skumanich2002}, and given the relationship between the Stokes~$I$ and $V$ profiles in the weak field approximation \citep{Unno1956,Stenflo1994}: 
\begin{equation}
V(\lambda) = -C \lambda^2_0 g_{\mrm{eff}} B_{\mrm{long}} \frac{\partial I(\lambda)}{\partial \lambda},
\label{Eq:StokesV}
\end{equation}
where $C = 4.67\cdot 10^{-13}\,\mrm{G^{-1}} $\AA$^{-1}$, $g_{\mrm{eff}}$ is the effective Land\'{e} factor of the mean line, $B_{\mrm{long}}$ the longitudinal magnetic field, i.e., the line-of-sight projected component of the magnetic field averaged over the visible stellar hemisphere, and $I(\lambda)$ the unpolarised Stokes~$I$ profile. We find that the PCA method works best for Stokes~$V$ profiles, that can be described as a combination of the first few derivatives of the unpolarised Stokes~$I$ profiles, which is the case even for magnetic field strengths above the weak field approximation limit of $\approx1\,\mrm{kG}$.

\subsection{Simulations}
\label{SubSec:SImulations}

In the first part of our study, we apply PCA to simulated Stokes~$V$ profiles.  For this purpose, we use subroutines of the Zeeman-Doppler-Imaging code implemented by \cite{DonatiBrown1997, Donati2006, Donati2014} to compute the Stokes~$V$ profiles associated with a given surface magnetic field map of a model star. The surface magnetic field maps of our model stars are described using the spherical harmonics based on the equations in section~5.1 of \citet{Donati2006}, but with the $\beta$ coefficient (initially describing the non-radial poloidal component of the field) now replaced by $\alpha + \beta$\footnote{The formulation of $\alpha + \beta$ was used for the majority of the ZDI results published by our team over the last decade.}, see appendix~\ref{App:MagField}. We recall in particular that $\beta=\gamma=0$ implies that the field is potential, $\gamma=0$ that the field is purely poloidal and $\alpha=\beta=0$ that the field is purely toroidal.

To simulate the Stokes~$V$ profiles, the visible stellar surface of our model star is decomposed into a mesh of 5000 cells and the local Stokes~$V$ profiles are computed using Unno-Rachkovsky's analytical solution of the polarised radiative transfer equations in a Milne-Eddington model atmosphere \citep{Landi2004}. For each observation, the local Stokes~$V$ profiles are Doppler-shifted and weighted according to the projected area of the corresponding cell and taking into account a linear limb darkening law for spectral type M1 \citep{Claret2000}. We assume solid-body rotation and set the inclination angle to $i=75^\circ$ and the projected equatorial velocity to $v_e \sin i = 10\,\kms$ for each model star unless otherwise indicated.

The average line profile is assumed to be located at a wavelength $\lambda_0 = 1750\,\mrm{nm}$ and to feature a Land\'{e} factor $g_{\mrm{eff}} = 1.2$, with a width and strength that mimics LSD profiles of slowly to moderately rotating M~dwarfs. The simulated LSD Stokes~$V$ profiles have a spectral resolving power of 65\,000 and consist of 45 velocity bins, with a velocity step of $\approx 2.0\,\kms$ mimicking spectropolarimetric observations in the near-infrared collected by the SpectroPolarim\`etre InfraRouge (SPIRou) at Canada-France-Hawaii Telescope (CFHT). 
We add a Gaussian noise corresponding to a SNR of 5\,000 to all simulated Stokes~$V$ profiles. 

\section{Diagnosing the axisymmetric large-scale field topology}
\label{Sec:DiagField}

We can separate the signal originating from the axisymmetric and the non-axisymmetric large-scale field by analysing the time-averaged Stokes~$V$ profile~$\overline{V}$ and the mean-subtracted Stokes~$V$ profiles, respectively.  

\begin{figure}
\begin{flushleft}
		\textbf{a.} \hspace{2.7cm} \textbf{b.} \hspace{2.4cm} \textbf{c.}
	\end{flushleft}
\centering
\includegraphics[height=5.8cm, angle=0, trim={0 0 0 0}, clip]{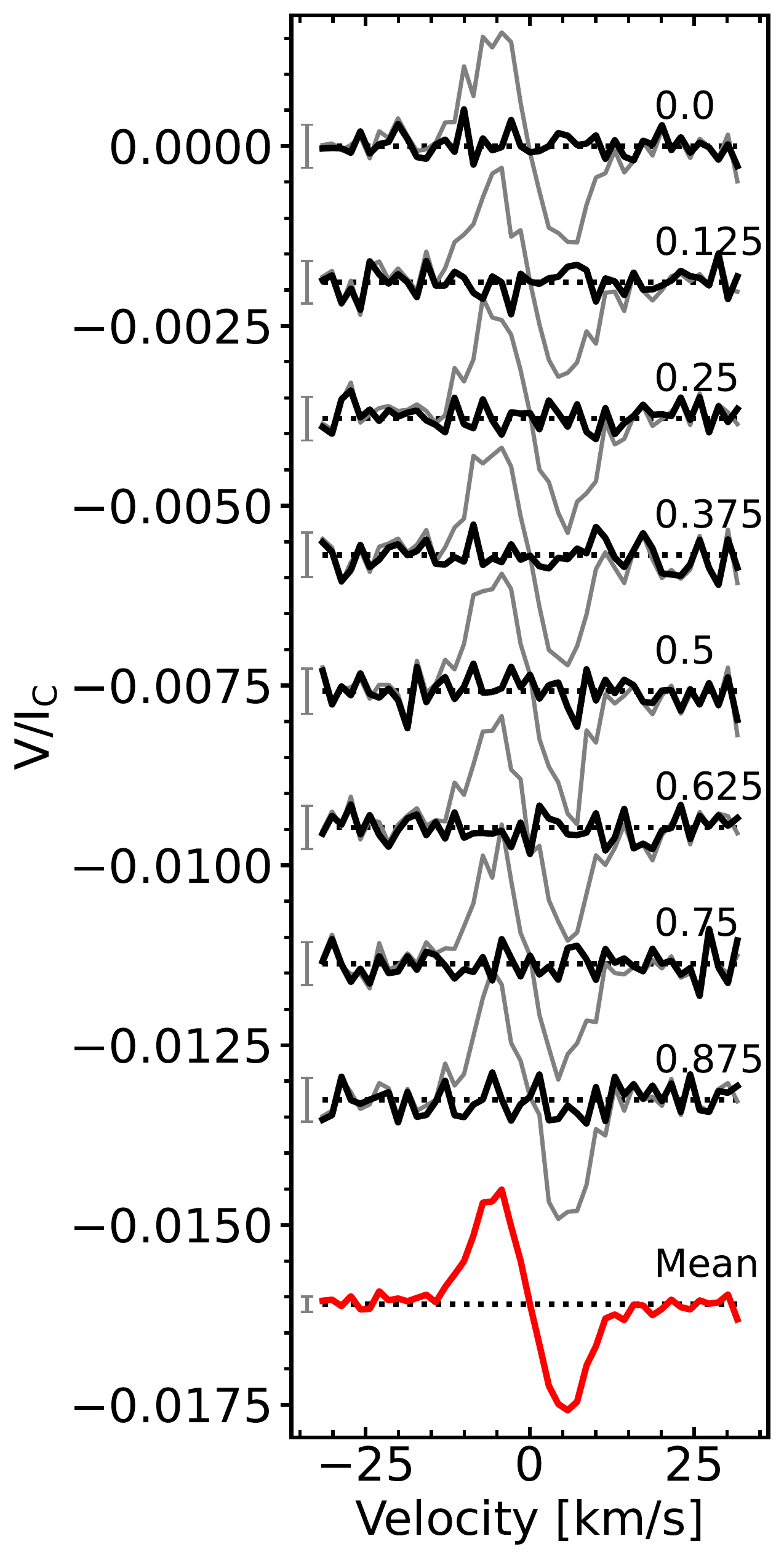}
\includegraphics[height=5.8cm, angle=0, trim={0 0 0 0}, clip]{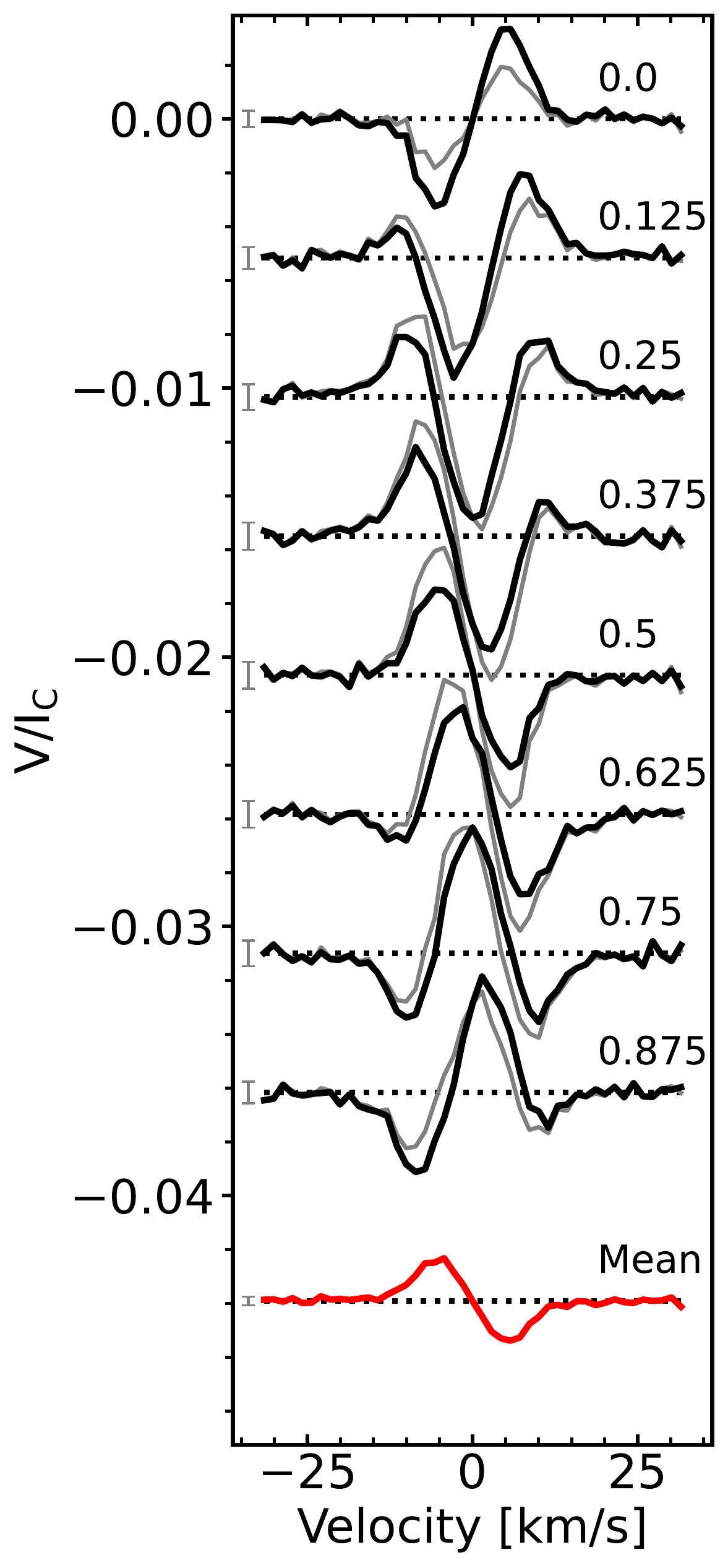}
\includegraphics[height=5.8cm, angle=0, trim={0 0 0 0}, clip]{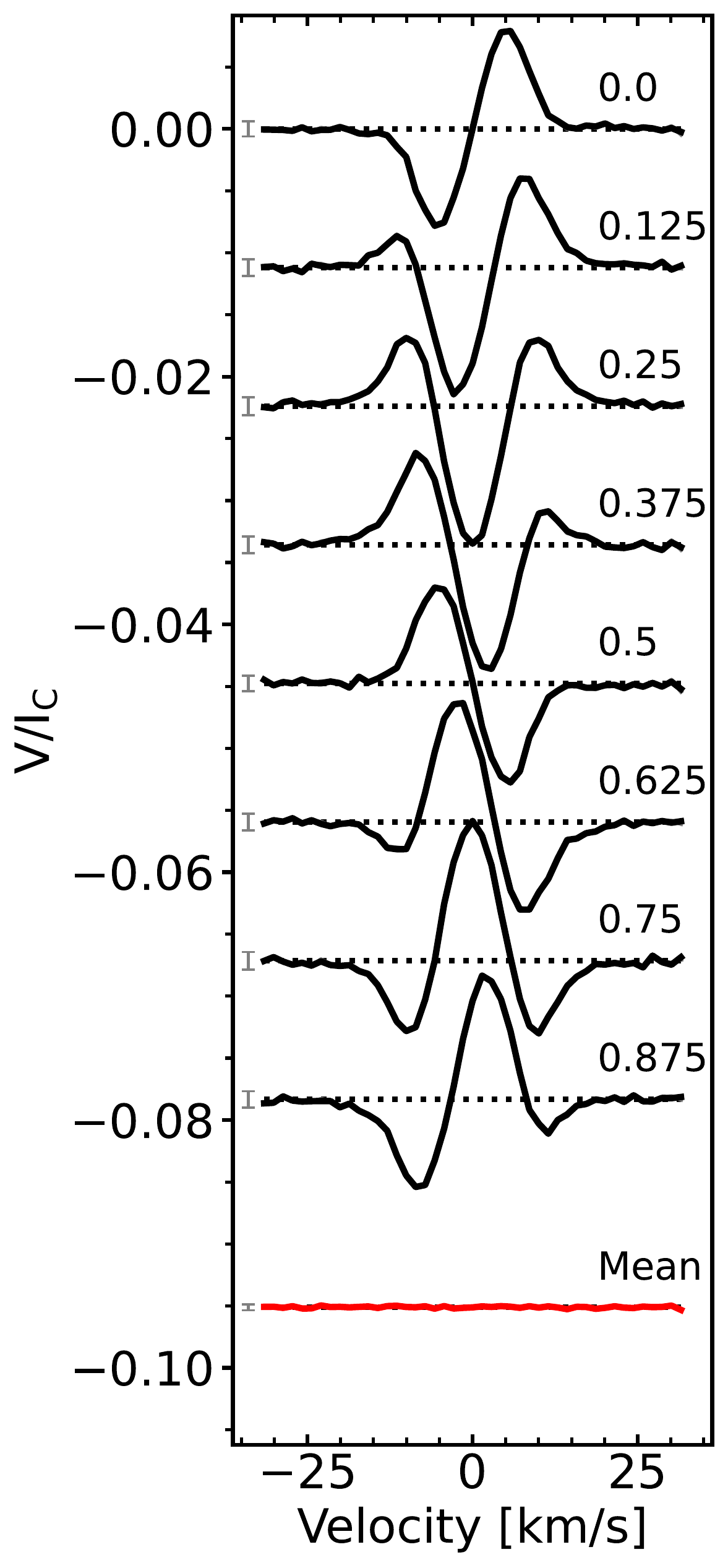}
\caption{Simulated Stokes~$V$ time series for three poloidal dipoles of different tilt angles $\psi$: a. $\psi = 0^\circ$, b. $\psi = 30^\circ$, c. $\psi = 90^\circ$. The Stokes~$V$ profiles are plotted in grey and the mean-subtracted Stokes~$V$ profiles in black, with the mean profile $\overline{V}$ of the time series being plotted in red at the bottom of each panel. The 1$\sigma$ error of the Stokes~$V$ profiles is indicated to the left of each profile and the rotational cycle to the right. The $V=0$ line is indicated by a dotted line. The rotation axis is assumed to be inclined at $75^\circ$ to the line-of-sight and the SNR is set to 5\,000 for all three cases. }
\label{Fig:AxisymMean}
\end{figure}

In the case of an axisymmetric field, the Stokes~$V$ signal is constant because the magnetic field is aligned with the rotation axis and its orientation with respect to the line-of-sight does not vary as the star rotates. 
Fig.~\ref{Fig:AxisymMean} shows the simulated Stokes~$V$ time series for three poloidal dipoles with different tilt angles $\psi$ between the magnetic and rotation axes. The rotation cycle is indicated to the right of each profiles and describes the fractional time span in which the star rotates once around its own axis. For a purely axisymmetric field configuration ($\psi=0^\circ$) each Stokes~$V$ profile (grey) is similar to the mean profile $\overline{V}$ (red) of the time series so that the mean-subtracted Stokes~$V$ profiles (black) are equal to zero except for the noise, see Fig.~\ref{Fig:AxisymMean}a. 

If the dipole field is tilted to the rotation axis, the Stokes~$V$ profiles are modulated as the star rotates, see Fig.~\ref{Fig:AxisymMean}b. The Stokes~$V$ profiles differ more and more from the mean profile $\overline{V}$ with growing tilt angle~$\psi$ until the mean profile $\overline{V}$ becomes zero for large tilt angles $\psi \approx 90^\circ$, see Fig.~\ref{Fig:AxisymMean}c. In the first two cases, Fig.~\ref{Fig:AxisymMean}a and b, the mean profile $\overline{V}$ is non zero and reflects the axisymmetric component of the field. The larger the mean profile $\overline{V}$ with respect to the mean-subtracted Stokes~$V$ profile, the more axisymmetric the large-scale field.

Furthermore, the mean profile $\overline{V}(v)$ can be decomposed into its antisymmetric and symmetric component with respect to the line centre, respectively related to the poloidal and toroidal components of the axisymmetric field,
\begin{align}
\text{antisymmetric (poloidal component):}& \ \frac{1}{2} \overline{V}(v) -  \overline{V}(-v) \\
\text{symmetric (toroidal component):}& \ \frac{1}{2} \overline{V}(v) + \overline{V}(-v). 
\end{align}
The symmetric component originates from the azimuthal magnetic field, which directly relates to the toroidal component for an axisymmetric field, see appendix~\ref{App:MagField}.
Fig.~\ref{fig:SymAntiSym} displays the mean profiles $\overline{V}$ (red line) and their decomposition into the antisymmetric (poloidal component, blue dot-dashed line) and symmetric part (toroidal component, yellow dashed line) for a purely poloidal field (Fig.~\ref{fig:SymAntiSym}a) and a mixed poloidal and toroidal field configuration (Fig.~\ref{fig:SymAntiSym}b), where the poloidal field dominates. We find, that the decomposition of the mean profile $\overline{V}$ allows us to estimate, which component is dominant or if an axisymmetric toroidal field is present at all. This is especially interesting as a strong toroidal field configuration is often axisymmetric \citep{See2015}.

For a reliable toroidal field detection, the amplitude of the symmetric component must be significantly larger than the error of the mean profile $\overline{V}$ as, e.g., small uncertainties in the determination of the spectral line centre can also cause a symmetric component.
Caution needs also to be taken if the phase coverage is sparse or highly uneven, as this could bias the mean profile $\overline{V}$ and its symmetric and antisymmetric components. We come back on this point in Sec.~\ref{Sec:EvoMaps} to mitigate this potential issue.
We recall that the rotation period needs to be known with reasonable accuracy, which should be easily achievable by, e.g., applying Gaussian Process Regression (GPR) to time series of the longitudinal field or of the PCA coefficients (see below).

\begin{figure}
	\begin{flushleft}
	\textbf{a.} \hspace{3.9cm} \textbf{b.}
	\end{flushleft}
	\centering
	\includegraphics[width=0.49\columnwidth, trim={0 0 0 0}, clip]{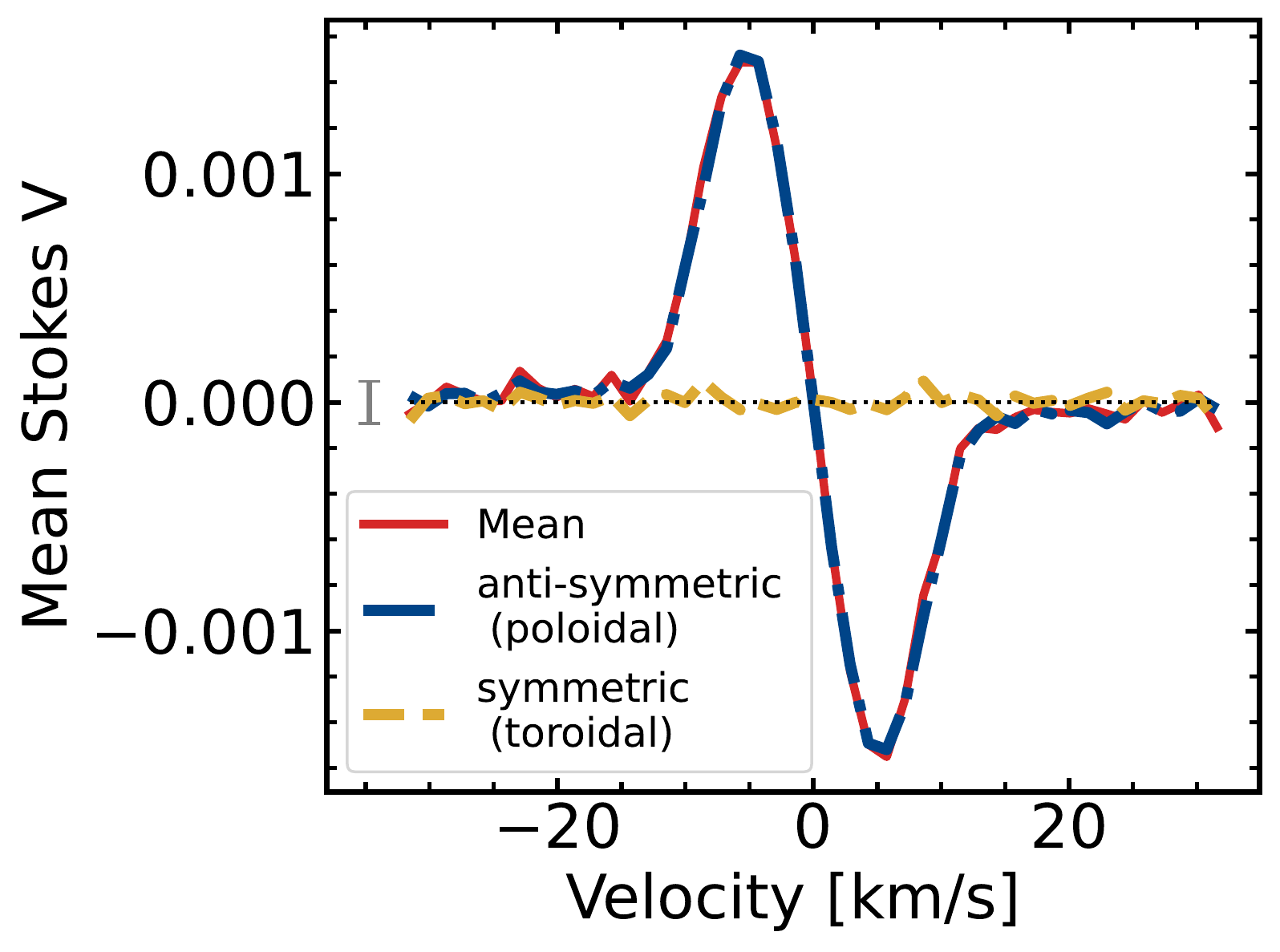} 
	\includegraphics[width=0.49\columnwidth, trim={0 0 0 0}, clip]{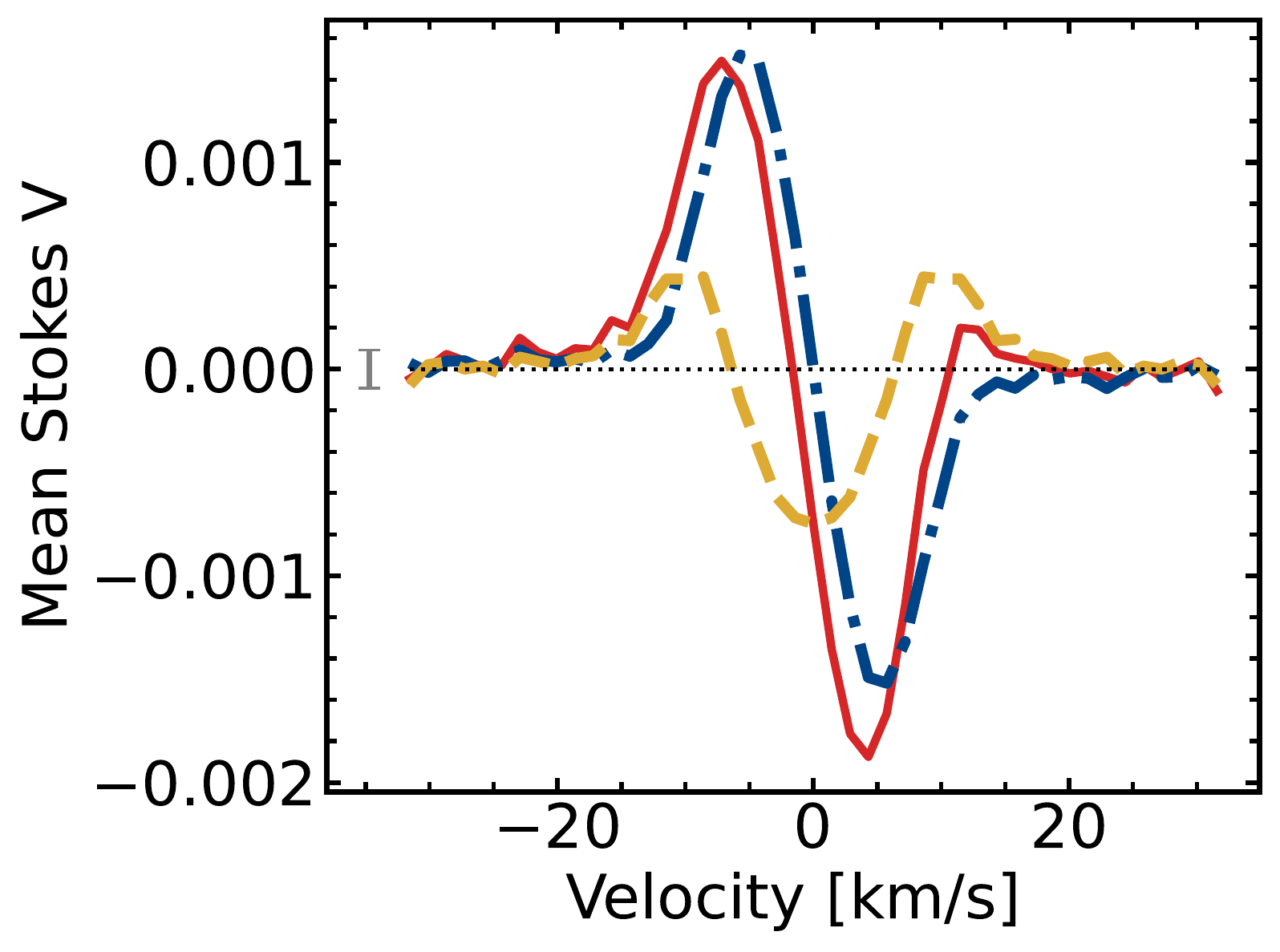} 
    \caption{Two mean profiles $\overline{V}$ (solid red line) and their antisymmetric (dash-dotted blue line) and symmetric (dashed yellow line) parts related to the poloidal and toroidal components of the axisymmetric large-scale field. The mean profile shown in a. corresponds to the poloidal dipole shown in Fig.~\ref{fig:Map_tiltDipol} and the mean profile in b. to the mixed field configuration in Fig.~\ref{fig:Map_mixDipol}, where the ratio between the toroidal and poloidal energy is 0.03. The rotation axis is assumed to be inclined at $75^\circ$ to the line-of-sight and the SNR is set to 5\,000 for both cases.}
    \label{fig:SymAntiSym}
\end{figure}

By subtracting the mean profile $\overline{V}$ from the Stokes~$V$ profiles, we removed the signal originating from the axisymmetric large-scale field and can now analyse the non-axisymmetric field via PCA. PCA captures the variation of the Stokes~$V$ profiles in one or more eigenvectors as soon as the magnetic field is not aligned with the rotation axis. Additionally, the corresponding coefficients will be phase-modulated.

\section{Simple case: A dipole field}
\label{Sec:Dipole}

\begin{figure*} 
	\begin{flushleft}
	\textbf{a.} \hspace{2.5cm} \textbf{b.} \hspace{3.6cm} \textbf{c.}
	\end{flushleft}
    \begin{minipage}{0.11\textwidth}
    \centering
    \includegraphics[height=0.85\columnwidth, angle=270, trim={140 0 0 29}, clip]{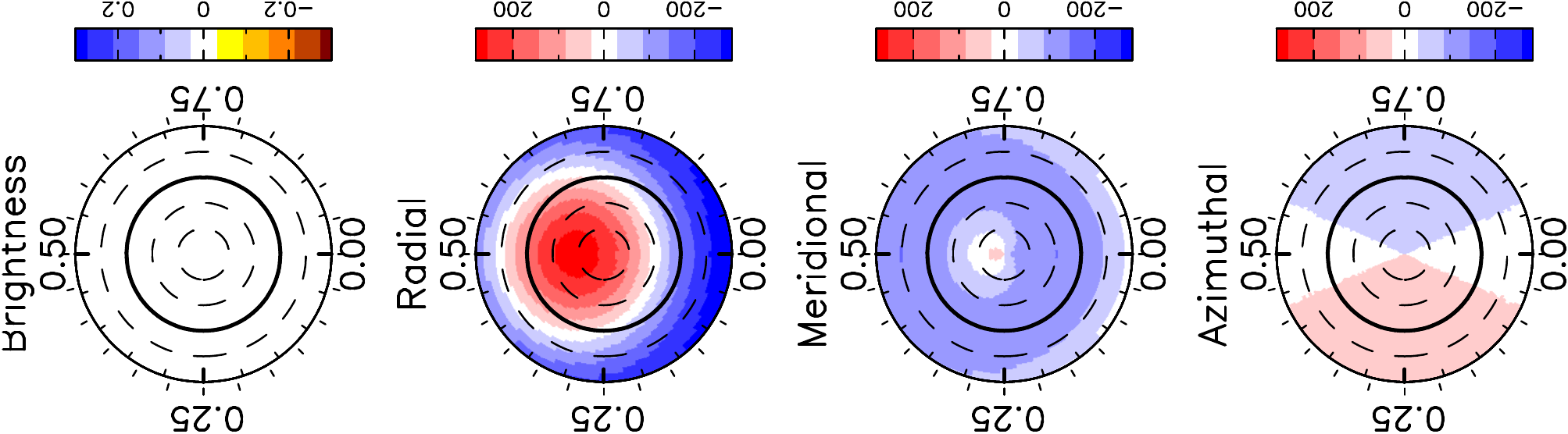} 
	\includegraphics[width=\columnwidth, angle=180, trim={470 130 12 0}, clip]{Figures/MapN2.pdf}
	\end{minipage}
    \begin{minipage}{0.25\textwidth}
    \centering
    \includegraphics[width=\columnwidth, angle=0, trim={0 0 0 0}, clip]{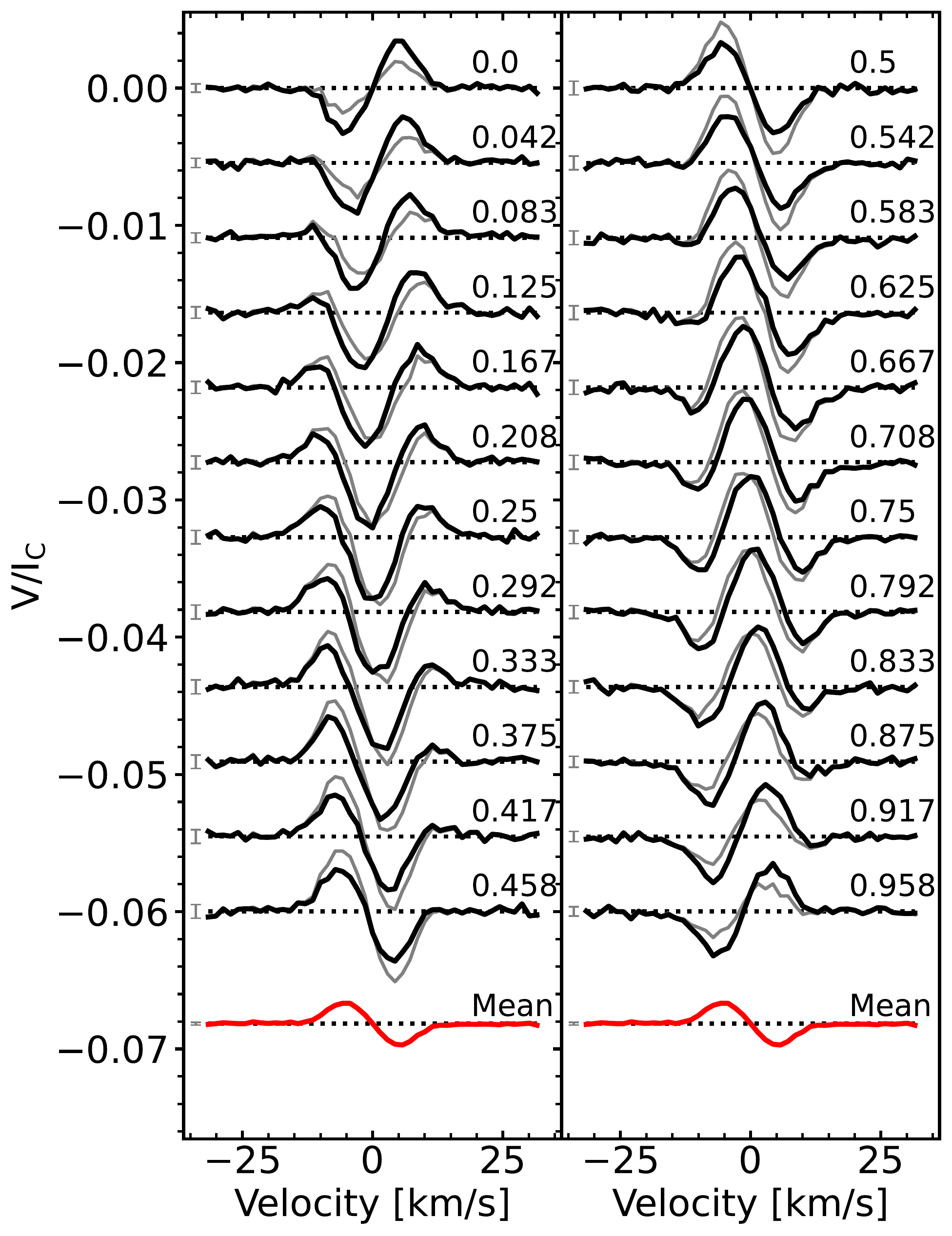} 
    	\end{minipage}
    \begin{minipage}{0.6\textwidth}
    \centering
    \includegraphics[width=\columnwidth, trim={0 400 0 0}, clip]{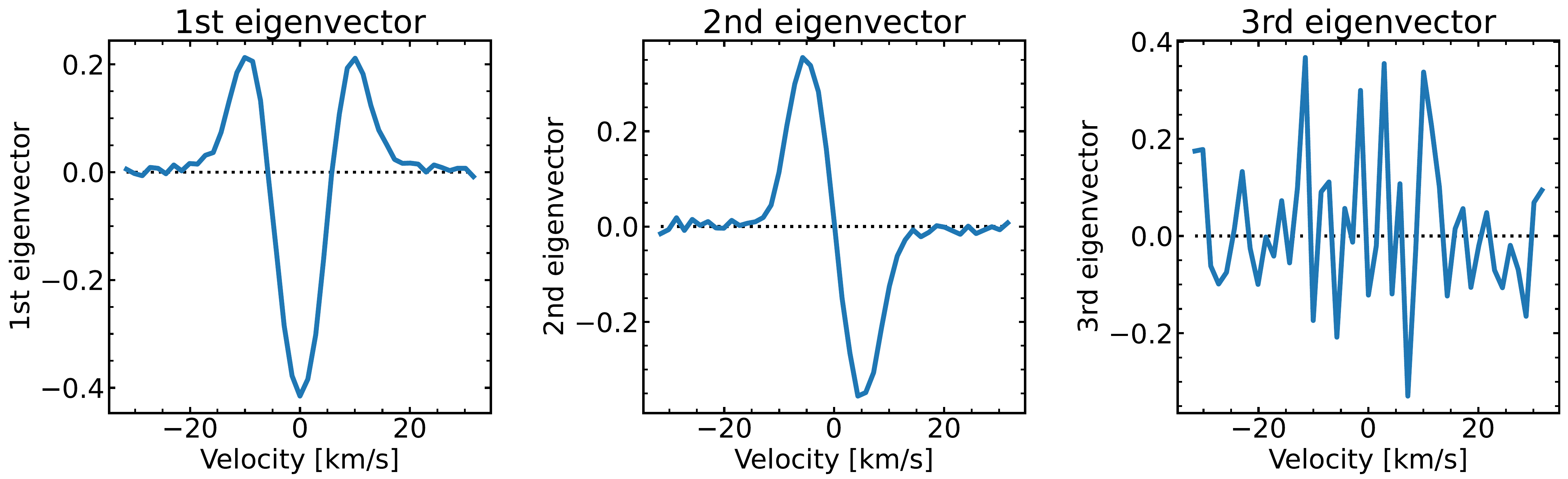} \\
    \includegraphics[width=\columnwidth, trim={0 400 0 0}, clip]{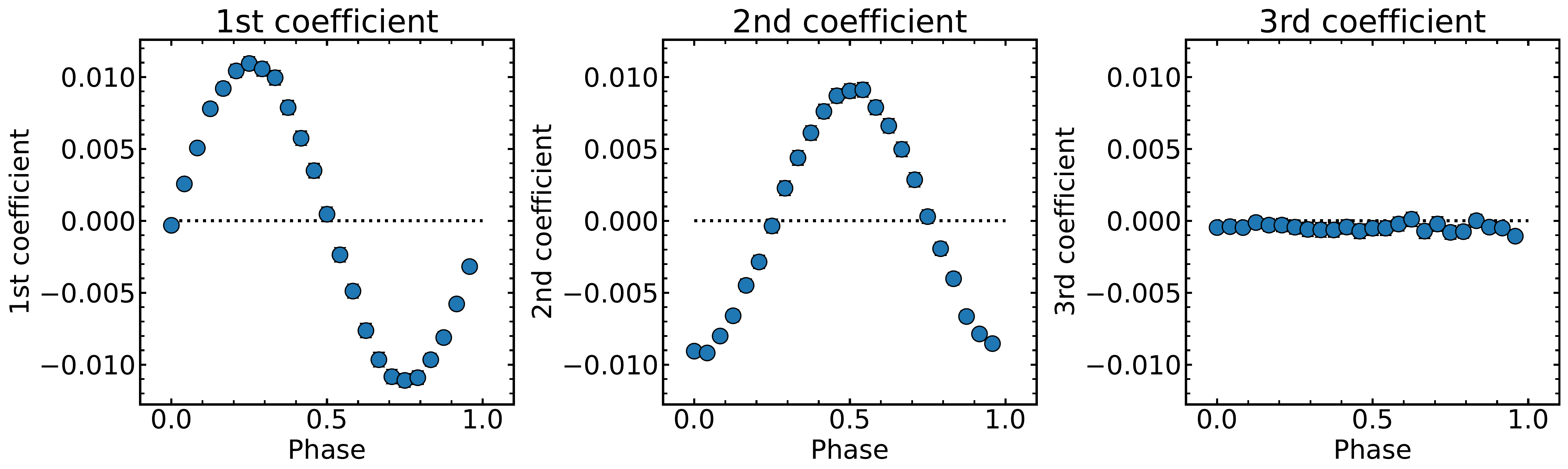} 
    	\end{minipage}
    \caption{a. The magnetic maps of the radial, meridional and azimuthal field components (top to bottom) are shown in a flattened polar view, for a titled poloidal dipole (tilt angle $\psi \approx30^{\circ}$).  The thick line depicts the equator and the dashed lines the latitudes in $30^{\circ}$ steps, whereas the ticks outside the plots illustrate the phases of observations. The colour bar below is used for all three maps and indicates the magnetic field strength in G. b. The corresponding simulated Stokes~$V$ time series is displayed in the same format as in Fig.~\ref{Fig:AxisymMean}. The mean-subtracted Stokes~$V$ profiles used as input for PCA are plotted in black.  c. The first three eigenvectors (top) and their corresponding phase-folded coefficients (bottom) of the PCA analysis.}
    \label{fig:Map_tiltDipol}
\end{figure*}

We start with the simple case of a dipole field for demonstrating the analysis of the non-axisymmetric field via PCA.

Fig.~\ref{fig:Map_tiltDipol}a shows the map of a poloidal dipole, whose magnetic axis is titled at $\psi = 30^{\circ}$. We first note that the mean profile $\overline{V}$ is, as expected, purely antisymmetric with respect to the line centre, see Fig.~\ref{fig:Map_tiltDipol}b or Fig.~\ref{fig:SymAntiSym}a, reflecting the purely poloidal axisymmetric large-scale field.

We apply PCA to the mean-subtracted Stokes~$V$ profiles and return eigenvectors and coefficients for the analysis of the non-axisymmetric field, see Fig.~\ref{fig:Map_tiltDipol}c. The first and second eigenvector are respectively found to be symmetric and antisymmetric with respect to the line centre, and proportional to the second and first derivative of the unpolarised Stokes~$I$ profile, see Fig.~\ref{fig:Map_tiltDipol}c top row.  In this example, the second eigenvector, proportional to the first derivative of the Stokes~$I$ profile, scales with the longitudinal magnetic field (see Eq.~\ref{Eq:StokesV}) and so does the corresponding coefficient, strongest at phase 0.5 and 0.0 when the positive and negative poles of the dipole come closest to the observer, see Fig.~\ref{fig:Map_tiltDipol}a. The following eigenvectors contain only noise, recognisable from the chaotic shape of the eigenvectors themselves, the much smaller amplitude of the corresponding coefficients and their smaller eigenvalues. Another approach consists in determining the reduced chi-squares:
\begin{equation}
\chi^2_r = \tfrac{\sum_i^{n} \left(\frac{\mathrm{coeff}_i} {\sigma_{\mathrm{coeff}_i}}\right)^2} {n-1}
\end{equation}
for each of the coefficient phase curves, to determine whether each coefficient phase curve significantly differs from zero. From the $\chi^2_r$ value and the number of points, one can compute the probability of whether a signal is detected. If no signal is detected (e.g., false alarm probability larger than 0.1\,\%), it implies that the eigenvector reduces to noise.

The first eigenvector, proportional to the second derivative of the Stokes~$I$ profile and thus to the first derivative of the second eigenvector, is what describes the temporal evolution of the Stokes~$V$ profiles between both longitudinal field maxima, with the corresponding coefficient reaching maxima at phases 0.25 and 0.75, see Fig.~\ref{fig:Map_tiltDipol}c. For a standard poloidal dipole with $\beta=\gamma=0$, the azimuthal and radial field component peak always 0.25 apart in phase from each other as seen in our example in Fig.~\ref{fig:Map_tiltDipol}. Note that, the symmetric eigenvector does not trace the azimuthal field exclusively, a pure radial field also returns a symmetric eigenvector, albeit with a weaker eigenvalue, see Fig.~\ref{fig:Map_radDipol} for the radial field restricted example of the topology shown in Fig.~\ref{fig:Map_tiltDipol}. Nevertheless, we find, that the symmetric eigenvector and its coefficients are dominated by the azimuthal field for most topologies.

For a moderately tilted dipole as in Fig.~\ref{fig:Map_tiltDipol}, two components are enough to describe the evolution of the Stokes~$V$ profiles;  if the dipole field is strongly tilted, more PCA eigenvectors are needed, see Fig.~\ref{fig:Map_VerytiltDipol} in the appendix. The more the magnetic axis is tilted to the rotation axis, the more eigenvectors contribute to the signal and have significantly non-zero coefficients, e.g., compare Fig.~\ref{fig:Map_tiltDipol} with Fig.~\ref{fig:Map_VerytiltDipol} and \ref{fig:Map_axisymDipol} in the appendix. It reflects that temporal variations of the Stokes~$V$ profile, which become larger when the dipole is more tilted, can only be reliably described by incorporating additional terms in the Taylor expansion.

\begin{figure*} 
	\begin{flushleft}
	\textbf{a.} \hspace{2.5cm} \textbf{b.} \hspace{3.6cm} \textbf{c.}
	\end{flushleft}
    \begin{minipage}{0.11\textwidth}
    \centering
    \includegraphics[height=0.85\columnwidth, angle=270, trim={140 0 0 29}, clip]{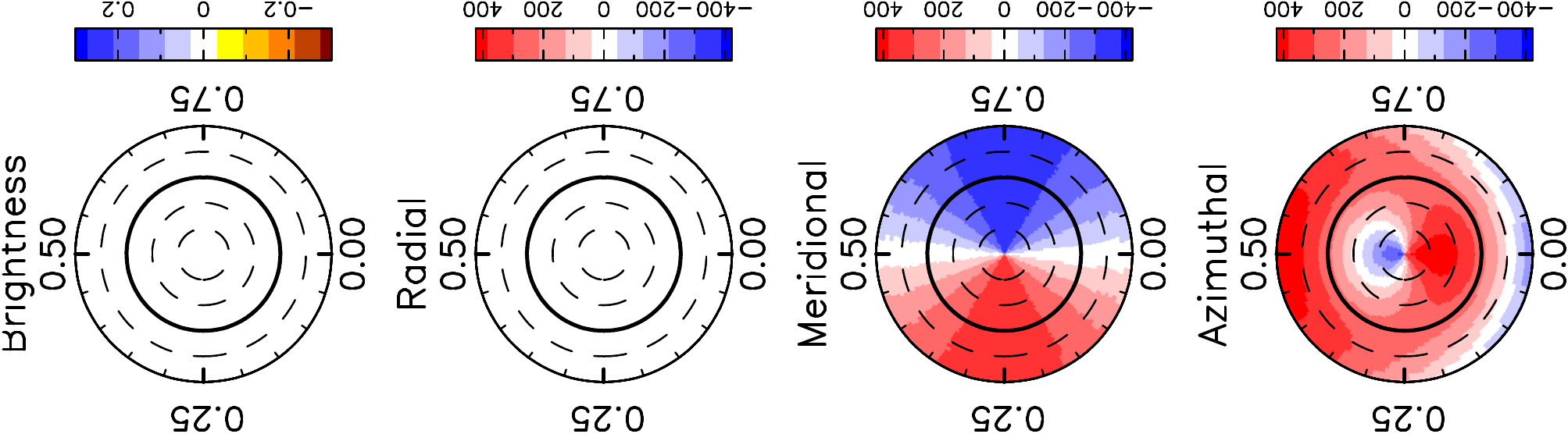} 
	\includegraphics[width=\columnwidth, angle=180, trim={470 130 12 0}, clip]{Figures/MapN6.pdf}
	\end{minipage}
    \begin{minipage}{0.25\textwidth}
    \centering
    \includegraphics[width=\columnwidth, angle=0, trim={0 0 0 0}, clip]{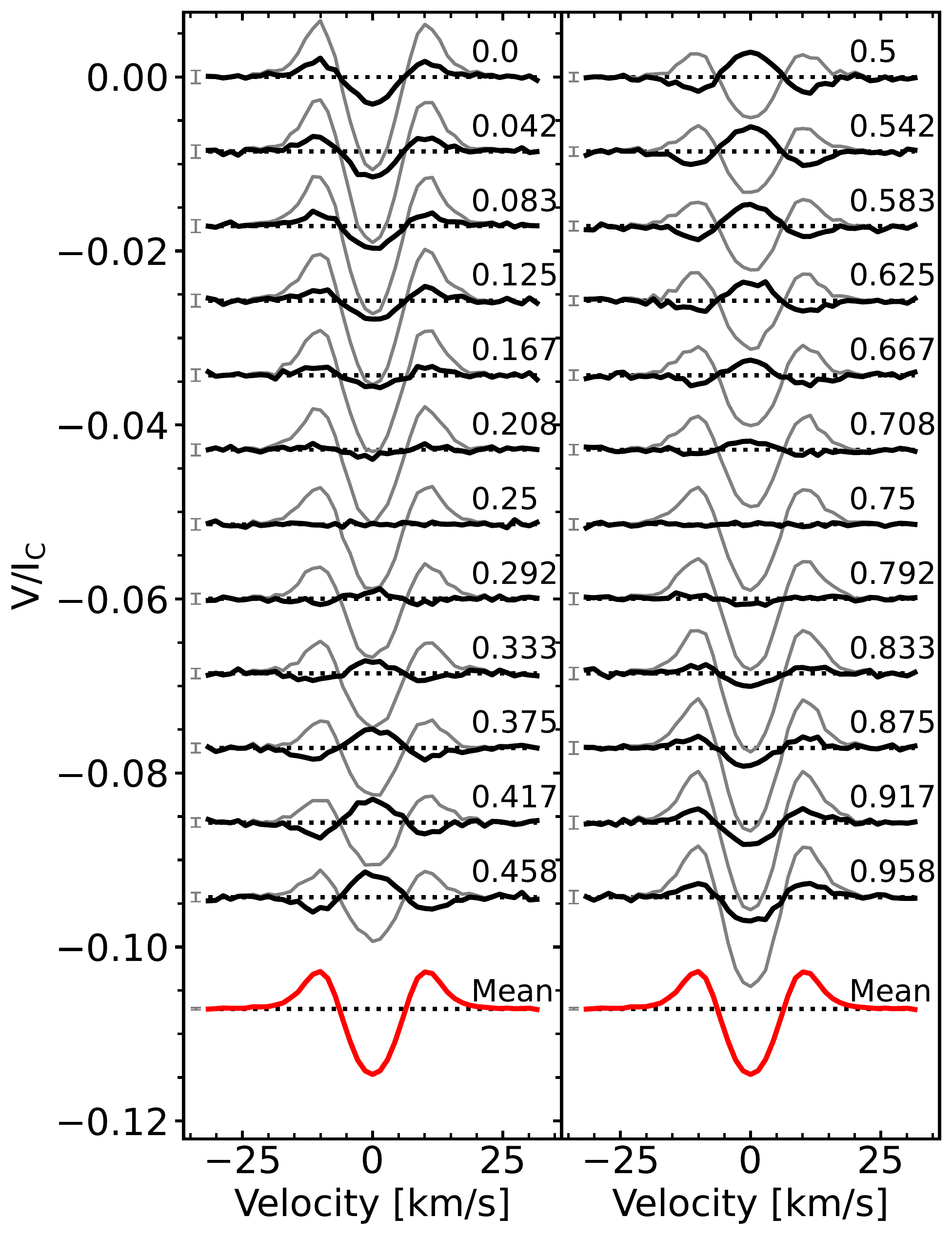} 
    	\end{minipage}
    \begin{minipage}{0.6\textwidth}
    \centering
    \includegraphics[width=\columnwidth, trim={0 400 0 0}, clip]{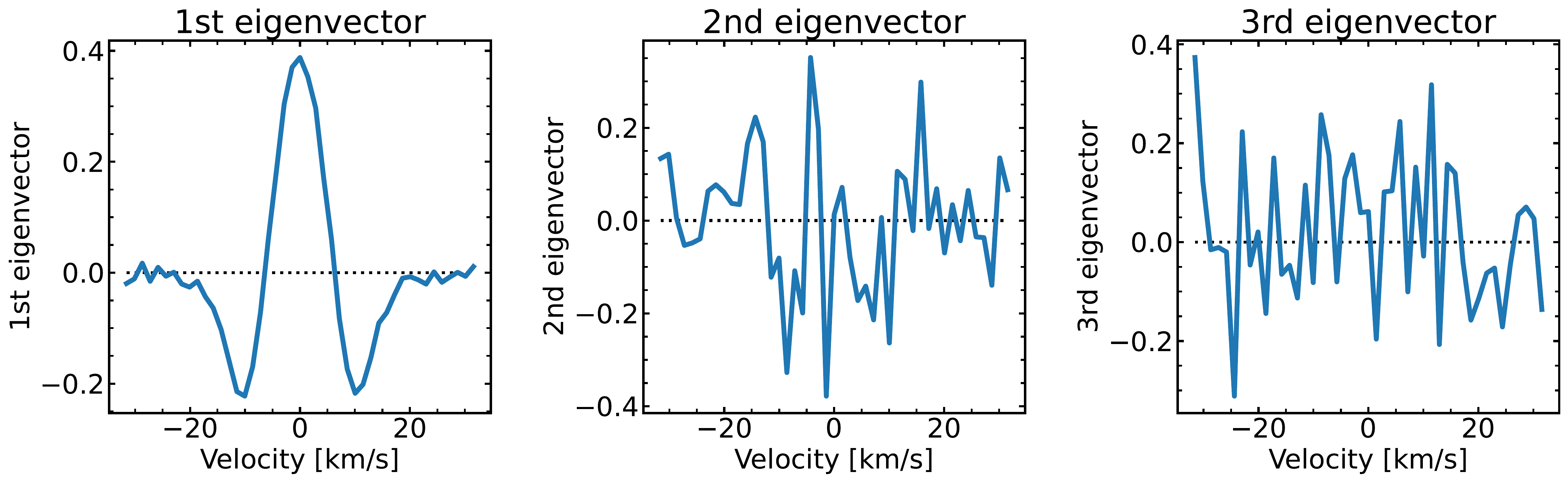} \\
    \includegraphics[width=\columnwidth, trim={0 400 0 0}, clip]{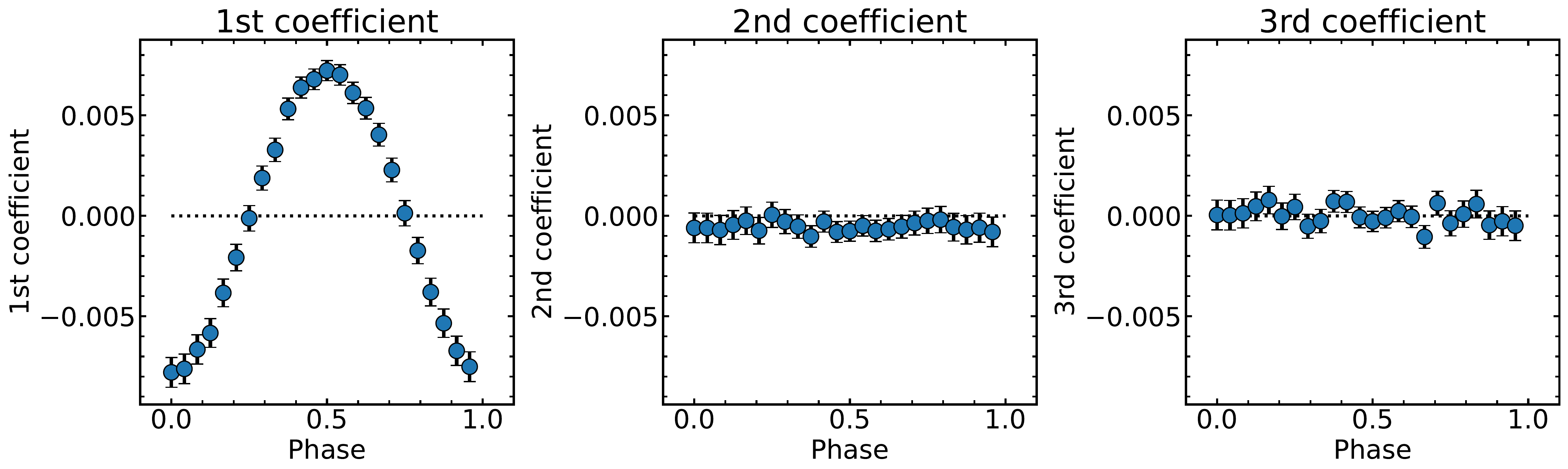} 
    	\end{minipage}
    \caption{The map (a.), Stokes~$V$ time series (b.) and PCA analysis of the mean-subtracted Stokes~$V$ profiles (c.) for a tilted toroidal dipole presented in the same format as in Fig.~\ref{fig:Map_tiltDipol}.}
    \label{fig:Map_torDipol}
\end{figure*}

\begin{figure*} 
	\begin{flushleft}
	\textbf{a.} \hspace{2.5cm} \textbf{b.} \hspace{3.6cm} \textbf{c.}
	\end{flushleft}
    \begin{minipage}{0.11\textwidth}
    \centering
    \includegraphics[height=0.85\columnwidth, angle=270, trim={140 0 0 29}, clip]{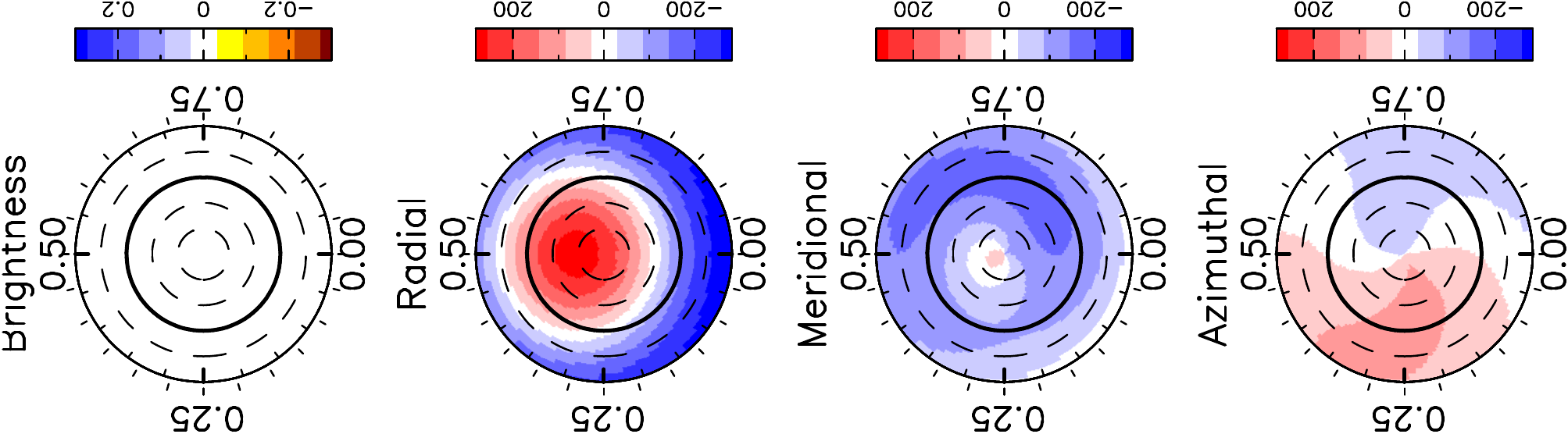} 
	\includegraphics[width=\columnwidth, angle=180, trim={470 130 12 0}, clip]{Figures/MapN3.pdf}
	\end{minipage}
    \begin{minipage}{0.25\textwidth}
    \centering
    \includegraphics[width=\columnwidth, angle=0, trim={0 0 0 0}, clip]{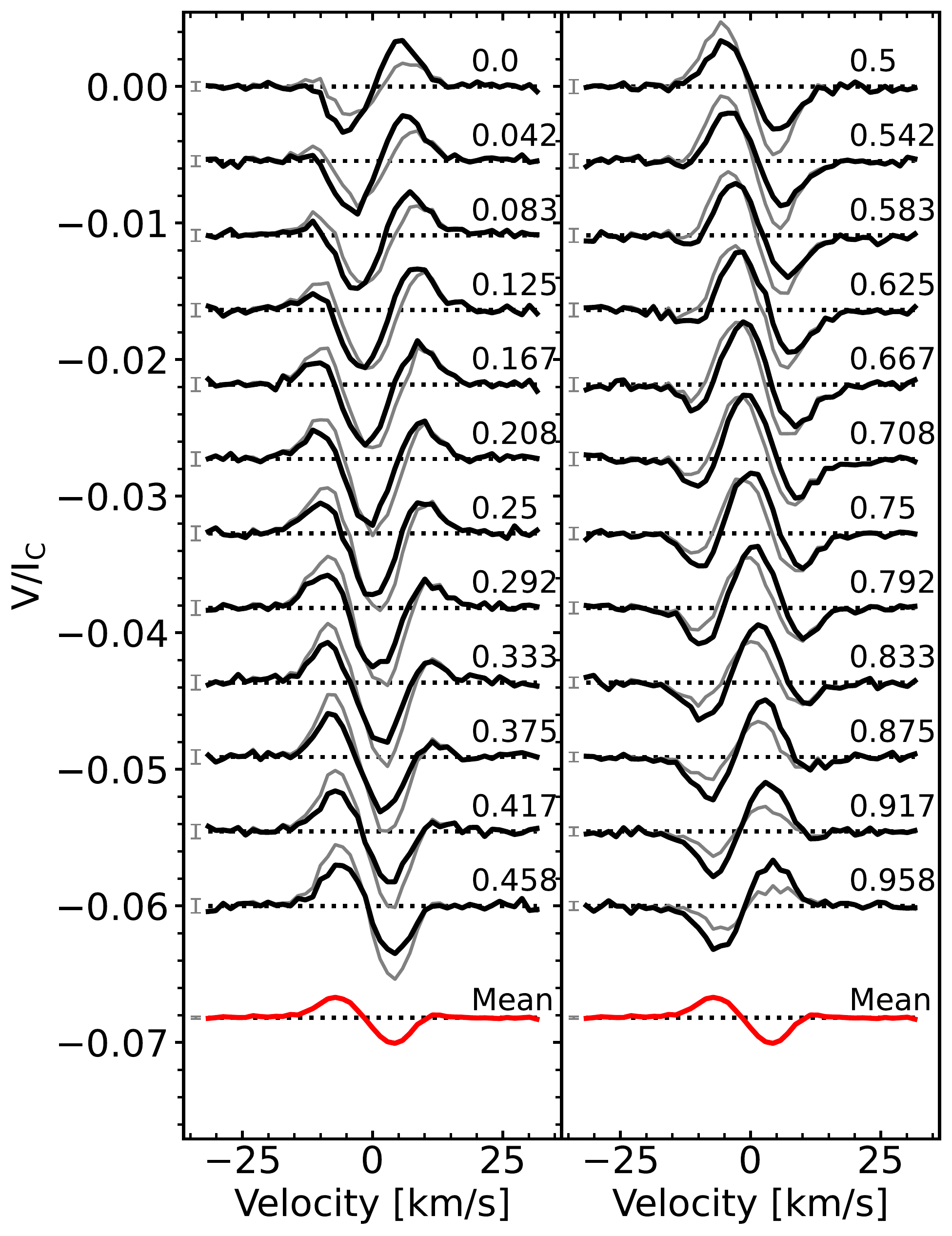} 
    	\end{minipage}
    \begin{minipage}{0.6\textwidth}
    \centering
    \includegraphics[width=\columnwidth, trim={0 400 0 0}, clip]{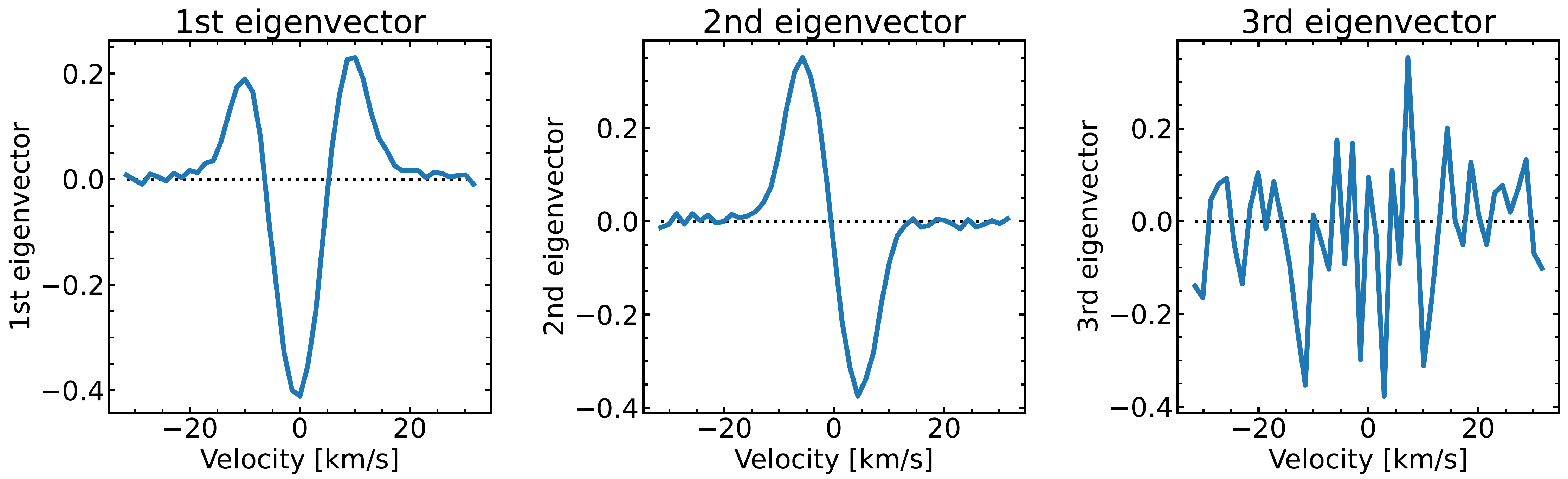} \\
    \includegraphics[width=\columnwidth, trim={0 400 0 0}, clip]{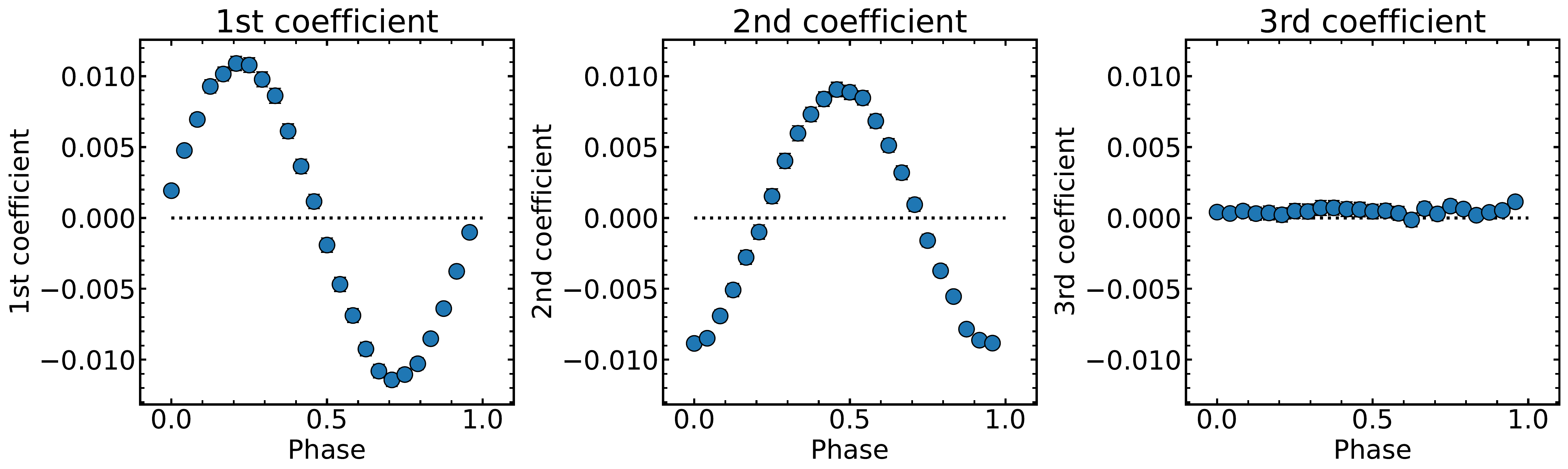} 
    	\end{minipage}
    \caption{The map (a.), Stokes~$V$ time series (b.) and PCA analysis of the mean-subtracted Stokes~$V$ profiles (c.) for a tilted dipole having a mixed poloidal and toroidal field shown in the same format as in Fig.~\ref{fig:Map_tiltDipol}. This dipole represents the sum of the tilted poloidal dipole shown in Fig.~\ref{fig:Map_tiltDipol} with one tenth of the tilted toroidal dipole displayed in Fig.~\ref{fig:Map_torDipol}.}
    \label{fig:Map_mixDipol}
\end{figure*}

Looking at the case of a purely toroidal large-scale field, tilted with respect to the rotation axis, see Fig.~\ref{fig:Map_torDipol}, one can see that the mean profile $\overline{V}$ is now symmetric with respect to the line centre as expected. The first eigenvector is symmetric and proportional to the second derivative of the Stokes~$I$ profiles. The further eigenvectors contain noise. For noise-free simulations, further derivatives of the Stokes~$I$ profiles appear, with the first derivative no longer showing up, except for highly titled toroidal configurations with tilt angles near $\psi = 90^\circ$. Strong non-axisymmetric toroidal dipoles are therefore easily detectable even if they are not likely to occur.

Let us now describe the case of a tilted dipolar field with a poloidal and toroidal component. For the example dipole shown in Fig.~\ref{fig:Map_mixDipol}, we sum the tilted poloidal field presented in Fig.~\ref{fig:Map_tiltDipol} with one tenth of the tilted toroidal field displayed in Fig.~\ref{fig:Map_torDipol}. 
When the toroidal field is added to the poloidal dipole the mean profile $\overline{V}$ and the eigenvectors are no longer purely symmetric or antisymmetric, and they become asymmetric. By decomposing the mean profile $\overline{V}$ into its symmetric and antisymmetric part, see Fig.~\ref{fig:Map_mixDipol}b or Fig.~\ref{fig:SymAntiSym}b in detail, we can conclude in this example that there is an axisymmetric toroidal component but the poloidal axisymmetric field is dominant. For the non-axisymmetric field, the initially symmetric eigenvector turns asymmetric, see Fig.~\ref{fig:Map_mixDipol}c. The azimuthal field peaks no longer 0.25 apart from the radial field in phase due to the additional toroidal field, see Fig.~\ref{fig:Map_mixDipol}a. If we increase the strength of the toroidal field, the initially antisymmetric mean profile and eigenvector becomes more and more symmetric (with a shape evolving from a first derivative to a second derivative of the Stokes~$I$) and the initially symmetric eigenvector turns antisymmetric (with a shape evolving from a second derivative to a first sometimes in combination with a third derivative of the Stokes~$I$ profile).  

An asymmetric eigenvector indicates that the dipole is no longer a standard poloidal dipole with $\beta = \gamma = 0$. This information is important for simulating the stellar wind using magnetic field extrapolations as some models made certain assumptions for $\beta$ and $\gamma$, e.g.\ the potential field source surface model \citep{Altschuler1969, Jardine2002}. Indications for $\beta$ and $\gamma$ from the PCA method become especially relevant, as the $\ell=1$ modes are the most important modes for the stellar wind at larger distances \citep{Jardine2017}.

In addition to the eigenvectors, the coefficients can also indicate if the field is indeed a standard poloidal dipole ($\beta=\gamma=0$). For these dipoles, the phase difference between the coefficient peaks of the antisymmetric and symmetric profile is always 0.25 apart in phase and the absolute values of the coefficient maxima and minima are equal $|\mrm{max}| = |\mrm{min}|$, see Fig.~\ref{fig:Map_tiltDipol}c bottom row. When $\beta$ and $\gamma$ depart from zero, the phase difference and discrepancy between $|\mrm{max}|$ and $|\mrm{min}|$ increases. (Note that all this only applies to field components with $\ell=1$ modes and not for more complex fields, discussed in the following section.) In our $\gamma \neq 0$ topology example of Fig.~\ref{fig:Map_mixDipol}, the differences are still very small as the ratio between the toroidal and poloidal energy is only 0.03.

The PCA analysis reported in this section is valid for several stars showing dipolar large-scale fields, for example: late and mid M~dwarfs, e.g. AD~Leo (Gl388, \citealt{Morin2008}) or WX~Uma (GJ~412B, \citealt{Morin2010}); or sun-like stars at activity minimum \citep{BoroSaikia2018, Lehmann2021}. In the next section, we will present the PCA analysis of more complex magnetic field topologies.

\section{General case: A complex field}
\label{Sec:Complex}

\begin{figure}
	   \textbf{a.}	
	\centering
	\includegraphics[height=0.85\columnwidth, angle=270, trim={0 180 0 0}, clip]{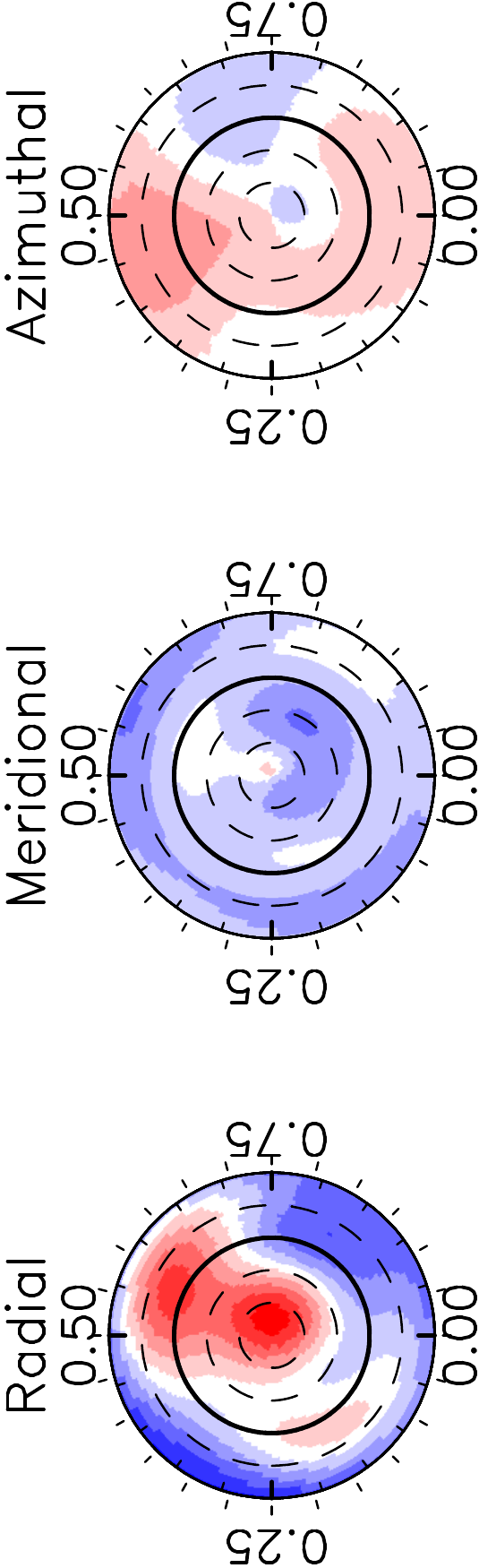} 
	\includegraphics[width=0.25\columnwidth, angle=270, trim={455 130 12 0}, clip]{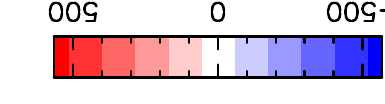} \\
       \vspace{0.1cm}
\hrule
     	\vspace{0.1cm}
    \includegraphics[width=\columnwidth, trim={0 400 0 0}, clip]{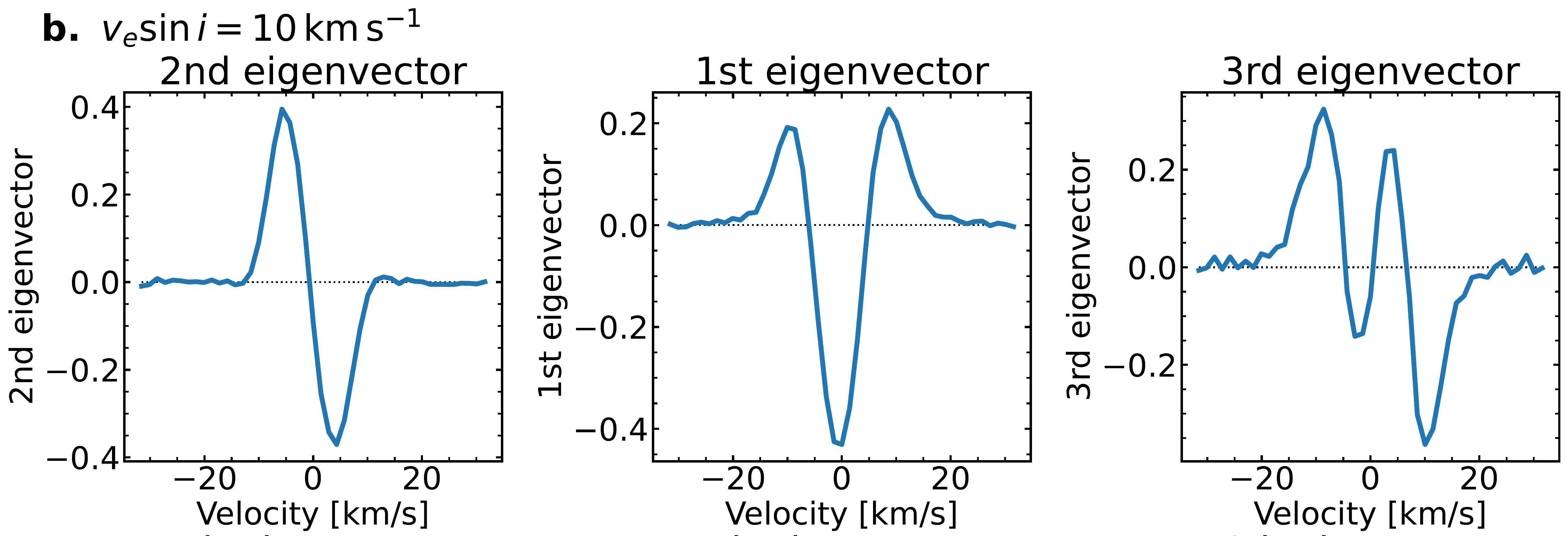} \\
    \includegraphics[width=\columnwidth, trim={0 400 0 0}, clip]{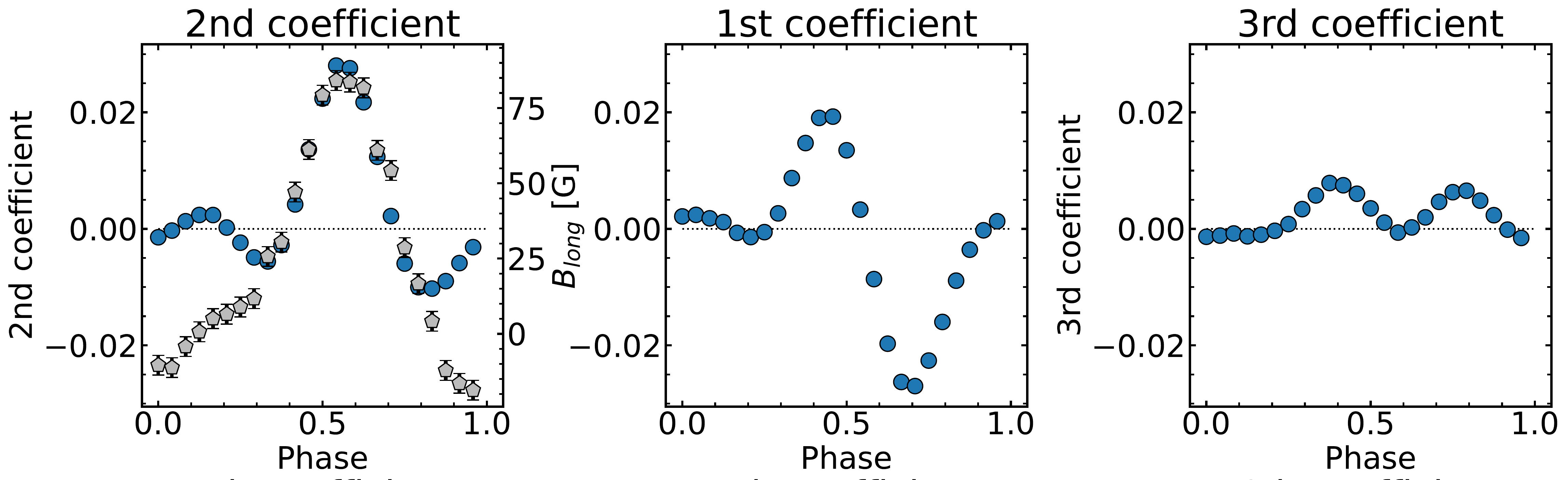} \\        
       \vspace{0.1cm}
\hrule
     	\vspace{0.1cm}
    \includegraphics[width=\columnwidth, trim={0 400 0 0}, clip]{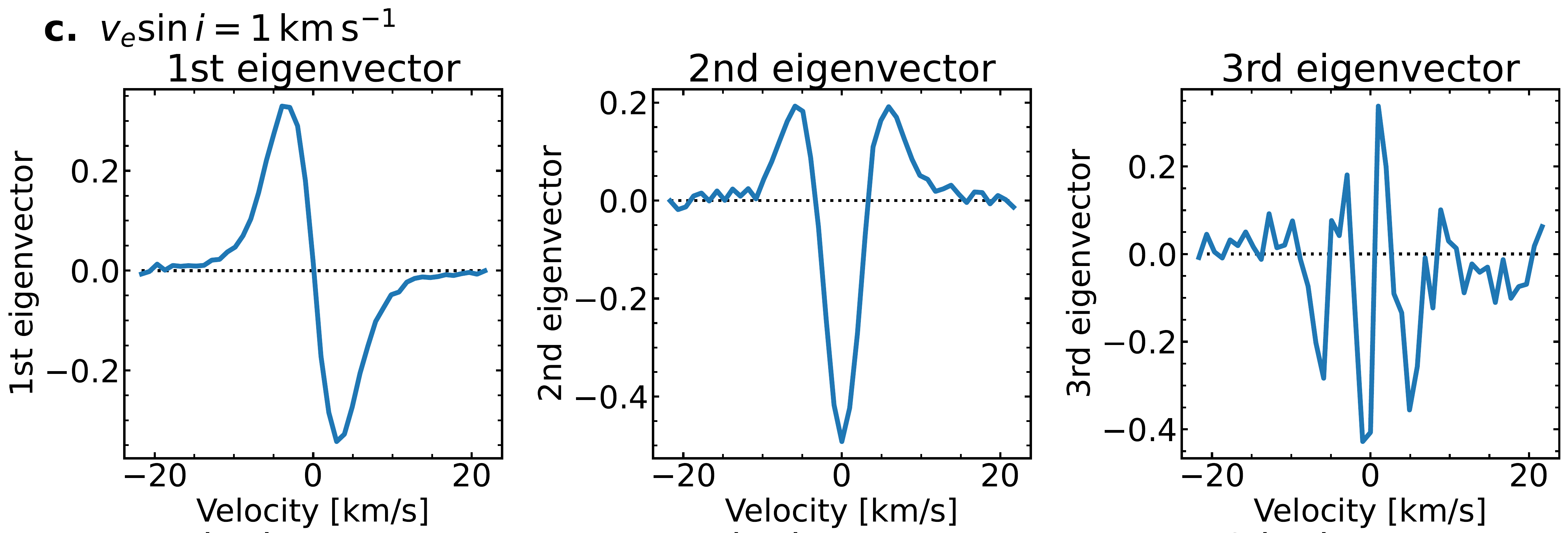} \\
    \includegraphics[width=\columnwidth, trim={0 400 0 0}, clip]{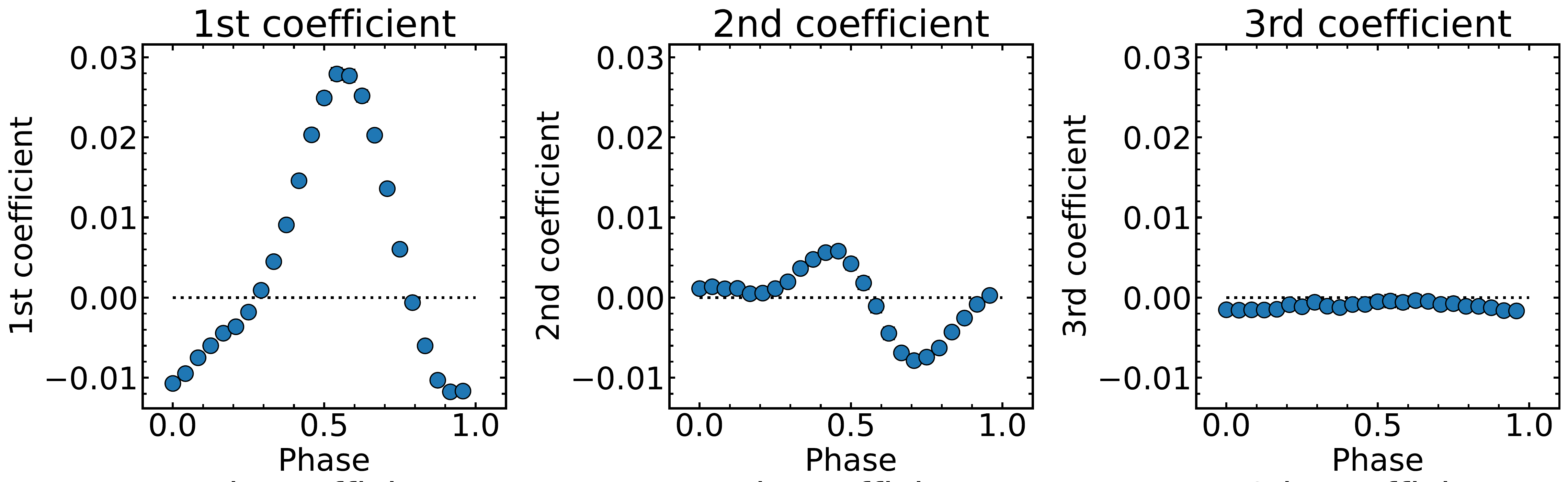} \\   
       \vspace{0.1cm}
\hrule
     	\vspace{0.1cm}
    \includegraphics[width=\columnwidth, trim={0 400 0 0}, clip]{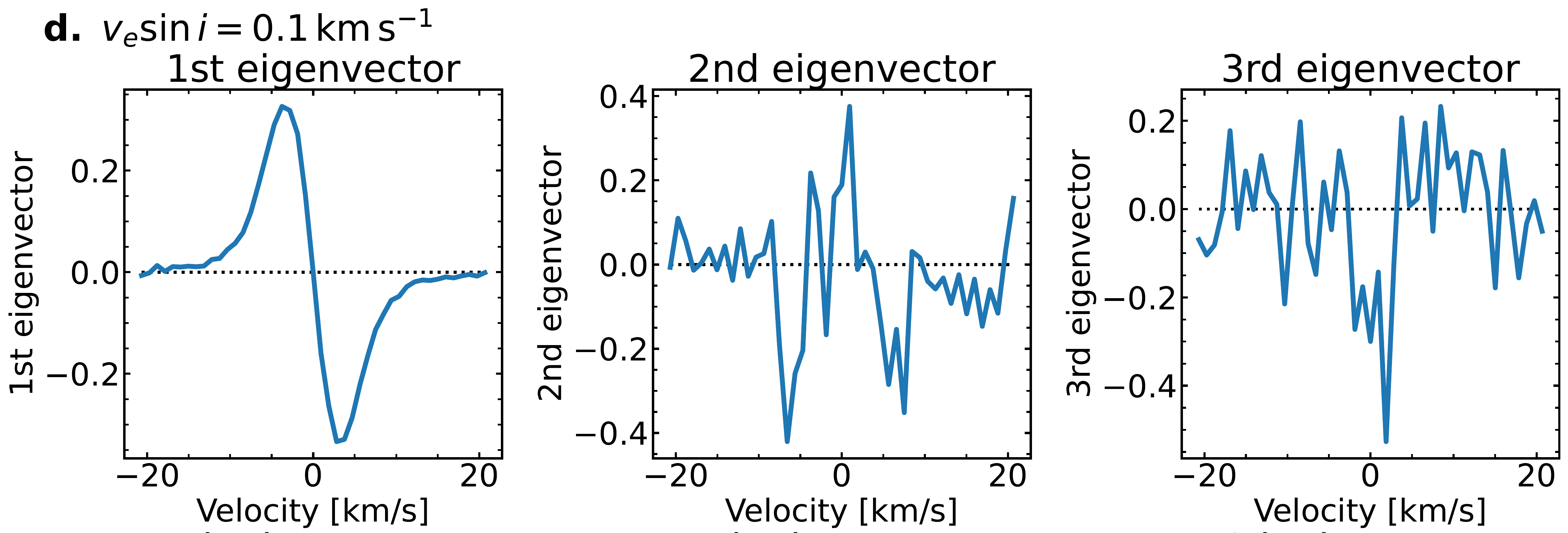} \\
    \includegraphics[width=\columnwidth, trim={0 400 0 0}, clip]{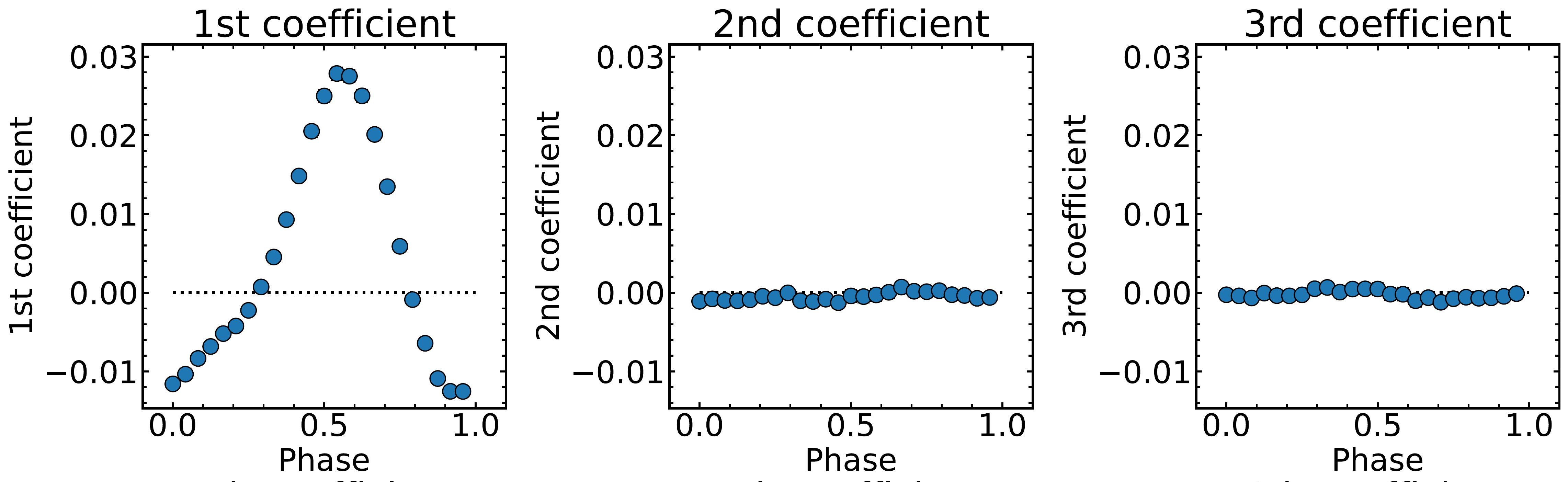} \\
       \vspace{0.1cm}
\hrule
     	\vspace{0.1cm}
	\includegraphics[width=0.32\columnwidth, angle=0, trim={0 0 0 0}, clip]{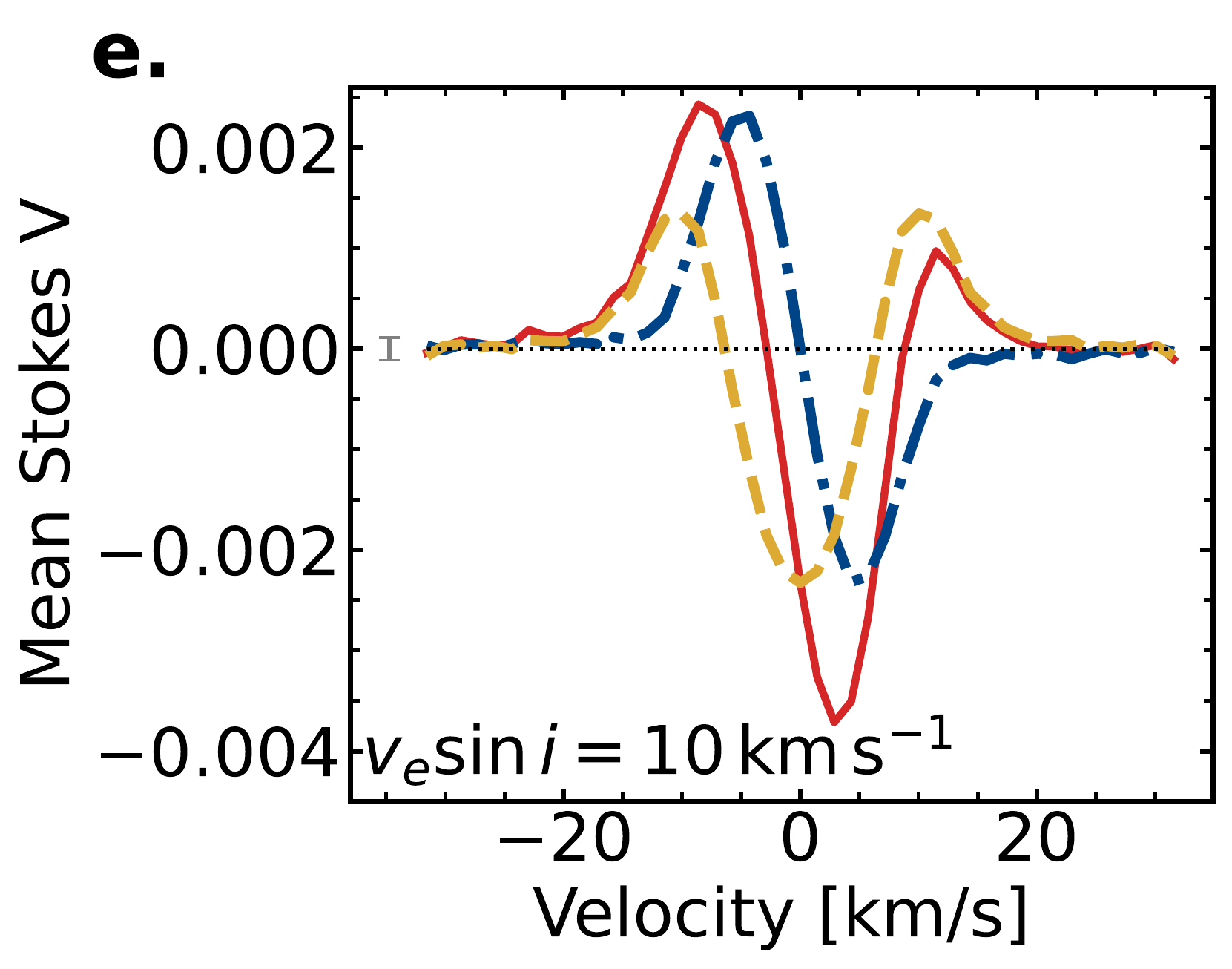}
	\includegraphics[width=0.32\columnwidth, angle=0, trim={0 0 0 0}, clip]{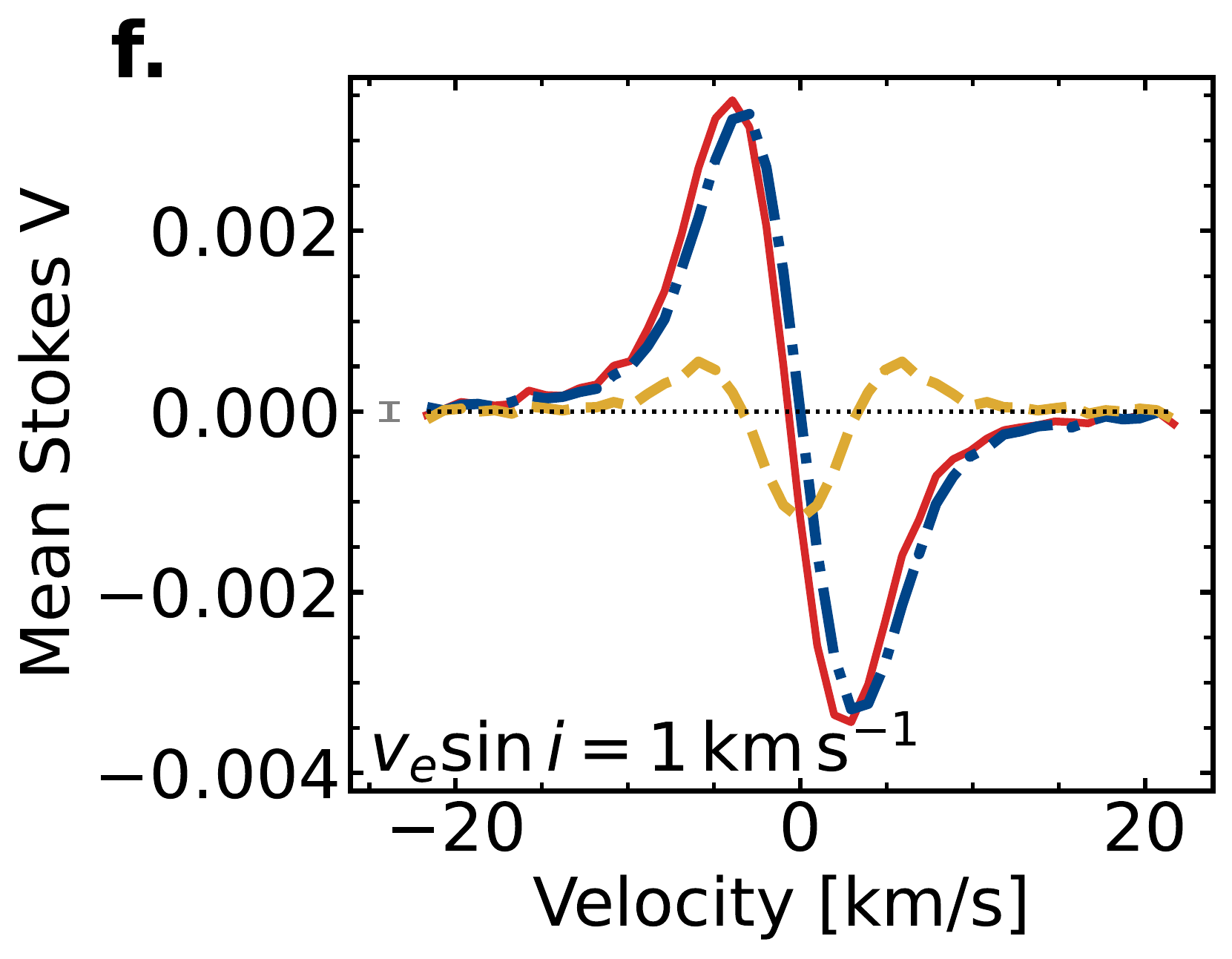}
	\includegraphics[width=0.32\columnwidth, angle=0, trim={0 0 0 0}, clip]{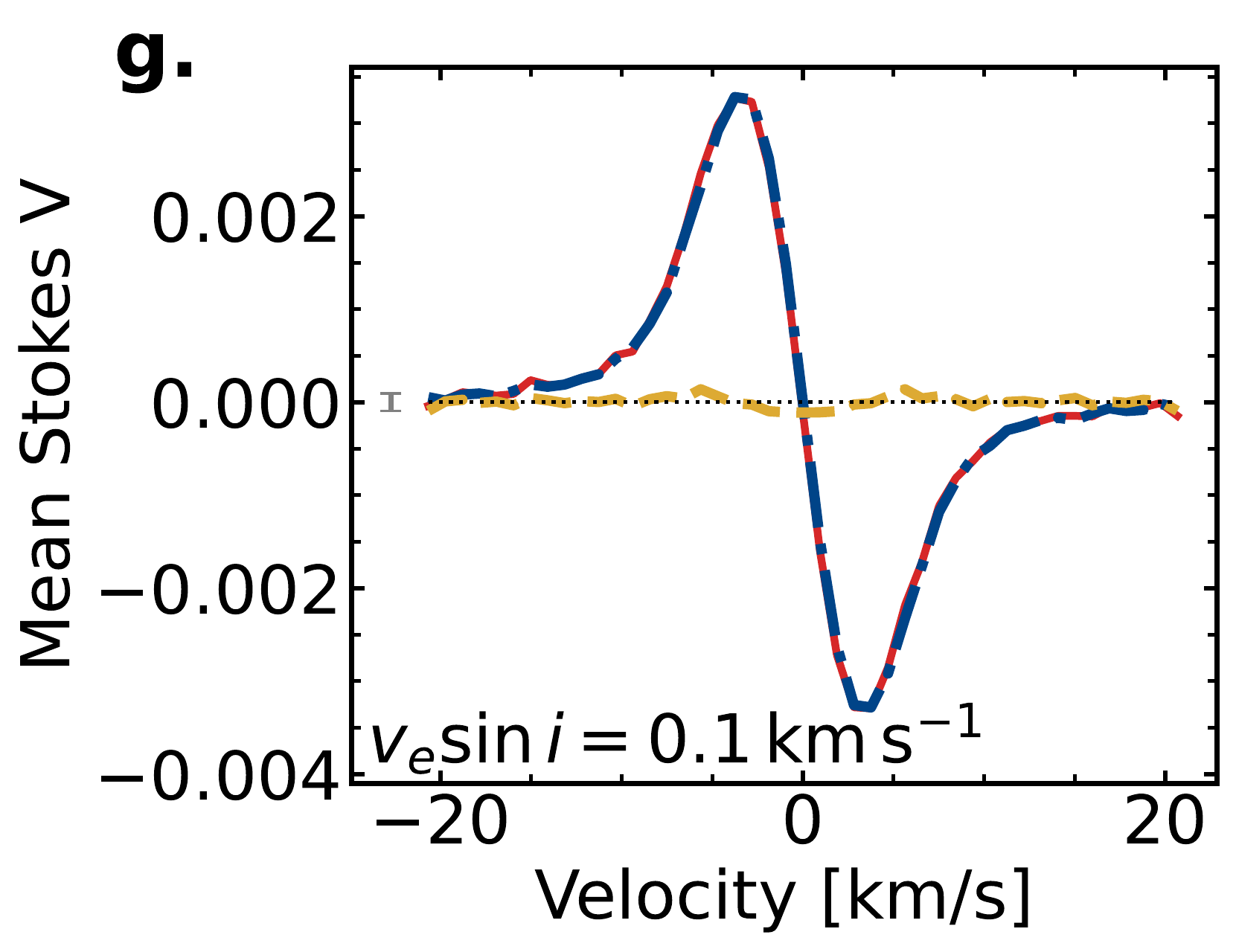}\\
    \caption{The map (a.), the PCA analysis for $v_e \sin i = 10\,\kms$ (b.), $1.0\,\kms$ (c.) and $0.1\,\kms$ (d.) and their mean profiles (e.-g.) using the same format as in  Fig.~\ref{fig:SymAntiSym} and \ref{fig:Map_tiltDipol}. For the PCA analysis of $v_e \sin i = 10\,\kms$ (b.) we swap the order of the 1st and 2nd eigenvector and coefficient. Additionally, the longitudinal field (grey pentagons) is shown in the panel of the 2nd coefficient. }
    \label{fig:ComplexMap}
\end{figure}

We ran the PCA analysis for about $200$ different field topologies ($\ell = 1-28$) including topologies based on real observations and completely randomly generated magnetic field maps. For a subsample of these topologies, we analysed the influence of different projected equatorial velocities ($v_e \sin i =0.001-100\,\kms$), inclinations ($i = 10^{\circ}-80^{\circ}$) and signal-to-noise ratios (SNR = 100-100\,000). In the following, we present our main results using one example field topology.

 PCA can reliably indicate whether the stellar magnetic field is more complex than a dipole (meaning several multipoles of different strengths and tilt angles). Fig.~\ref{fig:ComplexMap} presents the maps, PCA results, longitudinal field and mean profiles for a simulated complex magnetic field ($\ell,m \leq 3$) for different $v_e \sin i$ values. 

Fig.~\ref{fig:ComplexMap}b presents the eigenvectors and their coefficients\footnote{We swap here the first two eigenvectors and coefficients to better compare the antisymmetric and symmetric profiles of different $v_e \sin i$ in columns of Fig.~\ref{fig:ComplexMap}.} if the model star is observed with $v_e \sin i = 10\,\kms$. 
The first three eigenvectors are asymmetric. Unlike for the dipole field, asymmetric eigenvectors can occur for any complex topology even for magnetic field configurations with $\beta = \gamma = 0$, e.g., if different multipoles are shifted to each other, see Fig.~\ref{fig:Map_ComplexPol} in the appendix.

The coefficients can easily unveil if the magnetic field topology is dipolar-like or more complex. As soon as the coefficient trend with phase differs from a sinusoidal trend with two extrema 0.5 apart in phase, we can conclude that the field topology must be more complex than a dipole. In Fig.~\ref{fig:ComplexMap}b the coefficients show clearly three or more extrema indicating a complex field topology. The coefficient extrema are usually tracing individual, large magnetic features or spots, that are often related to the poles of the low order multipoles. In our example, the coefficient extrema of the first and second eigenvector correlate with the phases of the most dominant azimuthal and radial features.

With decreasing $v_e \sin i$, the coefficients become less sensitive to individual magnetic field features. In Fig.~\ref{fig:ComplexMap}c and d we display the PCA results for the same map (Fig.~\ref{fig:ComplexMap}a) but for a $v_e \sin i = 1\,\kms$ and $0.1\,\kms$ instead of $10\,\kms$. With decreasing $v_e \sin i$, fewer eigenvectors contain imformation and the coefficient show less variability with phase, see Fig.~\ref{fig:ComplexMap}b,c,d. The lower the $v_e \sin i$, the larger the resolution elements on the stellar surface, where magnetic features of opposite polarity cancel and therefore the fewer details of the magnetic topology can be seen. 

PCA can unveil the complexity of the non-axisymmetric field to lower $v_e \sin i$ values than, e.g., the longitudinal field. We display the longitudinal field (grey pentagons) in the panel of the second coefficient in Fig.~\ref{fig:ComplexMap}b bottom left. The longitudinal field is the same for all three $v_e \sin i$ values and shows a simple, more or less sinusoidal trend with two extrema, that we would expect to see for a dipolar field topology.
With decreasing $v_e \sin i$, the coefficient of the most antisymmetric eigenvector become more and more similar to the longitudinal field, see Fig.~\ref{fig:ComplexMap}b,c,d bottom left. For a ten times lower $v_e \sin i = 1\,\kms$, the coefficient of the first eigenvector already shows the same trend as the longitudinal field but the coefficient of the second eigenvector indicates, that the field must be complex, see Fig.~\ref{fig:ComplexMap}c bottom middle. It would even be possible to diagnose the field complexity if the phase coverage were worse than the one shown here. The PCA method has a higher sensitivity to the large-scale field topology compared to the longitudinal field. As a result, PCA can trace complex fields even for lower $v_e \sin i$ and SNR levels. PCA's sensitivity to the field complexity increases with the number of eigenvectors containing a signal and therefore for more non-axisymmetric field topologies. 
For a hundred times lower $v_e \sin i = 0.1\,\kms$, also PCA becomes blind for the signatures of the complex field in our example, see Fig.~\ref{fig:ComplexMap}d.

Furthermore, the detection of an axisymmetric toroidal field is affected by $v_e \sin i$, too. The mean profile for $v_e \sin i = 10\,\kms$ indicates straight away, that there is a significant axisymmetric toroidal field, see Fig.~\ref{fig:ComplexMap}e. For higher $v_e \sin i$ values, the antisymmetric and symmetric component of the mean profile will start to appear more complex than the first and second derivative of the Stokes~$I$ profile, but they are still related to the poloidal and toroidal component, respectively.
The symmetric (toroidal) component of the mean profile becomes weaker for decreasing $v_e \sin i$ as seen in Fig.~\ref{fig:ComplexMap}f, g. For very low $v_e \sin i$ values the number of resolution elements on the stellar surface gets too low to generate symmetric Stokes~$V$ signals, so that the toroidal field is no longer detectable.

For a simple quantification: we can detect the toroidal field shown in Fig.~\ref{fig:Map_torDipol} ($B_{\phi,\mrm{max}} = 420\,\mrm{G}, B_{\theta,\mrm{max}} = 350\,\mrm{G}$) down to a $v_e \sin i = 0.1\,\kms$ observed with an inclination angle $i = 75^\circ$ and a LSD profile SNR = 5\,000\footnote{This noise level corresponds to $\approx 3\,\%$ of the maximum Stokes~$V$ amplitude of the time series.}. A ten times weaker toroidal field ($B_{\phi,\mrm{max}} = 42\,\mrm{G}, B_{\theta,\mrm{max}} = 35\,\mrm{G}$), is detectable with a ten times higher $v_e \sin i = 1\,\kms$ in the mean profile assuming the same SNR. 

Determining the symmetric component of the mean profile is a fast and convenient method to detect the (axisymmetric) toroidal large-scale field, e.g.\ in large surveys like the SPIRou Legacy Survey (SLS, \citealt{Donati2020}), or in archives like the PolarBase \citep{PolarBasePetit2014}, especially as magnetic field topologies with high toroidal field fractions are also highly axisymmetric \citep{See2015}.

\section{Tracing evolving large-scale fields}
\label{Sec:EvoMaps}

\begin{figure*}
	\raggedright
	\textbf{a.} \hspace{1.7cm} \textbf{b.} \hspace{3.6cm} \textbf{c.} \hspace{1.3cm} \textbf{d.}\\
	\centering
    \begin{minipage}{0.09\textwidth}
    \centering
    \includegraphics[height=0.85\columnwidth, angle=270, trim={140 0 0 29}, clip]{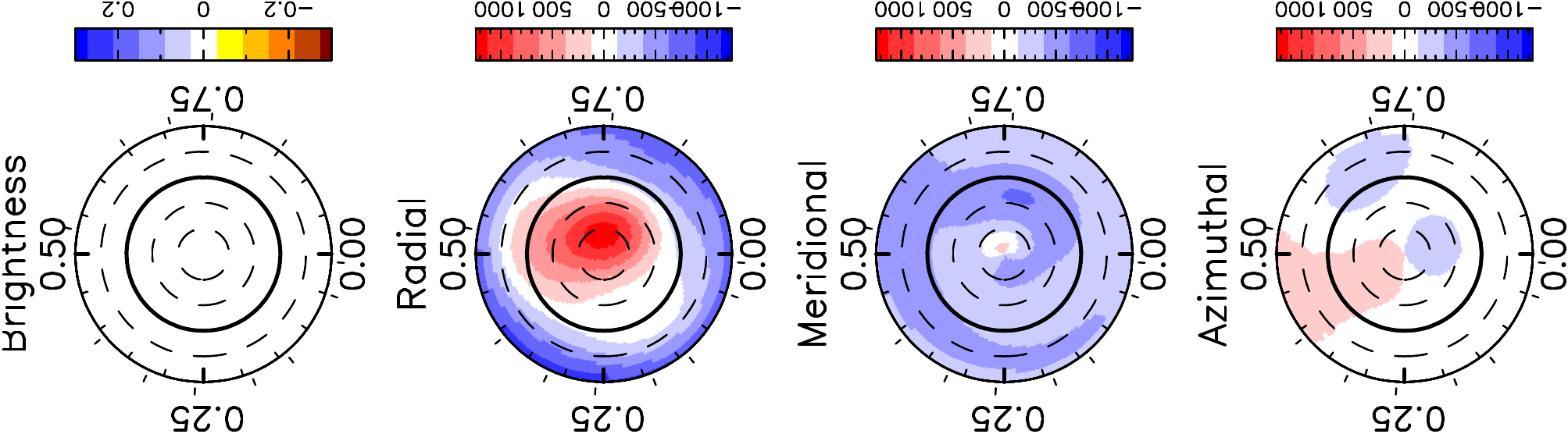} 
	\includegraphics[width=\columnwidth, angle=180, trim={470 130 12 0}, clip]{Figures/MapN9_1.pdf}
    \end{minipage}
    \begin{minipage}{0.22\textwidth}
    \centering
    \includegraphics[width=\columnwidth, angle=0, trim={0 0 0 0 0}, clip]{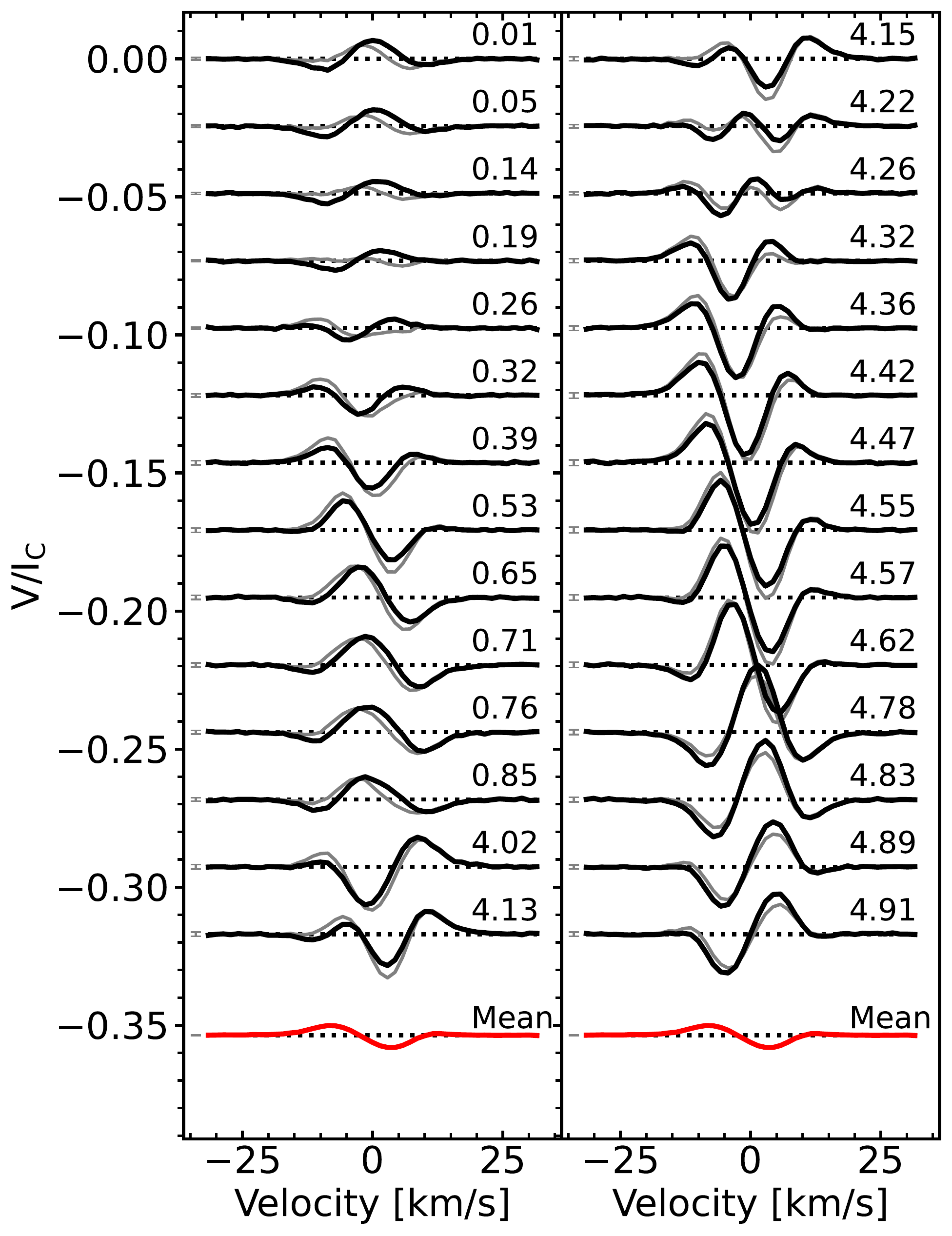}     
    	\end{minipage}
     \begin{minipage}{0.09\textwidth}
     \centering
    \includegraphics[height=0.85\columnwidth, angle=270, trim={140 0 0 29}, clip]{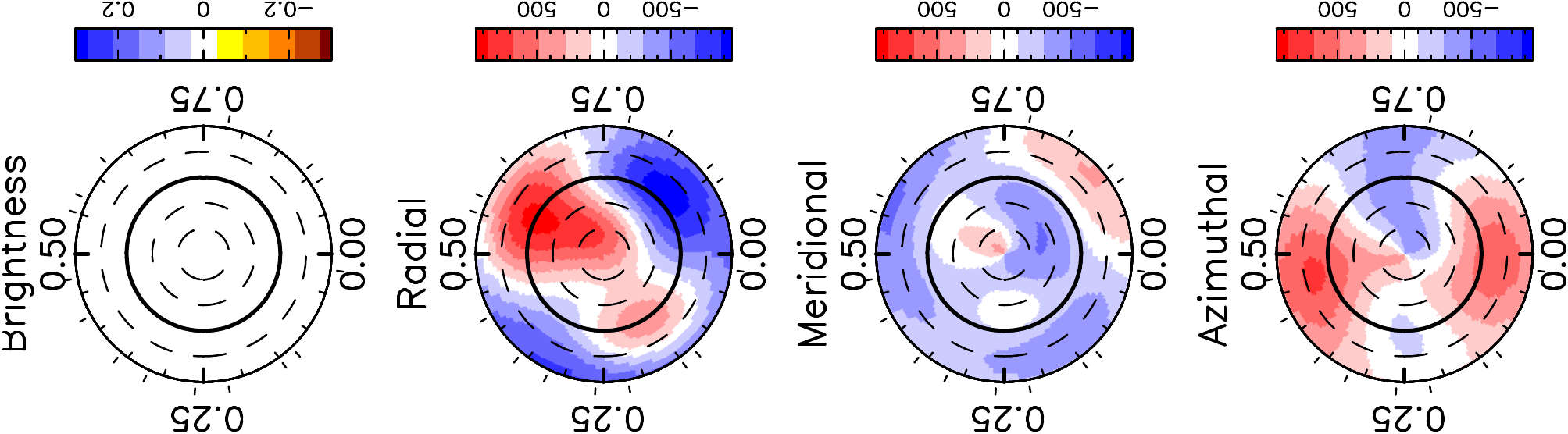} 
	\includegraphics[width=\columnwidth, angle=180, trim={470 130 12 0}, clip]{Figures/MapN9_2.pdf} 
	\end{minipage}
    \begin{minipage}{0.55\textwidth}
    \centering
    \includegraphics[width=\columnwidth, trim={0 400 0 0}, clip]{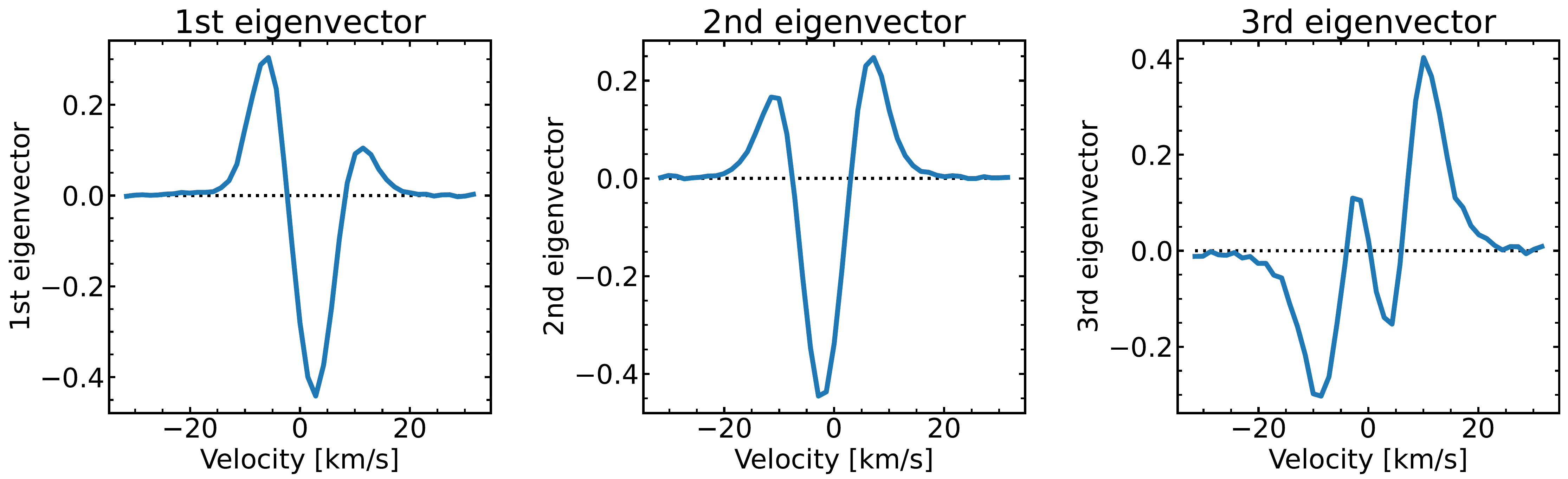} \\
    \includegraphics[width=\columnwidth, trim={0 400 0 0}, clip]{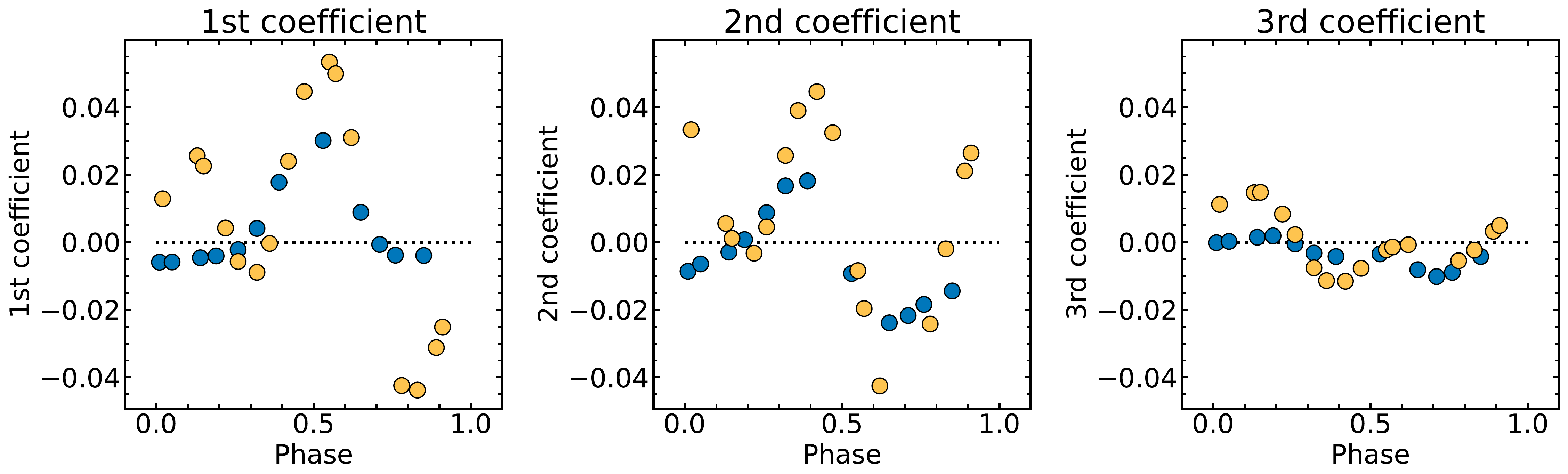} 
    	\end{minipage}
    \caption{A simulated example of an evolving field configuration: The magnetic field evolves from the map in panel a. to the map in panel c. The corresponding Stokes~$V$ profiles can be found in panel b. using the same format as in Fig.~\ref{Fig:AxisymMean}. Panel d. shows the first three eigenvectors (top row) and coefficients (bottom row) found by PCA for the complete sample of 28 Stokes~$V$ profiles shown in b. The coefficients are colour-coded by the rotation cycle. Blue corresponds to the map in panel a. and yellow to the map in panel c.}
    \label{fig:Map_EvoMap}
\end{figure*}

\begin{figure*}
	\begin{flushleft}
	\textbf{a.} \hspace{3.5cm}\textbf{b.} \hspace{3cm} \textbf{c.}  \hspace{5cm} \textbf{d.} \\
	\end{flushleft}
	 \begin{minipage}{0.17\textwidth}
     \centering
     \includegraphics[width=\columnwidth, trim={0 0 0 0}, clip]{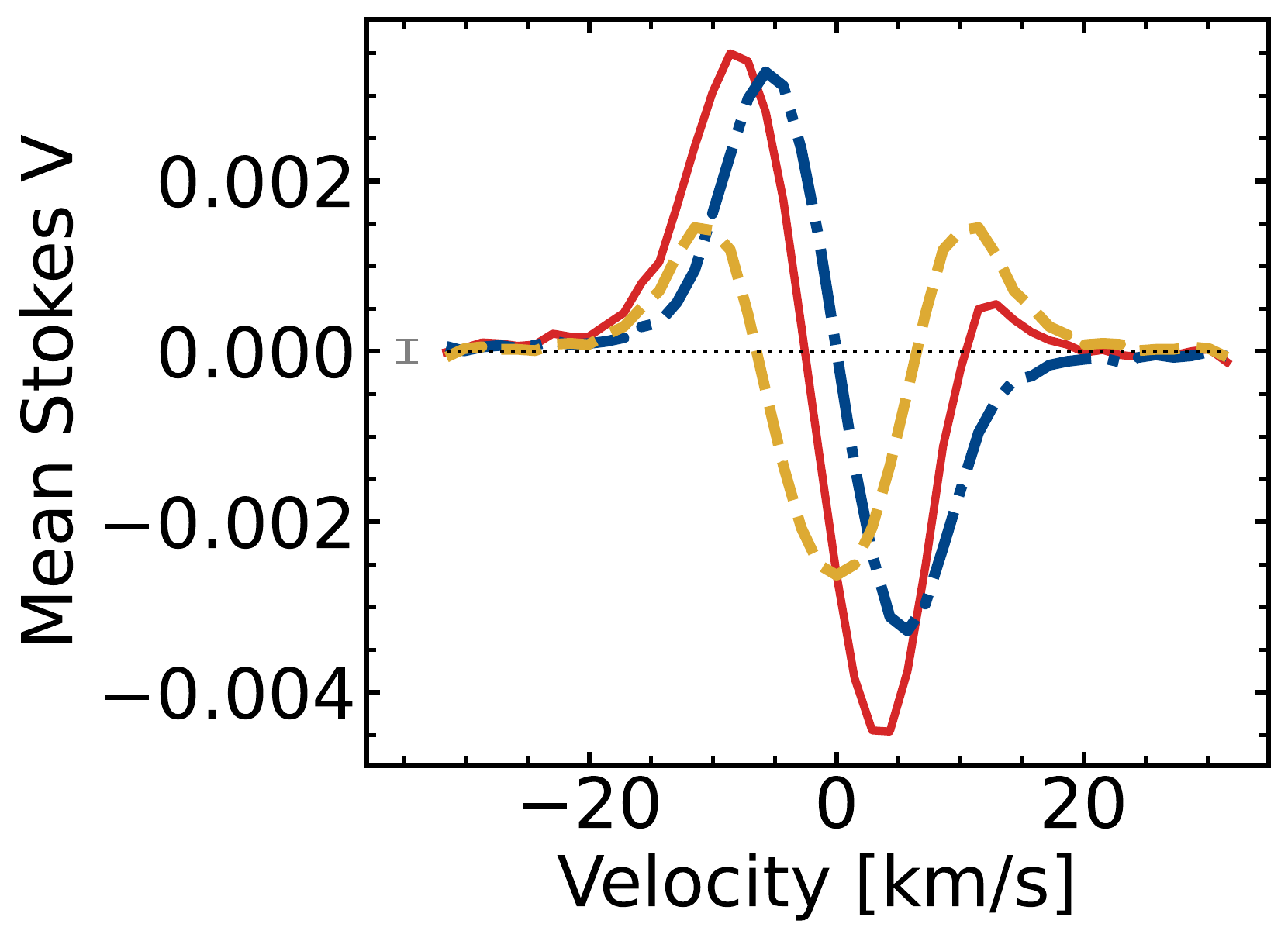} 
     \includegraphics[width=\columnwidth, trim={0 0 0 0}, clip]{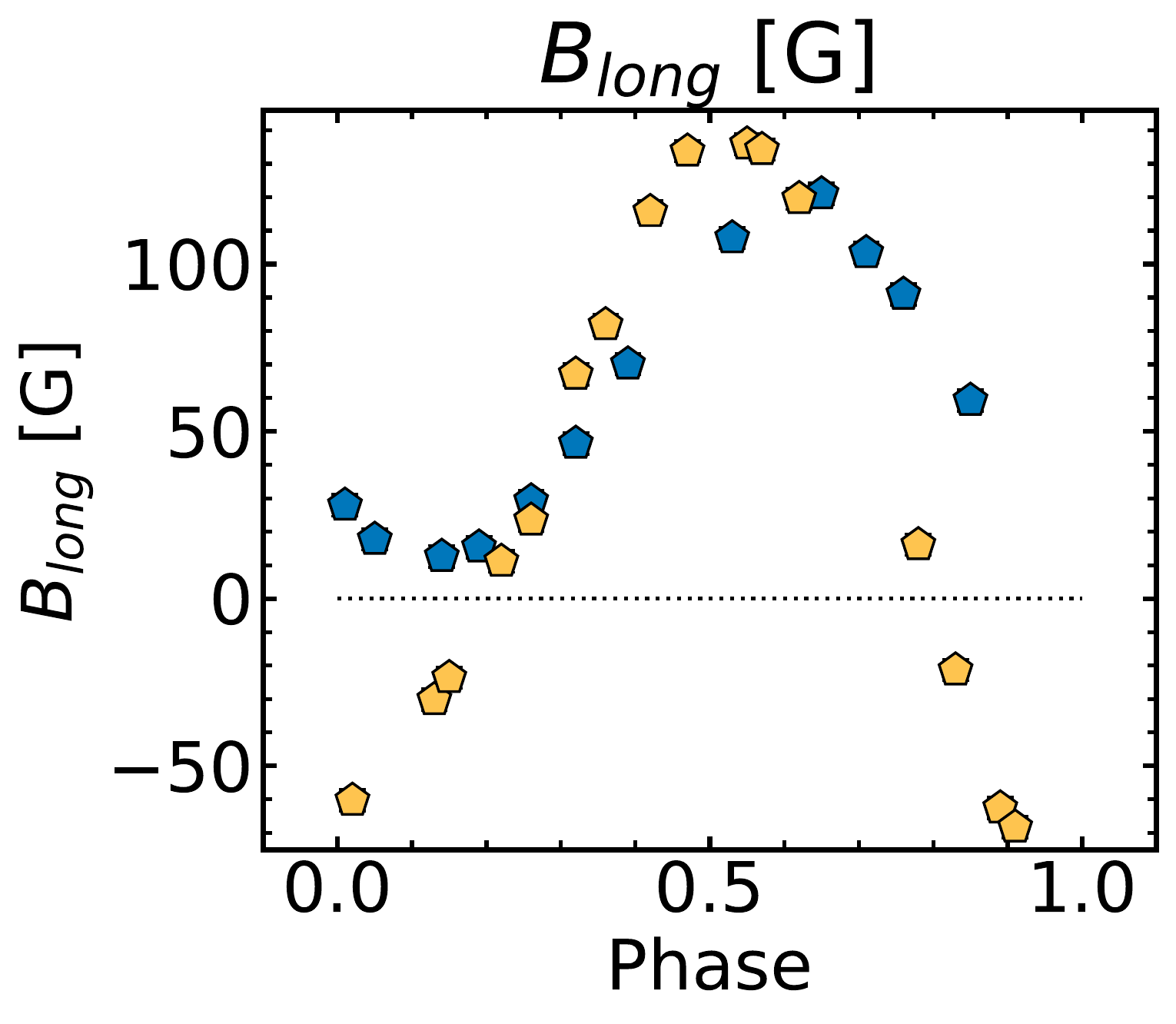} 
     \end{minipage}
        \hspace{0.3cm}
     	\vline
     \begin{minipage}{0.80\textwidth}
	\centering
	\includegraphics[width=0.21\columnwidth, trim={0 0 0 0}, clip]{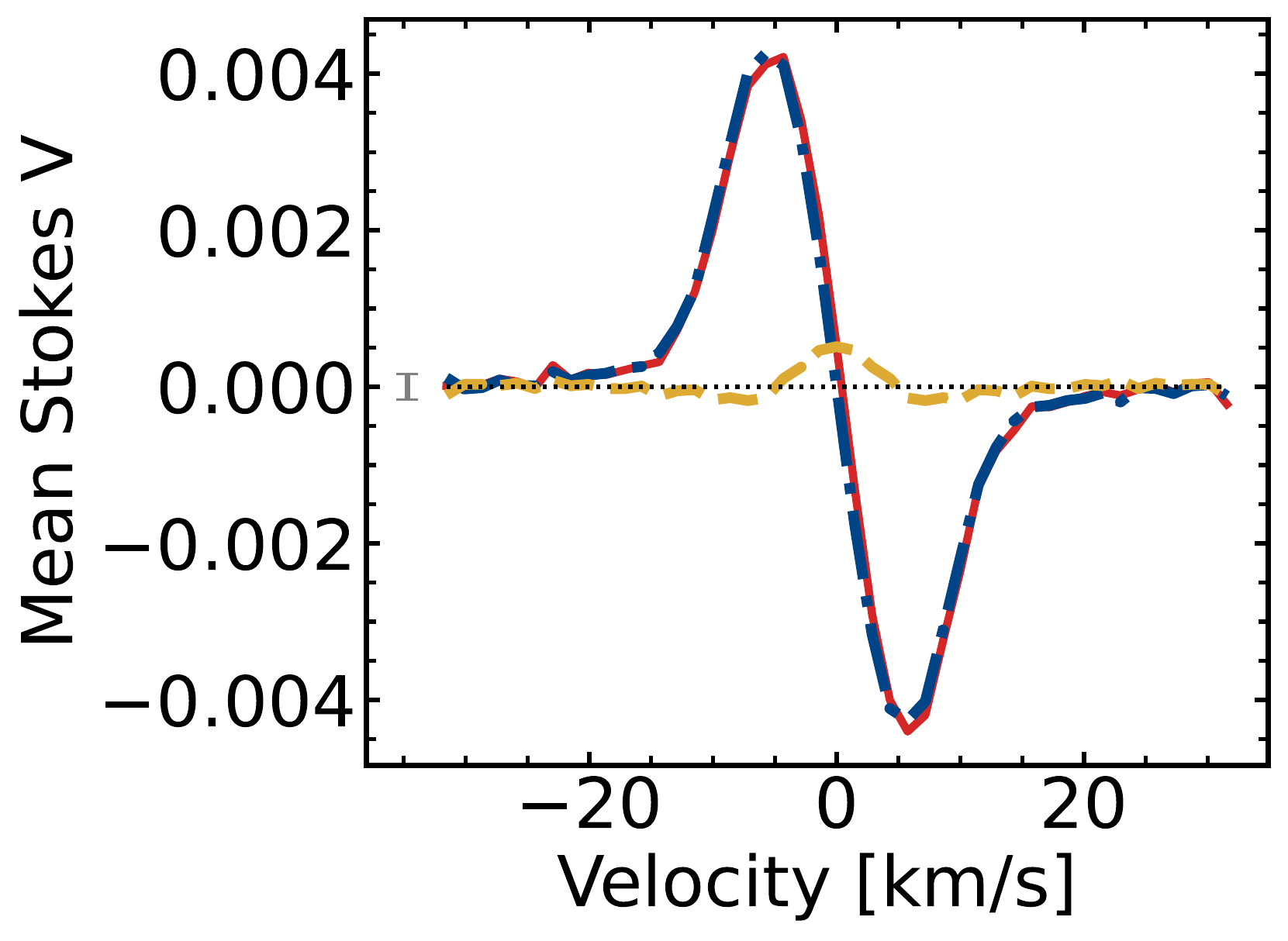} 
	\includegraphics[width=0.37\columnwidth, trim={0 400 440 0}, clip]{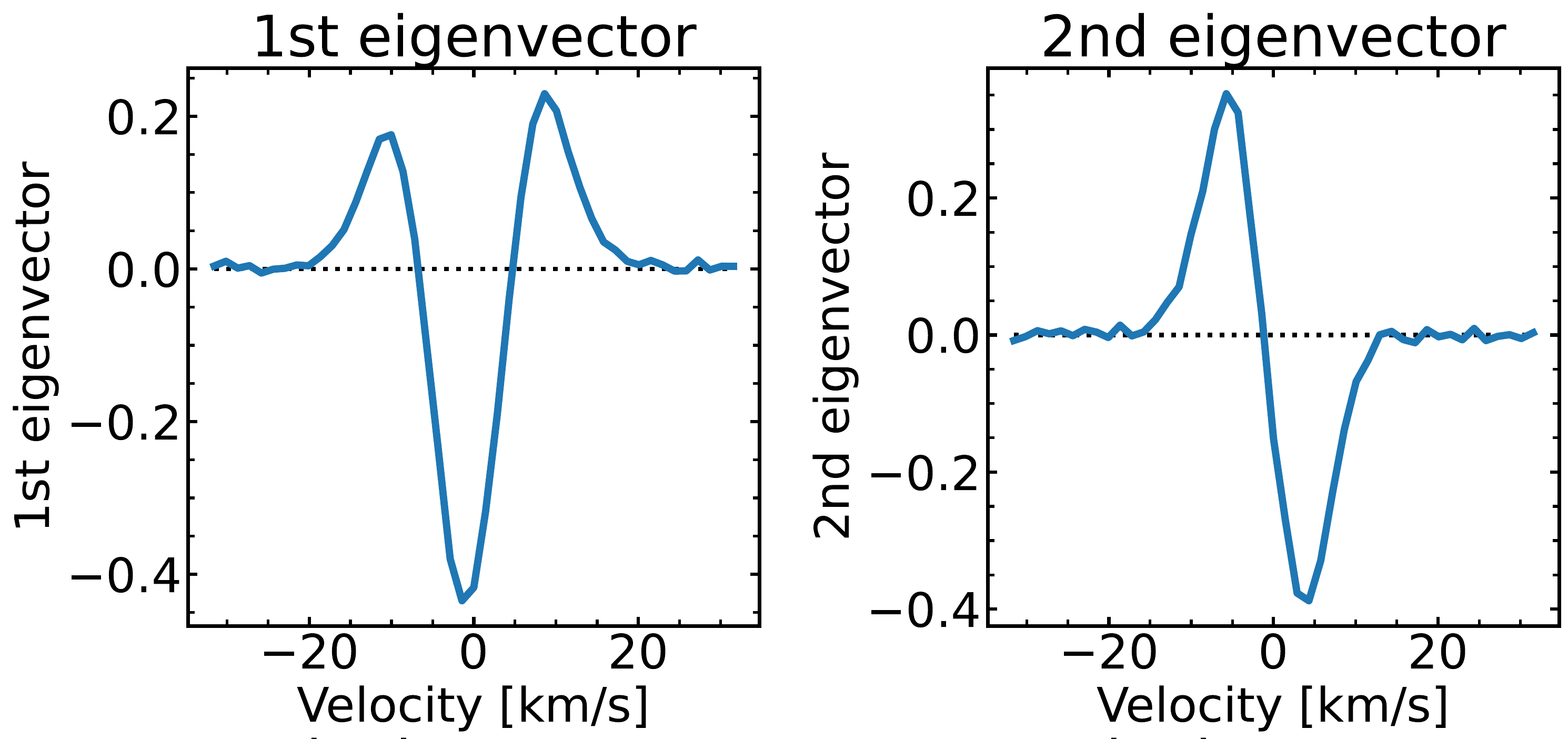} 
     \includegraphics[width=0.37\columnwidth, trim={0 400 440 0}, clip]{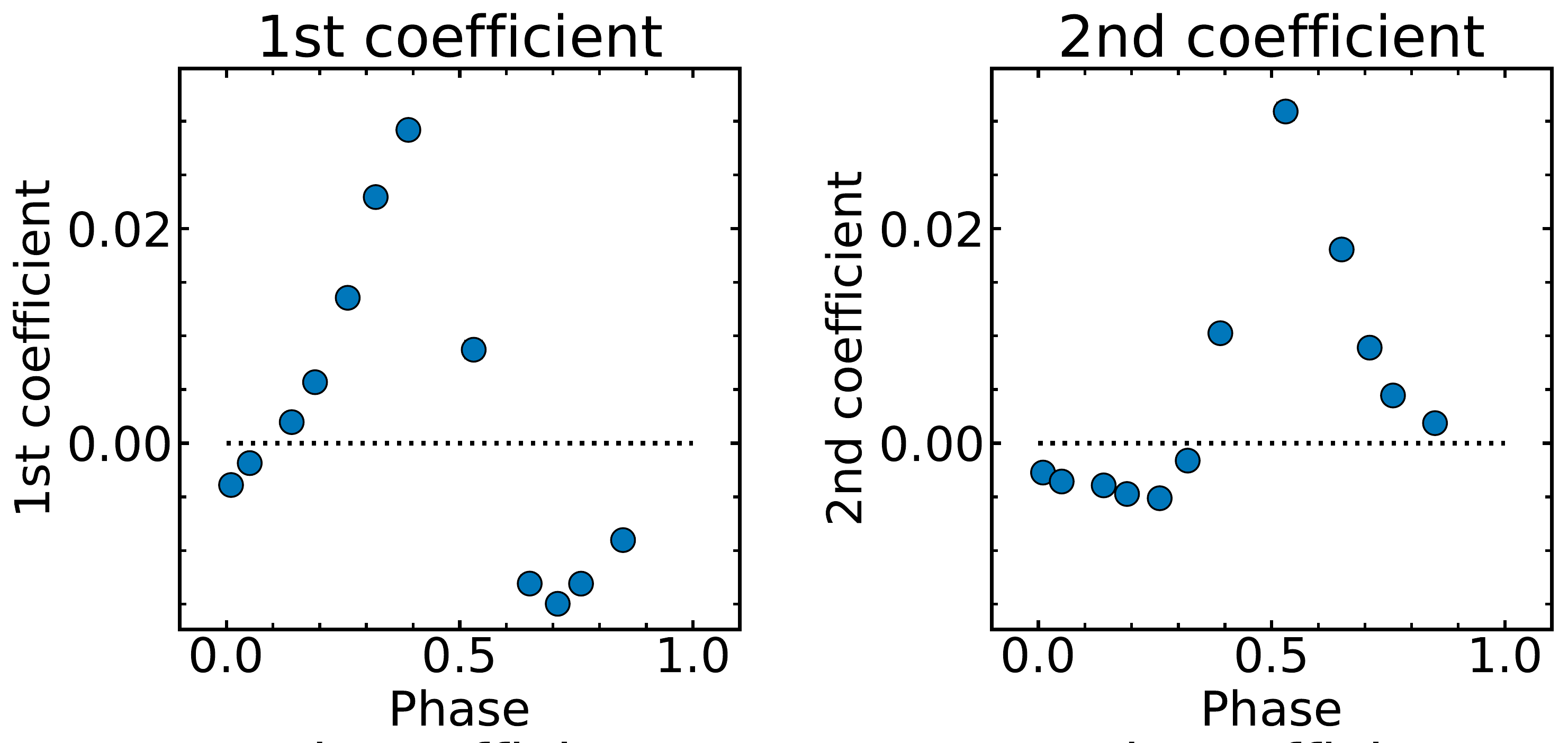} \\
	\includegraphics[width=0.21\columnwidth, trim={0 0 0 0}, clip]{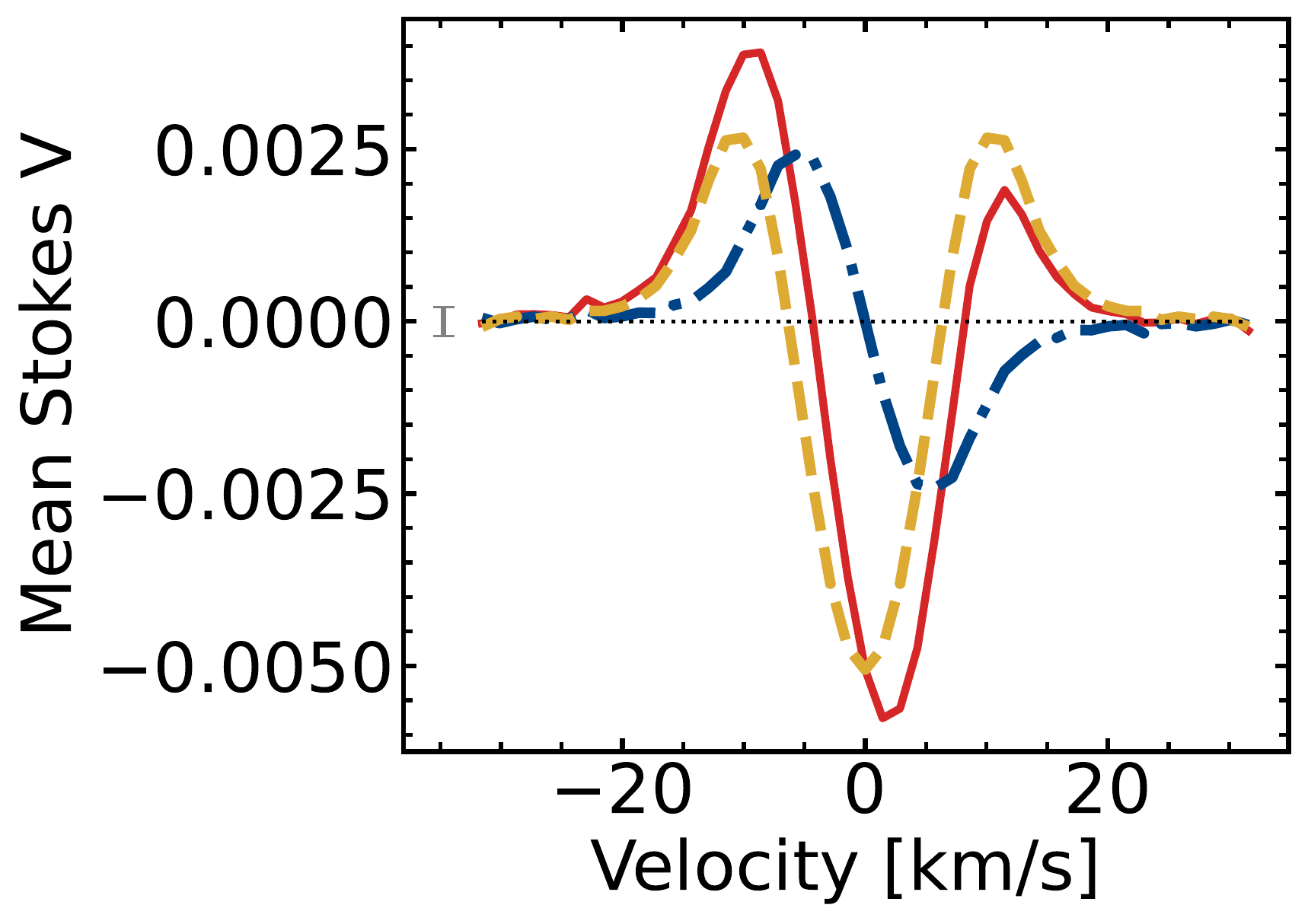} 
	\includegraphics[width=0.37\columnwidth, trim={0 400 440 0}, clip]{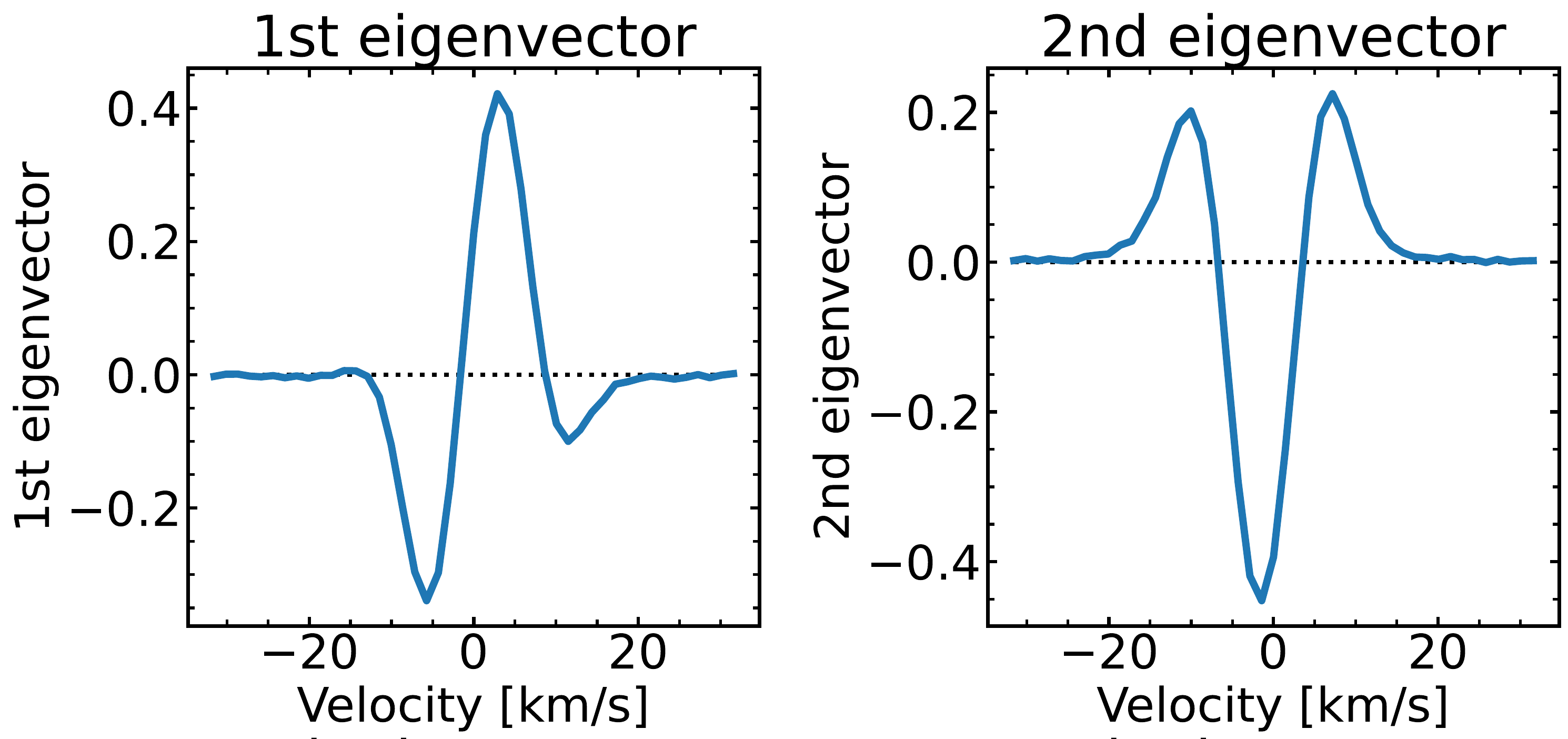} 
	\includegraphics[width=0.37\columnwidth, trim={0 400 440 0}, clip]{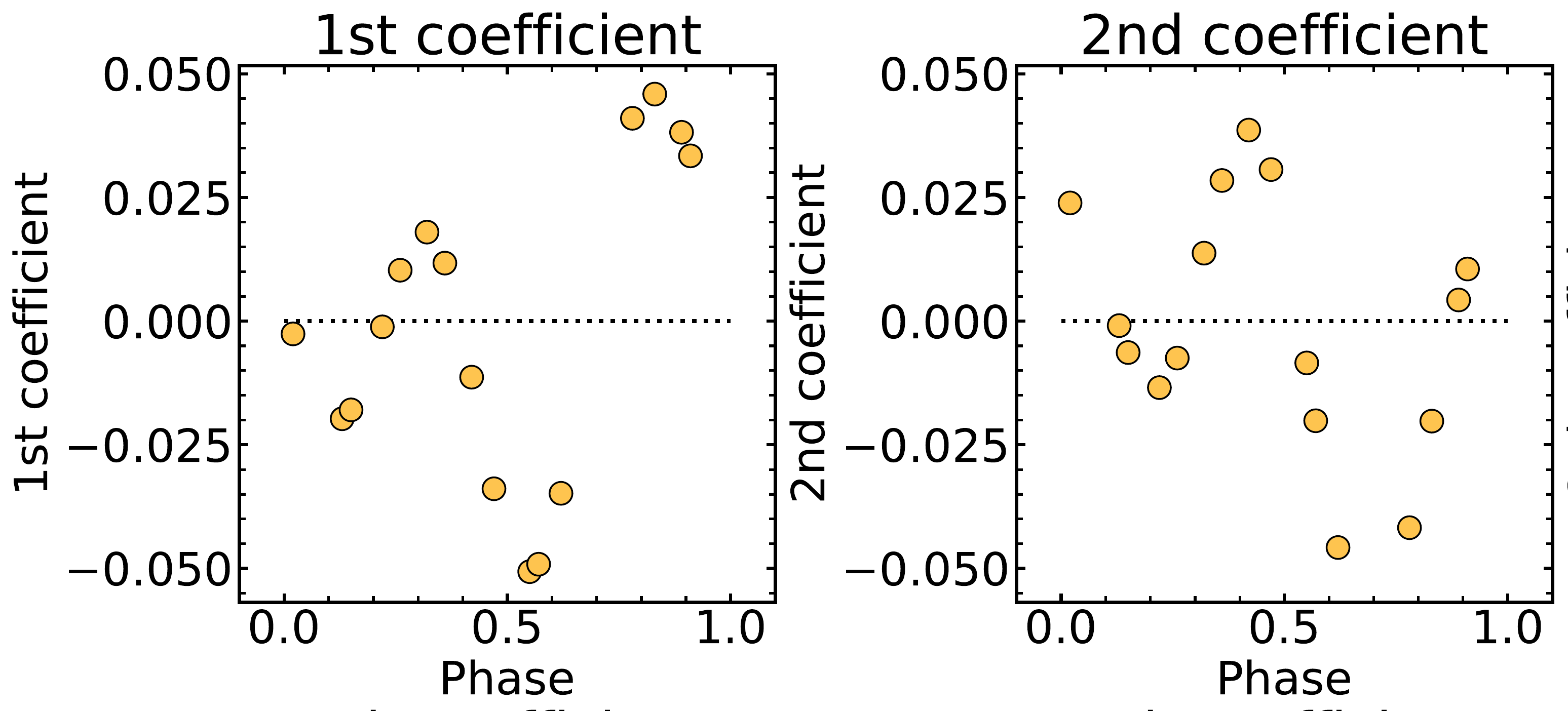} 
    \end{minipage}
    \caption{a.: Top: The mean profile and its decomposition for the complete Stokes~$V$ time series shown in Fig.~\ref{fig:Map_EvoMap}b of the simulated star with the evolving magnetic field using the same format as in Fig.~\ref{fig:SymAntiSym}. Bottom: The longitudinal field against phase colour-coded by the rotation cycle (blue for the first rotation cycle and yellow for the second). b. to d.: The mean profiles and their decomposition (b.), the first two eigenvectors (c.) and their corresponding coefficients (d.) for the first (top row) and second (bottom row) rotation cycle of the evolving map shown in Fig.~\ref{fig:Map_EvoMap} using the same format. }
    \label{fig:EvoMapSplit}
\end{figure*}

A PCA analysis of the Stokes~$V$ profiles can unveil changes in the large-scale magnetic field over the time scale of the observations. It can therefore reveal variations from one stellar rotation to another but it can also trace magnetic cycles or slow temporal variations if polarimetric time series are available on similar timescales. In the following, we will present an example, where PCA is able to diagnose magnetic field variations on time scales similar to the stellar rotation period. As ZDI usually assumes a steady magnetic field, it can be especially useful to check a polarimetric data set for evolving large-scale fields beforehand. The longitudinal field allows the detection of changes in field topology, too. However, we find, that the PCA method is more sensitive and can provide extra insights.

In Fig.~\ref{fig:Map_EvoMap}, we show a simulated star whose large-scale magnetic field evolves from one rotation cycle to a subsequent one observed four rotation cycles later. The star hosts a complex magnetic field including several low order multipoles ($\ell,m \leq 3$)  of different tilt angles observed with an inclination of $i=75^\circ$ and a SNR = 5\,000 but with an uneven phase coverage.
Fig.~\ref{fig:Map_EvoMap}a shows the magnetic field map corresponding to the first rotation cycle and Fig.~\ref{fig:Map_EvoMap}c the magnetic field map of the second. We assume a steady magnetic field over each rotation cycle but the field evolves in between over the time span of three rotation cycles.
The corresponding Stokes~$V$ profiles are shown in Fig.~\ref{fig:Map_EvoMap}b and Fig.~\ref{fig:Map_EvoMap}d shows the first three eigenvectors of the PCA analysis for the complete set of the 28 Stokes~$V$ observations on the top row and the corresponding coefficients below.

The mean profile, Fig.~\ref{fig:Map_EvoMap}b, is small compared to the mean-subtracted Stokes~$V$ profiles, which indicates a dominant non-axisymmetric field. Further, the mean profile is mainly antisymmetric but includes a symmetric component as well, see top panel of Fig.~\ref{fig:EvoMapSplit}a. We therefore expect to detect poloidal and weaker toroidal axisymmetric fields.

PCA confirms as well that the large-scale field must be non-axisymmetric as all three eigenvectors show a clear signal and phase-modulated coefficients, see Fig.~\ref{fig:Map_EvoMap}d. The several extrema seen for the coefficients clearly indicate a complex magnetic field so that the asymmetry of the eigenvectors cannot reveal further hints about $\alpha$, $\beta$ and $\gamma$ of the large-scale field. Nonetheless, we can be sure, that in this example the field is not purely poloidal as we already detected the axisymmetric toroidal field via the mean profile decomposition.  

By colour-coding the coefficients, we clearly see that the large-scale field evolves during the observation gap of three rotation cycles. The amplitude of the coefficients rises indicating a growing longitudinal field which could be the result of an increase in the magnetic field strength or due to a magnetic field feature that moved closer to line-of-sight. For the first two coefficients, we also notice, that new extrema appear, indicating a significant change in the magnetic field topology, likely due to new emerging field structures hosting different field components, that are individually tracked by the eigenvectors. 

If the number of observations is high enough, the Stokes~$V$ time series can be split and separately analysed via PCA, see Fig.~\ref{fig:EvoMapSplit}b-d. By comparing the mean profiles of the split data set, see Fig.~\ref{fig:EvoMapSplit}b, we find that the amount of axisymmetric toroidal field increases. The coefficients of the two rotation cycles separately confirm that the large-scale magnetic field is complex at any time.

We can finally conclude from the mean profile and PCA analysis, that the star has an evolving complex, non-axisymmetric field, that shows signs of an increasing toroidal field and of a changing topology on time scales of the stellar rotation period.
In contrast, from the longitudinal field, see bottom panel of Fig.~\ref{fig:EvoMapSplit}a, the topology appears dipolar and shows a temporal evolution corresponding to a growing dipole whose axis is shifted by $\approx 0.2$ in phase, while the PCA analysis clearly unveils the complex and widely evolving topology including different field components.

Our conclusions derived from the PCA analysis agree with the magnetic field maps shown in Fig.~\ref{fig:Map_EvoMap}a and c. For this example, we are also able to trace some of the evolving magnetic field structures in the radial and azimuthal fields via the first and second eigenvector coefficients separately. For example: we see an increase for the coefficients of the first eigenvector around phase 0.15 and 0.6 related to the appearance of a new radial spot at phase 0.15 and the evolving positive radial field structure at phase 0.6, which becomes larger and moves closer to the line-of-sight by decreasing in latitude. Further, the growing azimuthal field structures around phases 0.0 and 0.5 cause an increase in the coefficient for the second eigenvector. 

We find based on a sample of about $100$ randomly generated field topologies, that the most antisymmetric eigenvector is usually sensitive to the radial field component, while the most symmetric eigenvector captures the azimuthal field in most simulations that contain significant azimuthal field but are still dominated by the radial field component. However, as the first two eigenvectors and their corresponding coefficients are not always related to the radial and azimuthal non-axisymmetric field for complex configurations, it is ultimately safer to retrieve the corresponding ZDI map in order to reliably relate coefficient trends to magnetic features on the stellar surface. 

Sparse or highly uneven phase coverage can impact the mean profile and PCA diagnostic. One can mitigate the effects of highly uneven phase coverage by (i) determining the eigenvectors and coefficients from the Stokes~$V$ time series before subtracting the mean profile. (ii) applying a GPR to the time series for all coefficients and deriving a set of evenly spaced coefficient. (iii) determining an uniformly distributed Stokes~$V$ time series by calculating the dot product between the evenly spaced coefficients and the eigenvectors before finally (iv) computing the mean profile of the uniformly sampled synthetic Stokes~$V$ time series.

In addition, the PCA analysis of the Stokes~$V$ profiles can be used to confirm characteristic magnetic features seen in the ZDI maps. The PCA coefficients can detect them directly in the observed polarimetry time series, especially for moderate or higher $v_e \sin i$ values and non-axisymmetric field topologies. The PCA method provides therefore a convenient cross-check option, that can help to identify the best fitting parameter. 
For instance, one can (i) identify the few PCA eigenvectors associated with a real Stokes~$V$ signal using the method outlined in Sec.~\ref{Sec:Dipole}, then (ii) reconstruct a noise-free set of Stokes~$V$ profiles from these few PCA eigenvectors, and (iii) compute the reduced $\chi^2_r$ between the observed and noise-free Stokes~$V$ profiles to determine the level to which ZDI can realistically fit the input Stokes~$V$ data.

\section{Real Observational Examples: CE~Boo and DS~Leo}
\label{Sec:RealObs}

\begin{figure}
	 \raggedright
	\textbf{a.} \hspace{4.3cm} \textbf{b.} \\
    \centering
    	\includegraphics[height=2.8cm, trim={0 0 0 0}, clip]{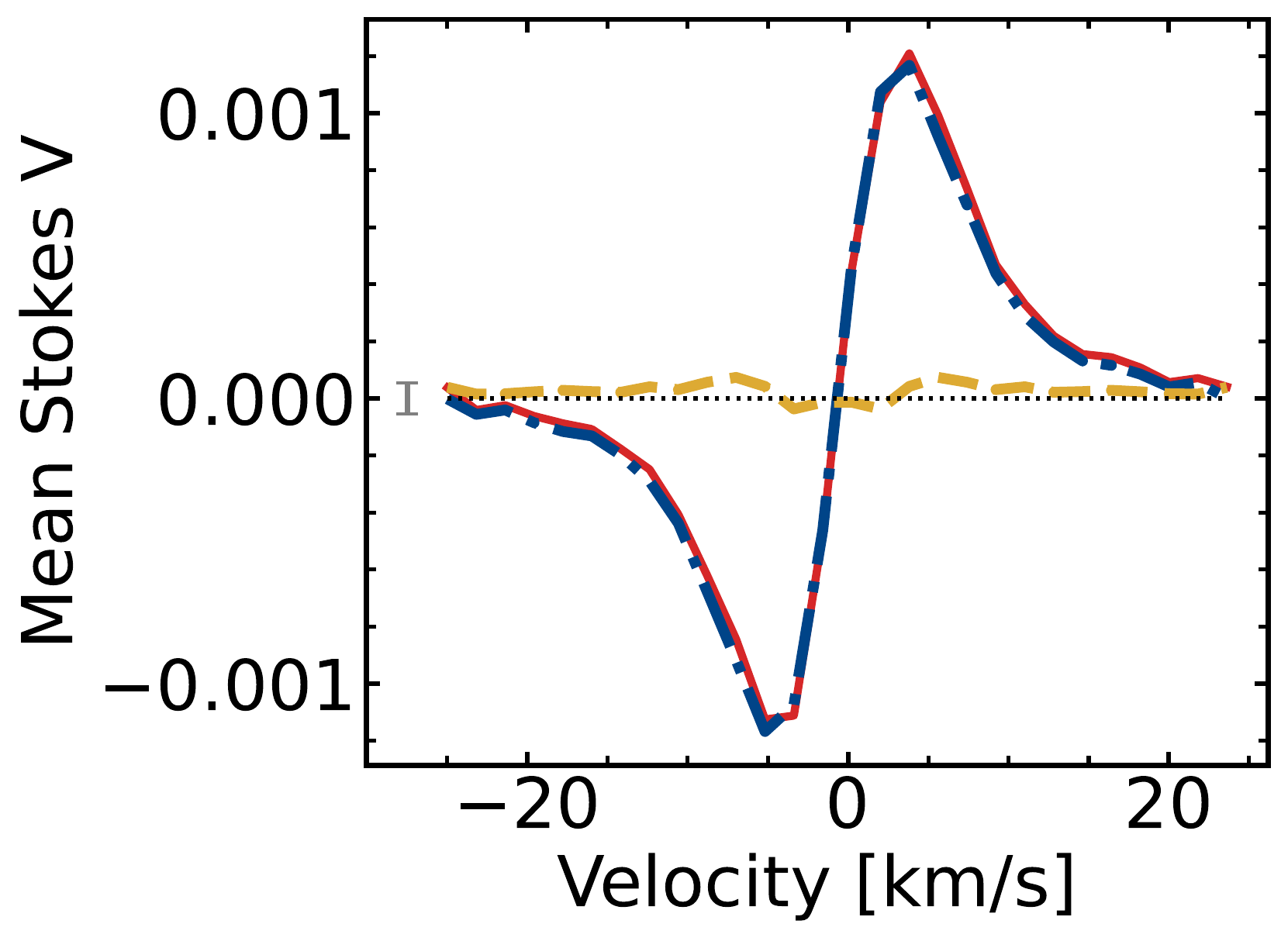} 
    	\hspace*{2ex}
		\includegraphics[height=3cm, trim={0 0 0 0}, angle=0, clip]{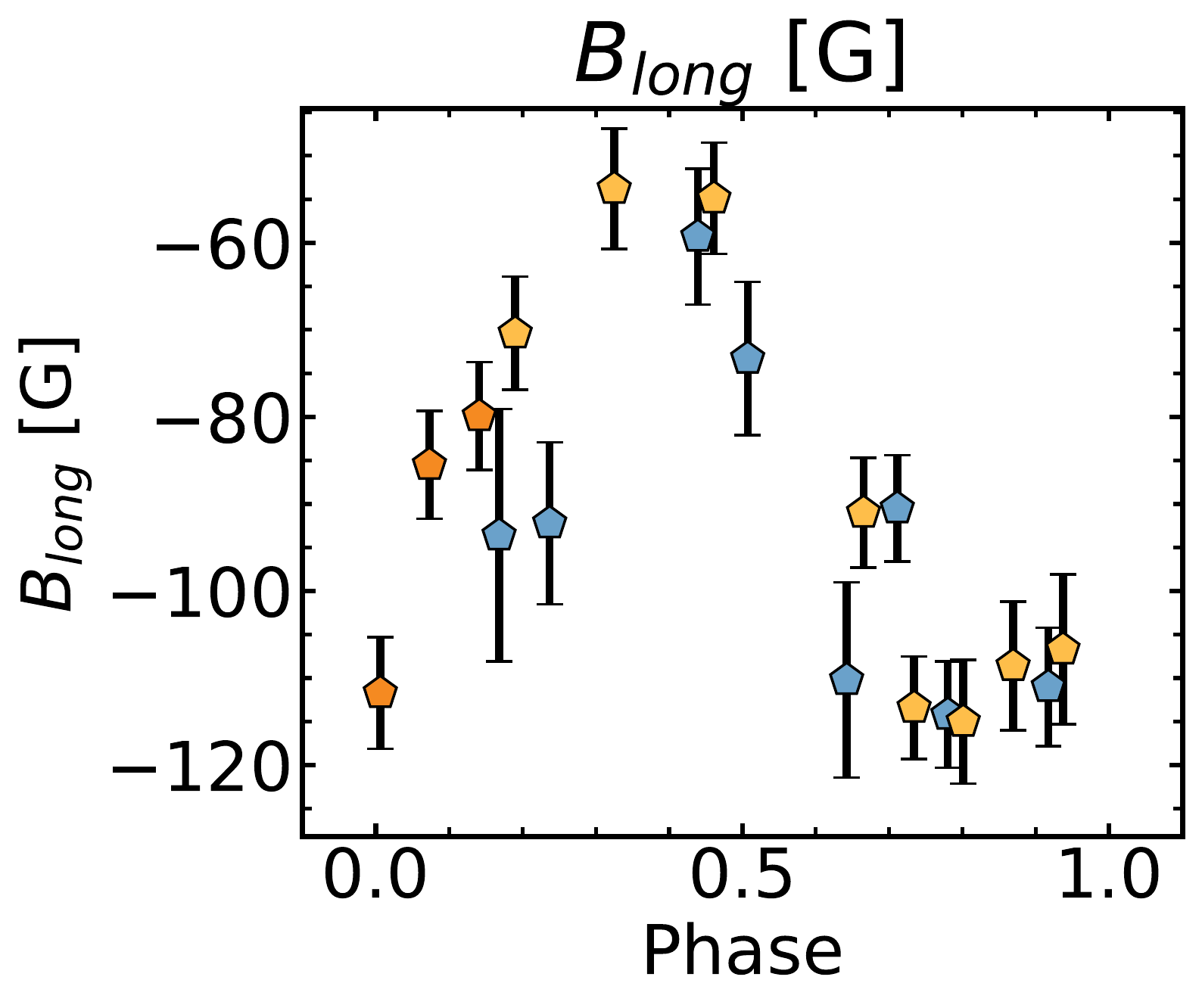} \\
     \raggedright
	\textbf{c.} \\
    \centering
    \includegraphics[width=\columnwidth, trim={0 400 0 0}, clip]{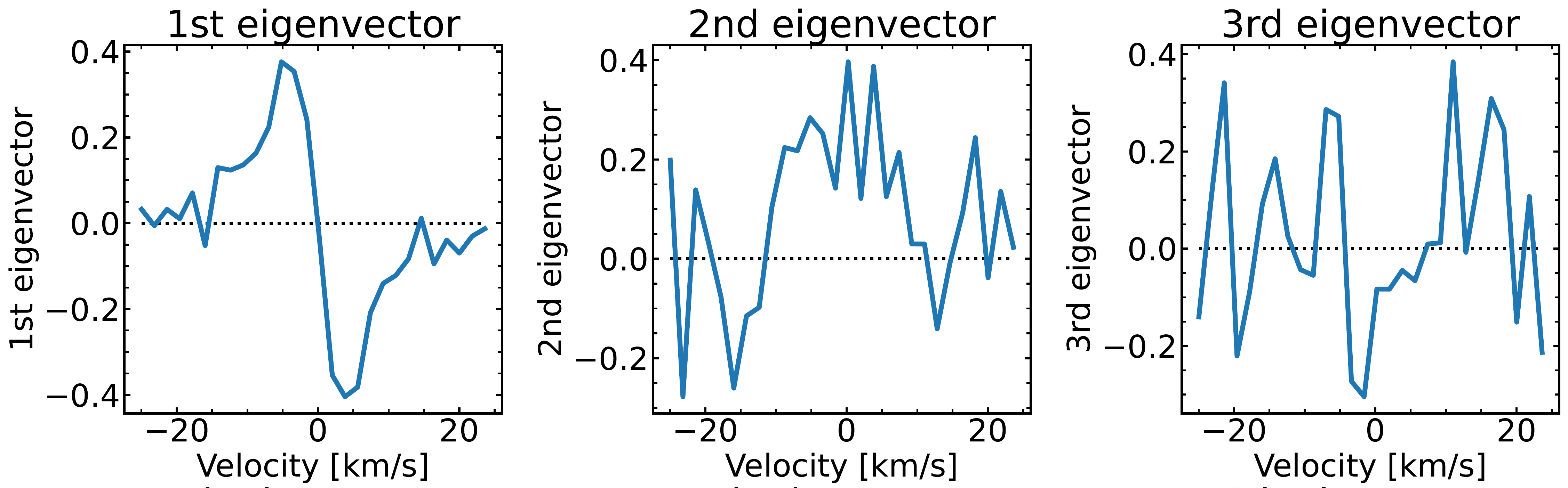} \\
    	 \raggedright
	\textbf{d.} \\
    \centering
    \includegraphics[width=\columnwidth, trim={0 400 0 0}, clip]{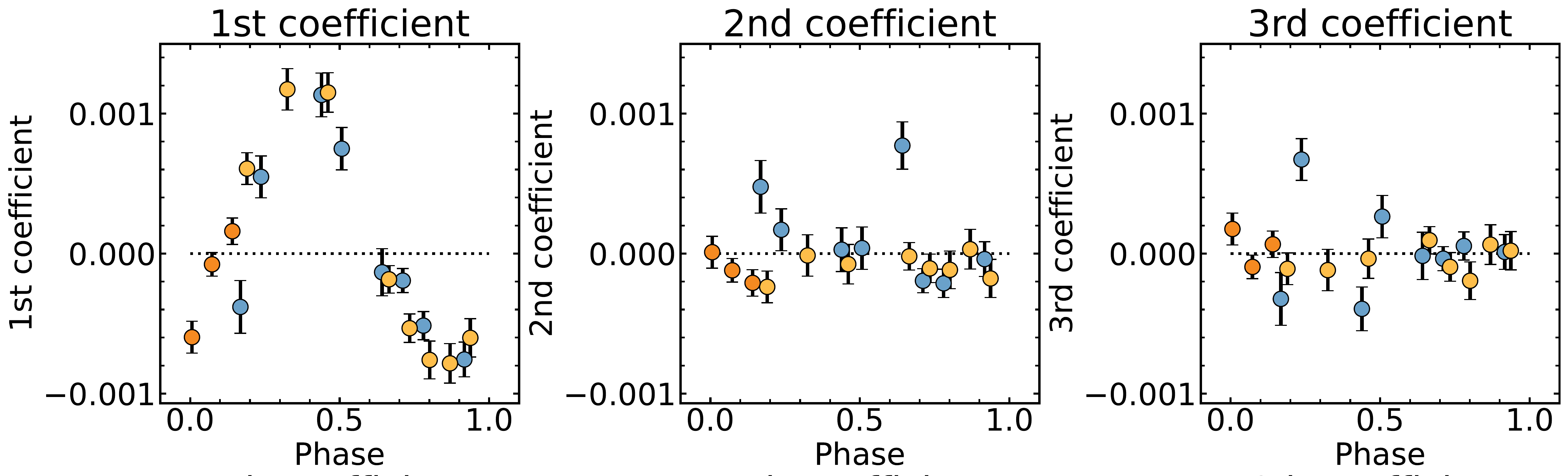} \\
    \caption{a. The mean profile of the 2008 observed Stokes~$V$ time series of CE Boo. b. CE~Boo's longitudinal field against phase for the 2008 epoch using on the corresponding longitudinal field data of \citet[Tab.~4]{Donati2008}. The eigenvectors (c.) and coefficients (d.) of the mean-subtracted Stokes~$V$ observations. The longitudinal field and coefficients are colour-coded by rotation cycle (blue, grey, orange).}
    \label{fig:CEBoo}
\end{figure}

\begin{figure} 
	 \raggedright
	\textbf{a.} \hspace{4.3cm} \textbf{b.} \\
    \centering
    	\includegraphics[height=2.8cm, trim={0 0 0 0}, clip]{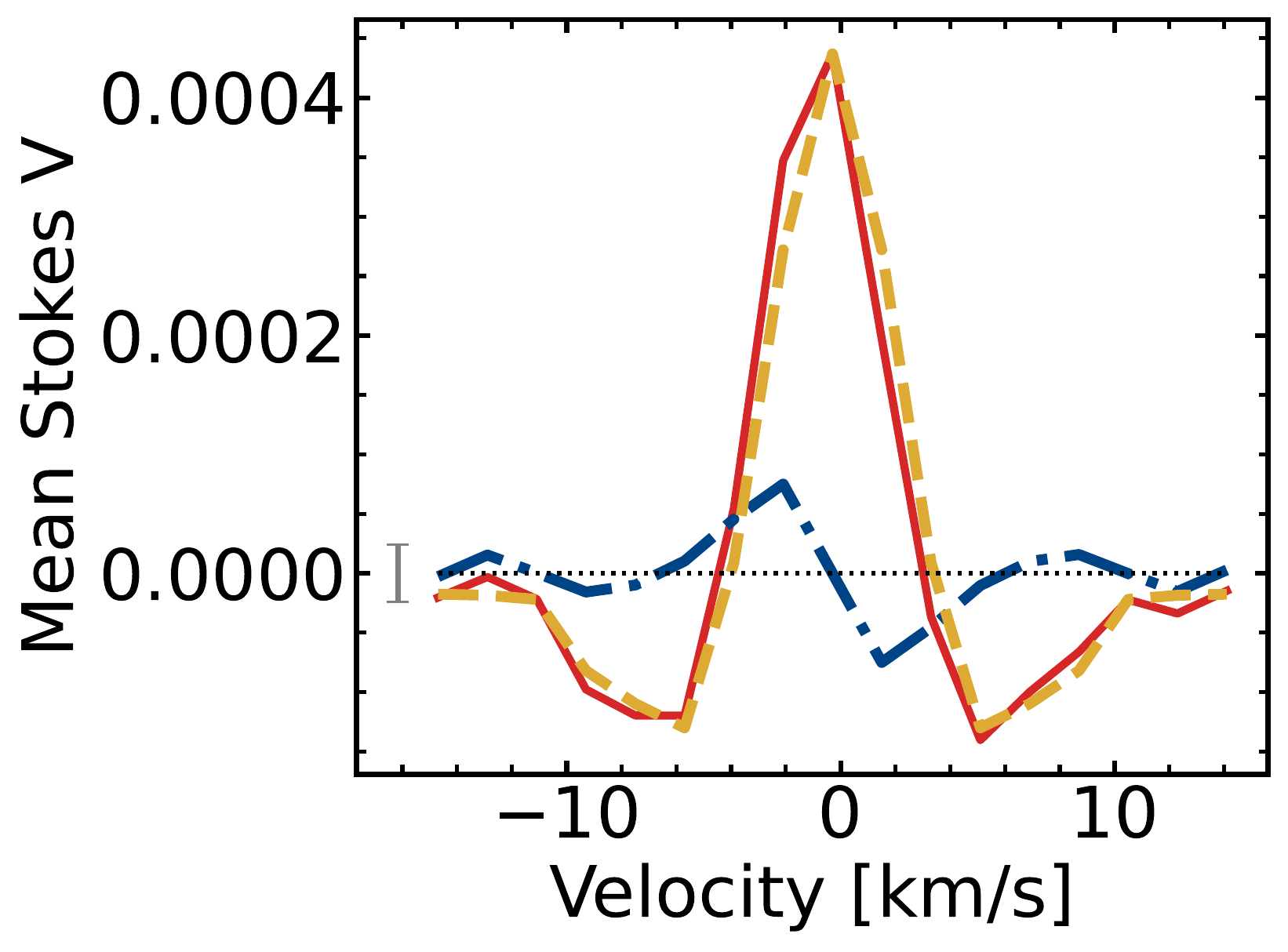} 
    	\hspace*{2ex}
		\includegraphics[height=3cm, trim={0 0 0 0}, angle=0, clip]{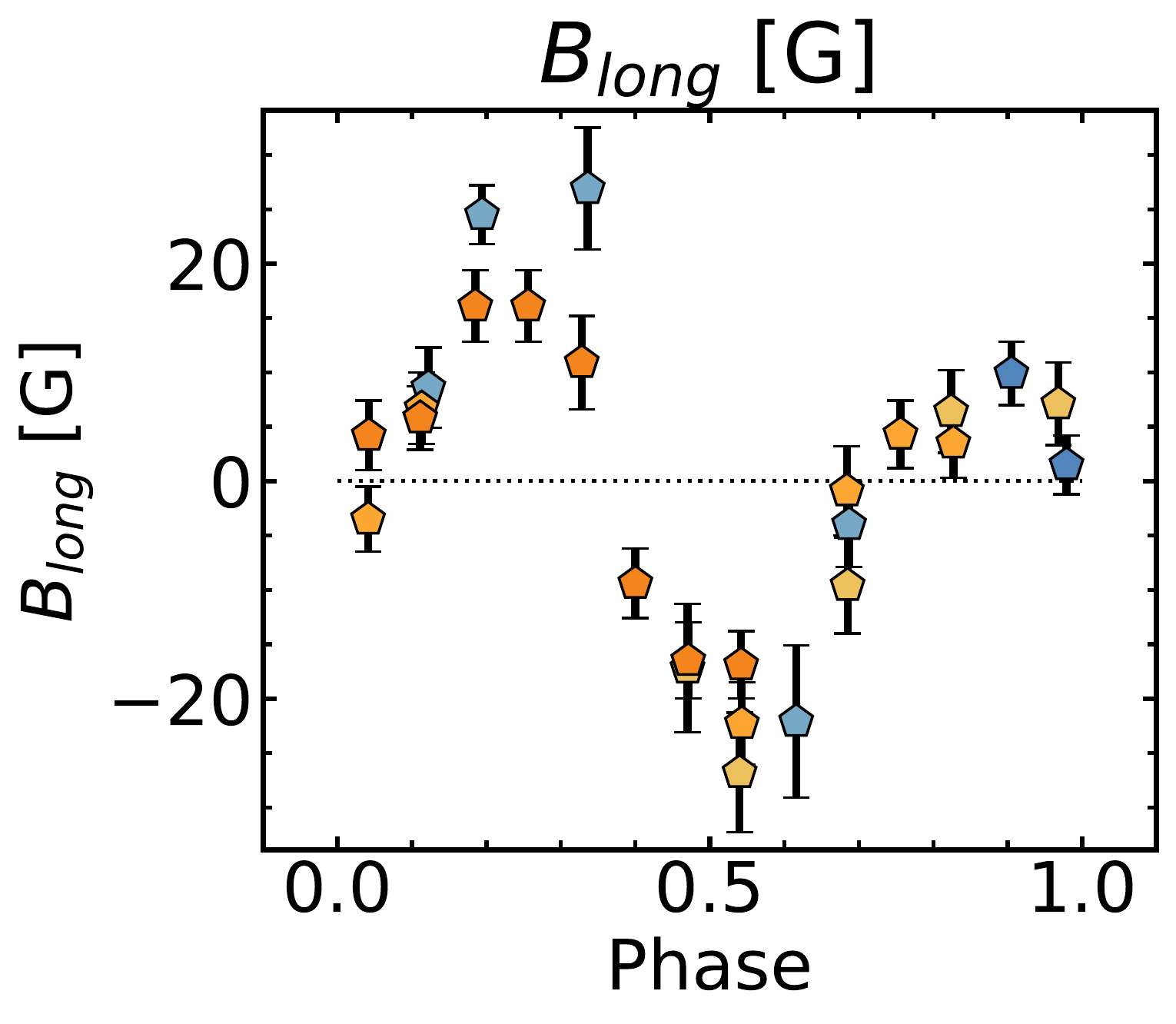} \\
     \raggedright
	\textbf{c.} \\
    \centering
    \includegraphics[width=\columnwidth, trim={0 400 0 0}, clip]{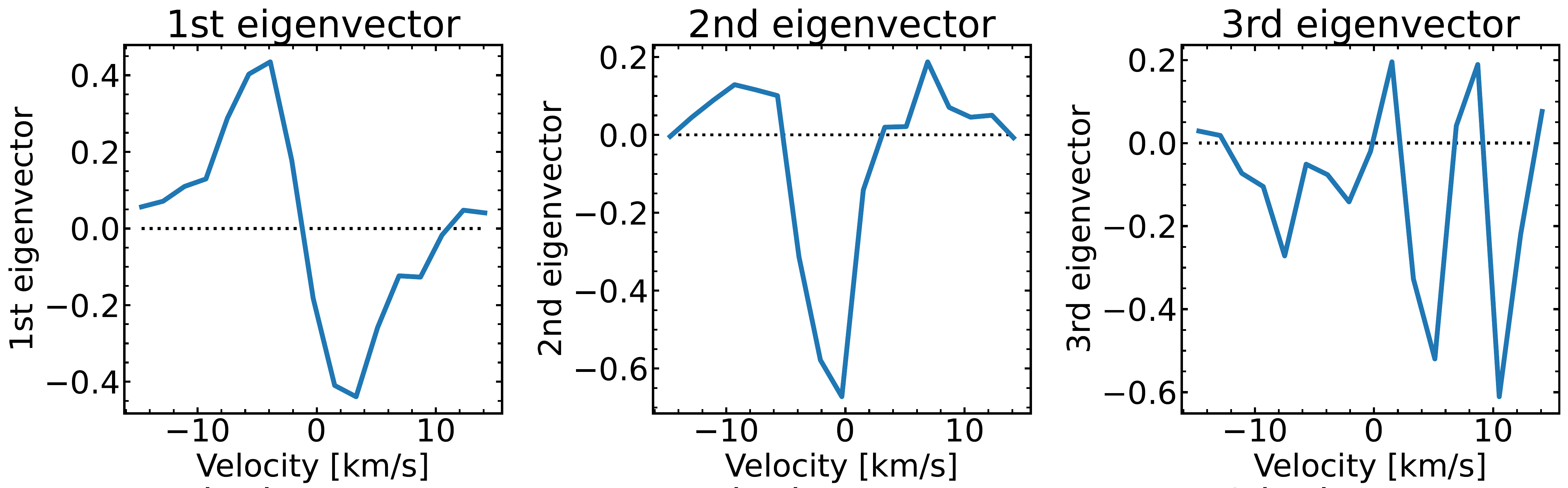} \\
    	 \raggedright
	\textbf{d.} \\
    \centering
    \includegraphics[width=\columnwidth, trim={0 400 0 0}, clip]{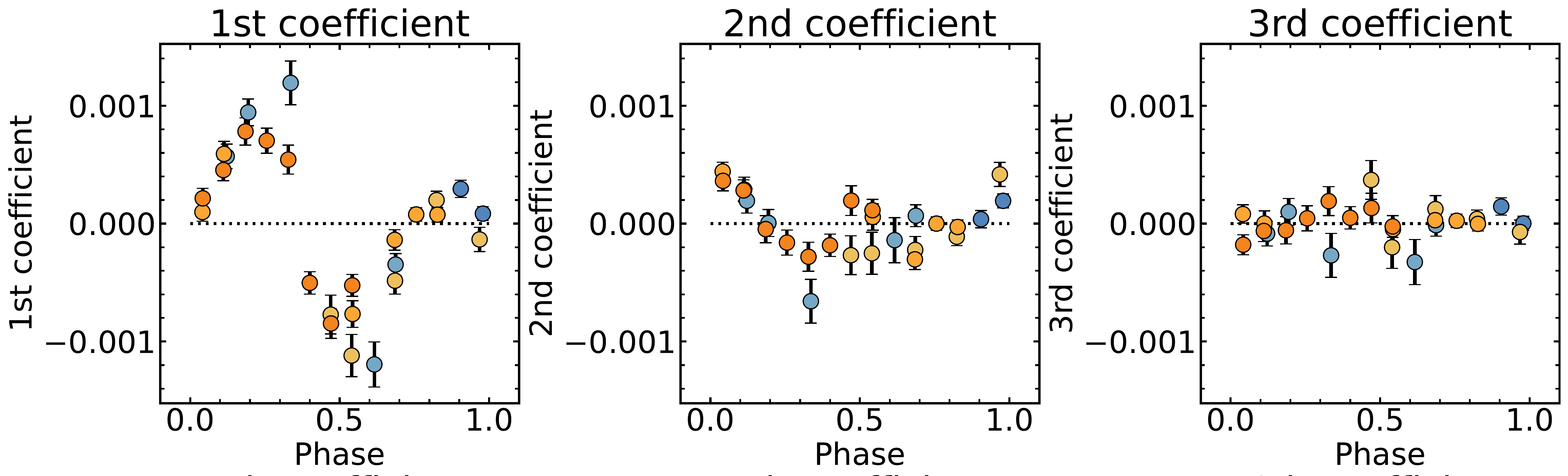} \\
    \caption{a. The mean profile of the 2008 observed Stokes~$V$ time series of DS~Leo. b. DS~Leo's longitudinal field against phase for the 2008 epoch using on the corresponding longitudinal field data of \citet[Tab.~3]{Donati2008}. The eigenvectors (c.) and coefficients (d.) of the mean-subtracted Stokes~$V$ observations. The longitudinal field and coefficients are colour-coded by the five continuously covered rotation cycle (dark blue, light blue, grey, yellow, orange).}
    \label{fig:DSLeo}
\end{figure}

In this section, we present how PCA performs on two real observed data sets. We test our PCA method on the Stokes~$V$ time series of CE~Boo (GJ569) and DS~Leo (GJ410) observed with NARVAL at the T\'elescope Bernard Lyot (TBL) published by \cite{Donati2008}. Both stars are early M~dwarfs observed in the optical between 2007--2008. The M~dwarfs have a mass of $0.48\,\mrm{M_{\odot}}$ and $0.58\,\mrm{M_{\odot}}$ and a rotation period of $P_{\mrm{rot}} = 14.7\,\mrm{d}$ and $14.0\,\mrm{d}$, respectively. They are observed at an inclination angle of $i = 45^\circ$ and $60^\circ$, respectively, with a similar $v_e \sin i$ of $1-2\,\mrm{km\,s^{-1}}$. However, they host two very different large-scale magnetic field configurations. Further details can be found in \cite{Donati2008}.

Fig.~\ref{fig:CEBoo}a displays the mean profile and its decomposition for the 19~Stokes~$V$ profiles observed for CE~Boo in 2008 Jan--Feb. The Stokes~$V$ profiles can be found in \citet[][Fig.~6]{Donati2008}. The mean profile is large compared to the mean-subtracted Stokes~$V$ profiles and shows a clear antisymmetric shape indicating a dominant axisymmetric poloidal field. Fig.~\ref{fig:CEBoo}b displays the longitudinal field measurements against phase as determined by \cite{Donati2008}.
Fig.~\ref{fig:CEBoo}c and d show the first three eigenvectors and the corresponding coefficients from the PCA analysis of CE~Boo's mean-subtracted Stokes~$V$ profiles. The first eigenvector displays a strong antisymmetric profile, which indicates, in connection with the typical sinusoidal trend seen for the corresponding coefficients, a dominantly poloidal dipole. As the following eigenvectors contain only noise, we predict a small tilt angle. 

In contrast to our simulations, the SNR of each LSD profile varies for real observations. For example, the SNR of the 19 LSD Stokes~$V$ profile for CE~Boo ranges from 3770--9720 (median = 8330). The first eigenvectors and their coefficients, which are used to analyse the non-axisymmetric field, are less affected by varying SNR as most of the varying noise is stored in the higher eigenvectors.
If the Stokes~$V$ time series is strongly affected by varying noise, one should consider using the weighted mean profile and applying a weighted PCA to the weighted mean subtracted Stokes~$V$ profiles. For the observed time series used in this section, a weighted mean and PCA show only minimal changes for the mean profile or the first eigenvectors and coefficients, reflecting the fact that only a few of our observed LSD profiles have significantly lower SNRs than the median value.  

Comparing the longitudinal field and the coefficients of the first eigenvector, we notice that the coefficients show a clearer trend with phase and are less affected by the noise, especially the first rotation cycle (blue symbols) benefits from the PCA analysis. The coefficients reveal the same trend with phase for the different rotation cycles so that the large-scale field did not evolve across the three stellar rotations, which is less obvious for the longitudinal field. Furthermore, $|\mrm{max}| \neq |\mrm{min}|$ for the coefficient of the first eigenvector also indicates that $\ell >1$ modes and / or $\beta$ coefficients are needed to fully describe the large-scale field of CE~Boo.

Our conclusion about a mainly axisymmetric poloidal dipole from the PCA analysis are in good agreement with CE~Boo's ZDI maps, see \citet[Fig.~7]{Donati2008}. The large-scale field of CE~Boo was found to be dominated by a poloidal dipole, that shows a very small tilt angle of a few degrees \citep{Donati2008}. Over 90\,\% of the magnetic energy is stored in the poloidal field and 96\,\% in the axisymmetric mode $m > \ell/2$. Additionally, the radial dipole is pointing at phase $\approx0.9$, where also the coefficients of the antisymmetric profile show the extrema of a negative pole.

Fig.~\ref{fig:DSLeo} shows the results of the PCA analysis for 26 Stokes~$V$ profiles of DS~Leo observed from 2007 Dec to Feb 2008\footnote{The Stokes~$V$ profile observed at the 2008 Jan 02 corresponding to phase 26.263 was not available in the data archive PolarBase and could not be included in our analysis but was used in the analysis of \cite{Donati2008}. The observed Stokes~$V$ profiles can be found in Fig.~4 of \cite{Donati2008}.
} using the same format as in Fig.~\ref{fig:CEBoo}. Their SNR ranges from 6450 to 15300 (median=12060.5).
The mean profile, Fig.~\ref{fig:DSLeo}a, shows a dominantly symmetric signal indicating a dominant toroidal axisymmetric field. The first two eigenvectors show a signal so that we can conclude that the large-scale field has a significant non-axisymmetric component. The coefficients related to the first eigenvector show that the non-axisymmetric poloidal component of the field must be more complex than a pure dipole due to the two minima. Furthermore, the amplitudes of the coefficients for the first eigenvector are weaker for the last observed cycles (the yellow and orange circles) than for the first observed rotation cycles (the dark blue, light blue and grey circles). This is a hint of an evolving field; in this specific case a weakening longitudinal field. The longitudinal field, Fig.~\ref{fig:DSLeo}b, shows the same trend as the coefficients of the first eigenvector although the errors are slightly bigger. The coefficients of the second eigenvector show some variation as well, especially around phase 0.5, indicating that the changes in the field topology affect different magnetic field components. The PCA analysis reveals that DS~Leo's topology evolves in a more complex manner than a simple weakening of the longitudinal field. 

Our conclusion of a non-axisymmetric, complex large-scale field with a strong toroidal component agrees with the ZDI map of DS~Leo, see \citet[Fig.~5 bottom row]{Donati2008}. \cite{Donati2008} found that about half of the energy is stored in $\ell > 1$ modes and 80\,\% of the magnetic energy is stored in the toroidal component. Further, the poloidal field is mainly non-axisymmetric. In comparison with the ZDI map, we can even recognise most of the radial field structures in the coefficient trends with phase related to the first eigenvector, although we caution again that the eigenvectors are not always straightforwardly related to individual field components for complex configurations but often do. Further, \cite{Donati2008} found that the large-scale magnetic field evolved from the first epoch observed in 2007 Jan to the observations 12 months later and analysed here. Our PCA analysis shows, that the field evolves on even smaller timescales in the order of the stellar rotation period of $\approx$14\,d.

\section{Summary}
\label{Sec:Conclusions}

We presented in this paper a fast and easy method to diagnose the large-scale field directly from the spectropolarimetric data, performing best for slowly to moderately rotating stars. The mean Stokes~$V$ profile and the PCA eigenvectors and coefficients determined from the mean-subtracted Stokes~$V$ profiles reveal hints about the axisymmetry and complexity of the large-scale field plus the main field components of the axisymmetric large-scale field.

PCA is a data-driven method working directly on the observations so that no assumptions are needed to get hints about the large-scale field topology. Our method allows a quick and easy first glance at key parameters of the stellar magnetic field topology before more advanced techniques like ZDI reconstruct a full map of the surface magnetic field. 
It can be conveniently applied to a large number of stellar targets in the already existing and upcoming databases and surveys, allowing to search for certain key parameters of the large-scale field, e.g., for toroidal fields. The quick analysis of the magnetic field topology will also be helpful for the target selection in polarimetric observing campaigns, e.g., to decide if further observing runs or epochs are of interest. Furthermore, our method can confirm the detection of characteristic magnetic field features seen in ZDI maps directly from the observations for mainly non-axisymmetric topologies observed with a moderate $v_e \sin i$ and can help to find the best fitting and stopping criteria for ZDI. 

In addition, PCA allows tracing the large-scale field evolution over the time range of the observations. We find, that the PCA method can unveil the complexity to lower $v_e \sin i$ values than the longitudinal field and can provide further insights into their topology, e.g., the evolution of individual field components or even magnetic features. Our results suggest, that especially the temporal evolution of the azimuthal field will be detectable to lower $v_e \sin i$ and SNR values. If the time series is large enough, our method can even trace the evolution of the axisymmetric and the non-axisymmetric field, separately.
Detecting evolving large-scale fields will be interesting 1) for long time series to discover stellar activity cycles, e.g., to find a solar analog \citep{Lehmann2021} or to unveil the nature of the bimodal field topology found for late M~dwarfs \citep{Morin2010}, and 2) it will be interesting for short time series to detect large-scale field variations on time scales of the stellar rotation period, whereas ZDI is so far limited by assuming a stable large-scale field though with variability coming from differential rotation.

Summarising our present method: the mean profile of an observed Stokes~$V$ series allows the analysis of the axisymmetric large-scale field. Its amplitude indicates the degree of axisymmetry and its shape unveils the presence of poloidal and/or toroidal fields and which component is dominant. The non-axisymmetric field can be analysed by applying PCA to the mean-subtracted Stokes~$V$ profiles. The more eigenvectors contain a signal, the less axisymmetric the large-scale field. The coefficients can very well unveil the large-scale field complexity and temporal variations. In the case of a pure dipole field, the coefficients of the first or of the first two eigenvectors show sine-like curves with two extrema $\approx$0.5 apart in phase. Any different coefficient trends with phase indicate more complex field topologies. The more non-axisymmetric the field, the better PCA exhibits the field complexity.

Our method is limited to Stokes~$V$ time series, where the Stokes~$V$ profiles show shapes, that can be described as a combination of the first few derivatives of the unpolarised Stokes~$I$ profiles. PCA works best for low to moderate $v_e \sin i$ ranges but even for higher $v_e \sin i$ ranges, the presented method provides insights into the complexity, the evolution of the large-scale field and the dominant component of the axisymmetric field by decomposing the magnetic field in its symmetric and antisymmetric component. As a result, PCA can be applied to the majority of the polarimetrically observed non-degenerated stars.

In addition, our method benefits from a decent phase coverage as the mean profile can be otherwise dominated by one hemisphere for a very uneven phase coverage. We also recommend having a fair number of $\approx 15$ or more observed Stokes~$V$ profiles, although the mean profile and the PCA analysis can still provide insights in case of small observing  samples, e.g. $\approx 5$ observations, too. As PCA is very sensitive to variations in the data, one needs to be especially careful, if systematics pollute the data. 

The PCA method presented here will be an important tool to explore stellar magnetism in the future, especially for upcoming large polarimetric surveys like the SLS, due to its quick and convenient analysis of the large-scale field key parameters and detection of stellar magnetic field evolutions on varies time scales to lower $v_e \sin i$ and SNR values than the longitudinal field.

\section*{Acknowledgements}

We acknowledge funding from the European Research Council under the H2020 research \& innovation programme (grant \#740651 NewWorlds).

\section*{Data Availability}

The simulated Stokes~$V$ profiles are available on reasonable request to the first author (lisa.lehmann[at]irap.omp.eu). The observational Stokes~$V$ profiles of CE~Boo and DS~Leo were collected from the \href{http://polarbase.irap.omp.eu}{PolarBase} archive, \citep{PolarBasePetit2014, PolarBaseDonati1997}.



\bibliographystyle{mnras}
\bibliography{Lehmann2022_PCAonStokes} 

\providecommand{\noopsort}[1]{}
\begin{thebibliography}{}
\makeatletter
\relax
\def\mn@urlcharsother{\let\do\@makeother \do\$\do\&\do\#\do\^\do\_\do\%\do\~}
\def\mn@doi{\begingroup\mn@urlcharsother \@ifnextchar [ {\mn@doi@}
  {\mn@doi@[]}}
\def\mn@doi@[#1]#2{\def\@tempa{#1}\ifx\@tempa\@empty \href
  {http://dx.doi.org/#2} {doi:#2}\else \href {http://dx.doi.org/#2} {#1}\fi
  \endgroup}
\def\mn@eprint#1#2{\mn@eprint@#1:#2::\@nil}
\def\mn@eprint@arXiv#1{\href {http://arxiv.org/abs/#1} {{\tt arXiv:#1}}}
\def\mn@eprint@dblp#1{\href {http://dblp.uni-trier.de/rec/bibtex/#1.xml}
  {dblp:#1}}
\def\mn@eprint@#1:#2:#3:#4\@nil{\def\@tempa {#1}\def\@tempb {#2}\def\@tempc
  {#3}\ifx \@tempc \@empty \let \@tempc \@tempb \let \@tempb \@tempa \fi \ifx
  \@tempb \@empty \def\@tempb {arXiv}\fi \@ifundefined
  {mn@eprint@\@tempb}{\@tempb:\@tempc}{\expandafter \expandafter \csname
  mn@eprint@\@tempb\endcsname \expandafter{\@tempc}}}

\bibitem[\protect\citeauthoryear{{Altschuler} \& {Newkirk}}{{Altschuler} \&
  {Newkirk}}{1969}]{Altschuler1969}
{Altschuler} M.~D.,  {Newkirk} G.,  1969, \mn@doi [\solphys]
  {10.1007/BF00145734}, \href
  {https://ui.adsabs.harvard.edu/abs/1969SoPh....9..131A} {9, 131}

\bibitem[\protect\citeauthoryear{{Boro Saikia} et~al.,}{{Boro Saikia}
  et~al.}{2018}]{BoroSaikia2018}
{Boro Saikia} S.,  et~al., 2018, \mn@doi [\aap] {10.1051/0004-6361/201834347},
  \href {https://ui.adsabs.harvard.edu/abs/2018A&A...620L..11B} {620, L11}

\bibitem[\protect\citeauthoryear{{Carroll}, {Kopf}, {Ilyin}  \&
  {Strassmeier}}{{Carroll} et~al.}{2007}]{Carroll2007}
{Carroll} T.~A.,  {Kopf} M.,  {Ilyin} I.,   {Strassmeier} K.~G.,  2007, \mn@doi
  [Astronomische Nachrichten] {10.1002/asna.200710884}, \href
  {https://ui.adsabs.harvard.edu/abs/2007AN....328.1043C} {328, 1043}

\bibitem[\protect\citeauthoryear{Carroll, Strassmeier, Rice  \&
  K{\"u}nstler}{Carroll et~al.}{2012}]{Carroll2012}
Carroll T.~A.,  Strassmeier K.~G.,  Rice J.~B.,   K{\"u}nstler A.,  2012,
  \mn@doi [\aap] {10.1051/0004-6361/201220215}, 548, A95

\bibitem[\protect\citeauthoryear{{Casini} \& {Li}}{{Casini} \&
  {Li}}{2019}]{Casini2019}
{Casini} R.,  {Li} W.,  2019, \mn@doi [\apj] {10.3847/1538-4357/ab0023}, \href
  {https://ui.adsabs.harvard.edu/abs/2019ApJ...872..173C} {872, 173}

\bibitem[\protect\citeauthoryear{{Chan}, {Jia}, {Gao}, {Lu}, {Zeng}  \&
  {Ma}}{{Chan} et~al.}{2015}]{Chan2015}
{Chan} T.-H.,  {Jia} K.,  {Gao} S.,  {Lu} J.,  {Zeng} Z.,   {Ma} Y.,  2015,
  \mn@doi [IEEE Transactions on Image Processing] {10.1109/TIP.2015.2475625},
  \href {https://ui.adsabs.harvard.edu/abs/2015ITIP...24.5017C} {24, 5017}

\bibitem[\protect\citeauthoryear{Chandrasekhar}{Chandrasekhar}{1961}]{Chandrasekhar1961}
Chandrasekhar S.,  1961, Hydrodynamic and Hydromagnetic Stability.
International Series of Monographs on Physics, Oxford: Clarendon

\bibitem[\protect\citeauthoryear{{Claret}}{{Claret}}{2000}]{Claret2000}
{Claret} A.,  2000, \aap, \href
  {https://ui.adsabs.harvard.edu/abs/2000A&A...363.1081C} {363, 1081}

\bibitem[\protect\citeauthoryear{{Donati} \& {Brown}}{{Donati} \&
  {Brown}}{1997}]{DonatiBrown1997}
{Donati} J.-F.,  {Brown} S.~F.,  1997, \aap, \href
  {http://adsabs.harvard.edu/abs/1997A\%26A...326.1135D} {326, 1135}

\bibitem[\protect\citeauthoryear{Donati, Semel, Carter, Rees  \&
  Collier~Cameron}{Donati et~al.}{1997a}]{Donati1997b}
Donati J.-F.,  Semel M.,  Carter B.~D.,  Rees D.~E.,   Collier~Cameron A.,
  1997a, \mn@doi [\mnras] {10.1093/mnras/291.4.658}, 291, 658

\bibitem[\protect\citeauthoryear{{Donati}, {Semel}, {Carter}, {Rees}  \&
  {Collier Cameron}}{{Donati} et~al.}{1997b}]{PolarBaseDonati1997}
{Donati} J.~F.,  {Semel} M.,  {Carter} B.~D.,  {Rees} D.~E.,   {Collier
  Cameron} A.,  1997b, \mn@doi [\mnras] {10.1093/mnras/291.4.658}, \href
  {https://ui.adsabs.harvard.edu/abs/1997MNRAS.291..658D} {291, 658}

\bibitem[\protect\citeauthoryear{Donati, Forveille, Collier~Cameron, Barnes,
  Delfosse, Jardine  \& Valenti}{Donati et~al.}{2006}]{Donati2006}
Donati J.-F.,  Forveille T.,  Collier~Cameron A.,  Barnes J.~R.,  Delfosse X.,
  Jardine M.~M.,   Valenti J.~A.,  2006, \mn@doi [Science]
  {10.1126/science.1121102}, 311, 633

\bibitem[\protect\citeauthoryear{Donati et~al.,}{Donati
  et~al.}{2008}]{Donati2008}
Donati J.-F.,  et~al., 2008, \mn@doi [\mnras]
  {10.1111/j.1365-2966.2008.13799.x}, 390, 545

\bibitem[\protect\citeauthoryear{Donati et~al.,}{Donati
  et~al.}{2014}]{Donati2014}
Donati J.-F.,  et~al., 2014, \mn@doi [\mnras] {10.1093/mnras/stu1679}, 444,
  3220

\bibitem[\protect\citeauthoryear{{Donati} et~al.,}{{Donati}
  et~al.}{2020}]{Donati2020}
{Donati} J.~F.,  et~al., 2020, \mn@doi [\mnras] {10.1093/mnras/staa2569}, \href
  {https://ui.adsabs.harvard.edu/abs/2020MNRAS.498.5684D} {498, 5684}

\bibitem[\protect\citeauthoryear{Elsasser}{Elsasser}{1946}]{Elsasser1946}
Elsasser W.~M.,  1946, \mn@doi [Physical Review] {10.1103/PhysRev.69.106}, 69,
  106

\bibitem[\protect\citeauthoryear{{Eydenberg}, {Balasubramaniam}  \& {L{\'o}pez
  Ariste}}{{Eydenberg} et~al.}{2005}]{Eydenberg2005}
{Eydenberg} M.~S.,  {Balasubramaniam} K.~S.,   {L{\'o}pez Ariste} A.,  2005,
  \mn@doi [\apj] {10.1086/426703}, \href
  {https://ui.adsabs.harvard.edu/abs/2005ApJ...619.1167E} {619, 1167}

\bibitem[\protect\citeauthoryear{{Fang}, {Luo}, {Zhang}, {Zhang}, {Nolan}  \&
  {Naidu}}{{Fang} et~al.}{2022}]{Fang2022}
{Fang} C.,  {Luo} Y.,  {Zhang} X.,  {Zhang} H.,  {Nolan} A.,   {Naidu} R.,
  2022, \mn@doi [Chemosphere] {10.1016/j.chemosphere.2021.131736}, \href
  {https://ui.adsabs.harvard.edu/abs/2022Chmsp.286m1736F} {286, 131736}

\bibitem[\protect\citeauthoryear{Folsom et~al.,}{Folsom
  et~al.}{2016}]{Folsom2016}
Folsom C.~P.,  et~al., 2016, \mn@doi [\mnras] {10.1093/mnras/stv2924}, 457, 580

\bibitem[\protect\citeauthoryear{{Glazebrook}, {Offer}  \&
  {Deeley}}{{Glazebrook} et~al.}{1998}]{Glazebrook1998}
{Glazebrook} K.,  {Offer} A.~R.,   {Deeley} K.,  1998, \mn@doi [\apj]
  {10.1086/305039}, \href
  {https://ui.adsabs.harvard.edu/abs/1998ApJ...492...98G} {492, 98}

\bibitem[\protect\citeauthoryear{{Hussain} et~al.,}{{Hussain}
  et~al.}{2016}]{Hussain2016}
{Hussain} G.~A.~J.,  et~al., 2016, \mn@doi [\aap]
  {10.1051/0004-6361/201526595}, \href
  {http://adsabs.harvard.edu/abs/2016A\%26A...585A..77H} {585, A77}

\bibitem[\protect\citeauthoryear{Jardine, Collier~Cameron  \& Donati}{Jardine
  et~al.}{2002}]{Jardine2002}
Jardine M.,  Collier~Cameron A.,   Donati J.-F.,  2002, \mn@doi [\mnras]
  {10.1046/j.1365-8711.2002.05394.x}, 333, 339

\bibitem[\protect\citeauthoryear{{Jardine}, {Vidotto}  \& {See}}{{Jardine}
  et~al.}{2017}]{Jardine2017}
{Jardine} M.,  {Vidotto} A.~A.,   {See} V.,  2017, \mn@doi [\mnras]
  {10.1093/mnrasl/slw206}, \href
  {https://ui.adsabs.harvard.edu/abs/2017MNRAS.465L..25J} {465, L25}

\bibitem[\protect\citeauthoryear{{Kiefer}, {Bohn}, {Quanz}, {Kenworthy}  \&
  {Stolker}}{{Kiefer} et~al.}{2021}]{Kiefer2021}
{Kiefer} S.,  {Bohn} A.~J.,  {Quanz} S.~P.,  {Kenworthy} M.,   {Stolker} T.,
  2021, \mn@doi [\aap] {10.1051/0004-6361/202140285}, \href
  {https://ui.adsabs.harvard.edu/abs/2021A&A...652A..33K} {652, A33}

\bibitem[\protect\citeauthoryear{Kochukhov, Bagnulo, Wade, Sangalli, Piskunov,
  Landstreet, Petit  \& Sigut}{Kochukhov et~al.}{2004}]{Kochukhov2004}
Kochukhov O.,  Bagnulo S.,  Wade G.~A.,  Sangalli L.,  Piskunov N.,  Landstreet
  J.~D.,  Petit P.,   Sigut T. A.~A.,  2004, \mn@doi [\aap]
  {10.1051/0004-6361:20031595}, 414, 613

\bibitem[\protect\citeauthoryear{Kochukhov, Makaganiuk  \& Piskunov}{Kochukhov
  et~al.}{2010}]{Kochukhov2010}
Kochukhov O.,  Makaganiuk V.,   Piskunov N.,  2010, \mn@doi [\aap]
  {10.1051/0004-6361/201015429}, 524, A5

\bibitem[\protect\citeauthoryear{{Landi Degl'Innocenti} \& {Landolfi}}{{Landi
  Degl'Innocenti} \& {Landolfi}}{2004}]{Landi2004}
{Landi Degl'Innocenti} E.,  {Landolfi} M.,  2004, {Polarization in Spectral
  Lines}.
~ Vol. 307, \mn@doi{10.1007/978-1-4020-2415-3, }

\bibitem[\protect\citeauthoryear{Lehmann, K{\"u}nstler, Carroll  \&
  Strassmeier}{Lehmann et~al.}{2015}]{Lehmann2015}
Lehmann L.~T.,  K{\"u}nstler A.,  Carroll T.~A.,   Strassmeier K.~G.,  2015,
  \mn@doi [Astron. Nachrichten] {10.1002/asna.201412162}, 336, 258

\bibitem[\protect\citeauthoryear{{Lehmann}, {Hussain}, {Vidotto}, {Jardine}  \&
  {Mackay}}{{Lehmann} et~al.}{2021}]{Lehmann2021}
{Lehmann} L.~T.,  {Hussain} G.~A.~J.,  {Vidotto} A.~A.,  {Jardine} M.~M.,
  {Mackay} D.~H.,  2021, \mn@doi [\mnras] {10.1093/mnras/staa3284}, \href
  {https://ui.adsabs.harvard.edu/abs/2021MNRAS.500.1243L} {500, 1243}

\bibitem[\protect\citeauthoryear{Morin et~al.,}{Morin et~al.}{2008}]{Morin2008}
Morin J.,  et~al., 2008, \mn@doi [\mnras] {10.1111/j.1365-2966.2007.12709.x},
  384, 77

\bibitem[\protect\citeauthoryear{Morin, Donati, Petit, Delfosse, Forveille  \&
  Jardine}{Morin et~al.}{2010}]{Morin2010}
Morin J.,  Donati J.-F.,  Petit P.,  Delfosse X.,  Forveille T.,   Jardine
  M.~M.,  2010, \mn@doi [\mnras] {10.1111/j.1365-2966.2010.17101.x}, 407, 2269

\bibitem[\protect\citeauthoryear{{Murtagh} \& {Heck}}{{Murtagh} \&
  {Heck}}{1987}]{Murtagh1987}
{Murtagh} F.,  {Heck} A.,  1987, {Multivariate Data Analysis}.
~ Vol. 131, \mn@doi{10.1007/978-94-009-3789-5, }

\bibitem[\protect\citeauthoryear{{Petit}, {Louge}, {Th{\'e}ado}, {Paletou},
  {Manset}, {Morin}, {Marsden}  \& {Jeffers}}{{Petit}
  et~al.}{2014}]{PolarBasePetit2014}
{Petit} P.,  {Louge} T.,  {Th{\'e}ado} S.,  {Paletou} F.,  {Manset} N.,
  {Morin} J.,  {Marsden} S.~C.,   {Jeffers} S.~V.,  2014, \mn@doi [\pasp]
  {10.1086/676976}, \href
  {https://ui.adsabs.harvard.edu/abs/2014PASP..126..469P} {126, 469}

\bibitem[\protect\citeauthoryear{{Rees}, {L{\'o}pez Ariste}, {Thatcher}  \&
  {Semel}}{{Rees} et~al.}{2000}]{Rees2000}
{Rees} D.~E.,  {L{\'o}pez Ariste} A.,  {Thatcher} J.,   {Semel} M.,  2000,
  \aap, \href {https://ui.adsabs.harvard.edu/abs/2000A&A...355..759R} {355,
  759}

\bibitem[\protect\citeauthoryear{Reiners \& Basri}{Reiners \&
  Basri}{2006}]{Reiners2006}
Reiners A.,  Basri G.,  2006, \mn@doi [\apj] {10.1086/503324}, 644, 497

\bibitem[\protect\citeauthoryear{Robinson, Worden  \& Harvey}{Robinson
  et~al.}{1980}]{Robinson1980}
Robinson R.~D.,  Worden S.~P.,   Harvey J.~W.,  1980, \mn@doi [\apjl]
  {10.1086/183217}, 236, L155

\bibitem[\protect\citeauthoryear{Saar}{Saar}{1988}]{Saar1988}
Saar S.~H.,  1988, \mn@doi [\apj] {10.1086/165907}, 324, 441

\bibitem[\protect\citeauthoryear{See et~al.,}{See et~al.}{2015}]{See2015}
See V.,  et~al., 2015, \mn@doi [\mnras] {10.1093/mnras/stv1925}, 453, 4301

\bibitem[\protect\citeauthoryear{{Skumanich} \& {L{\'o}pez Ariste}}{{Skumanich}
  \& {L{\'o}pez Ariste}}{2002}]{Skumanich2002}
{Skumanich} A.,  {L{\'o}pez Ariste} A.,  2002, \mn@doi [\apj] {10.1086/339503},
  \href {https://ui.adsabs.harvard.edu/abs/2002ApJ...570..379S} {570, 379}

\bibitem[\protect\citeauthoryear{{Stenflo}}{{Stenflo}}{1994}]{Stenflo1994}
{Stenflo} J.,  1994, in Astrophysics and Space Science Library. ,
  \mn@doi{10.1007/978-94-015-8246-9}

\bibitem[\protect\citeauthoryear{{Unno}}{{Unno}}{1956}]{Unno1956}
{Unno} W.,  1956, \pasj, \href
  {https://ui.adsabs.harvard.edu/abs/1956PASJ....8..108U} {8, 108}

\makeatother
\end{thebibliography}



\appendix

\section{Extra figures: diagnosing the large-scale field }

The first four figures show further example magnetic field topologies and their PCA analysis and are presented in the same format as in Fig.~\ref{fig:Map_tiltDipol}. 

\begin{figure*} 
	\begin{flushleft}
	\textbf{a.} \hspace{2.5cm} \textbf{b.} \hspace{3.6cm} \textbf{c.}
	\end{flushleft}
    \begin{minipage}{0.11\textwidth}
    \centering
    \includegraphics[height=0.85\columnwidth, angle=270, trim={140 0 0 29}, clip]{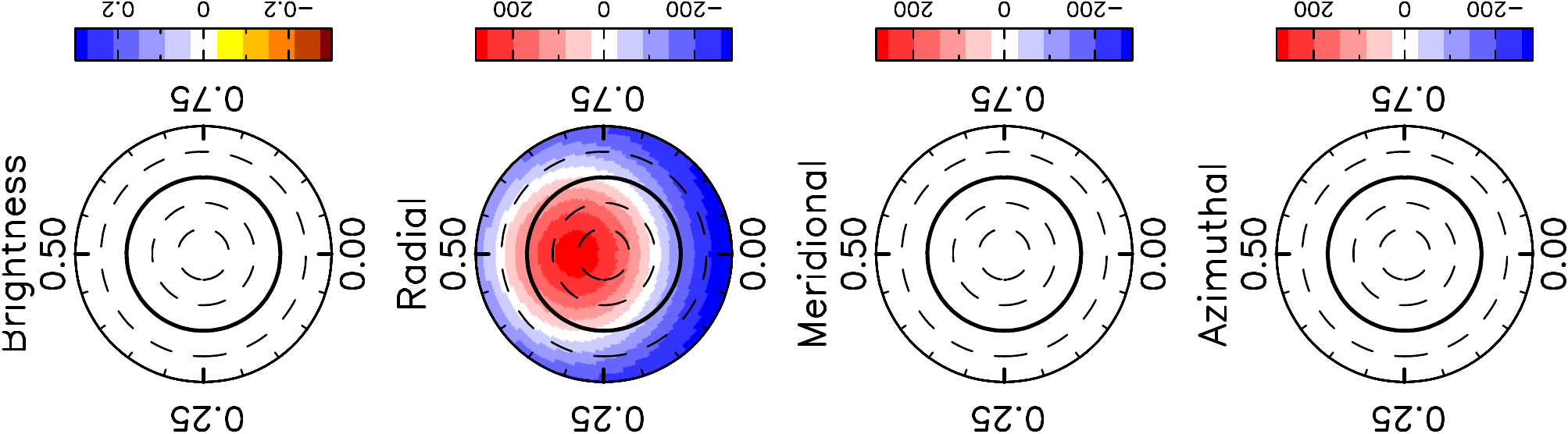} 
	\includegraphics[width=\columnwidth, angle=180, trim={470 130 12 0}, clip]{Figures/MapN2rad.pdf}
	\end{minipage}
    \begin{minipage}{0.25\textwidth}
    \centering
    \includegraphics[width=\columnwidth, angle=0, trim={0 0 0 0}, clip]{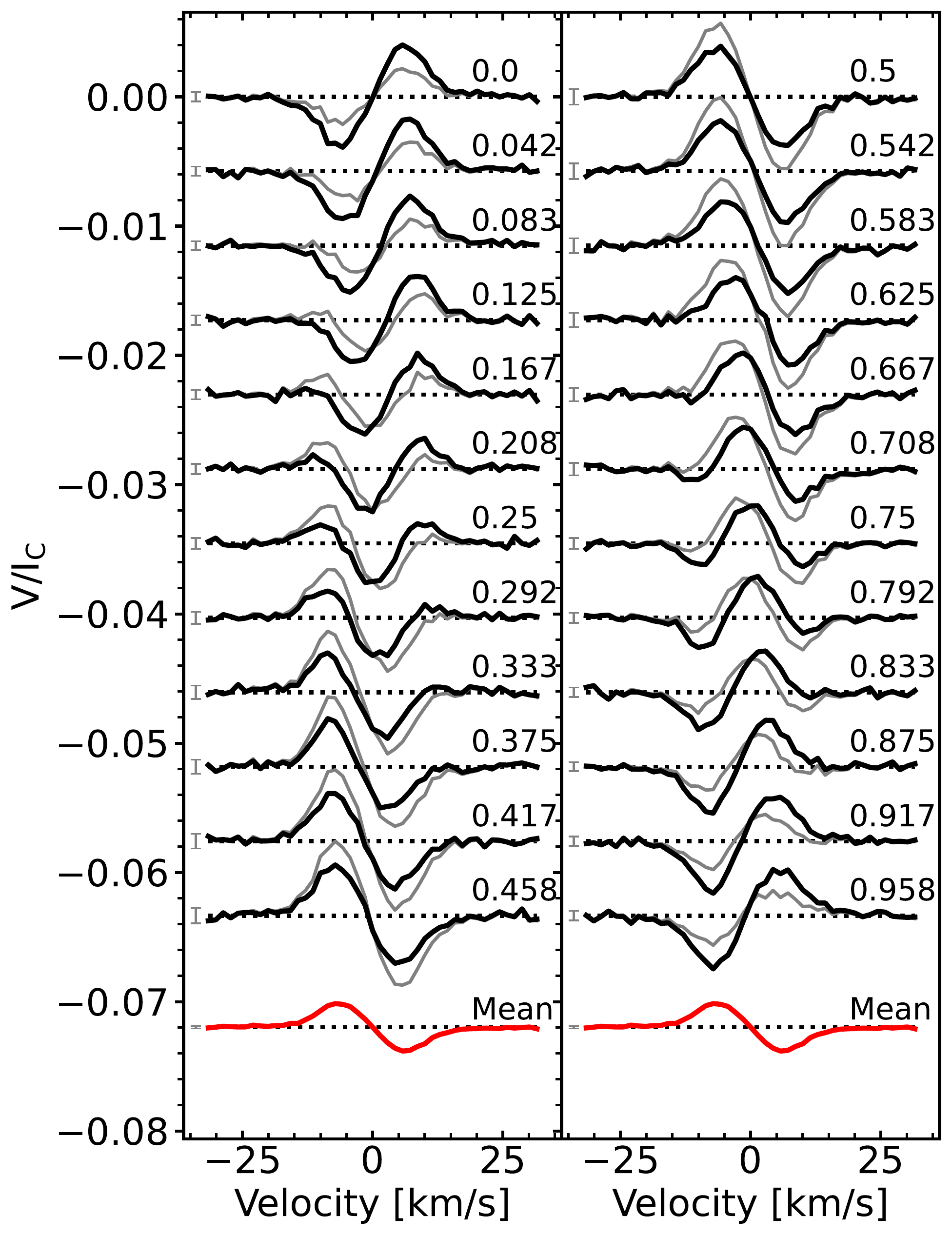} 
    	\end{minipage}
    \begin{minipage}{0.6\textwidth}
    \centering
    \includegraphics[width=\columnwidth, trim={0 400 0 0}, clip]{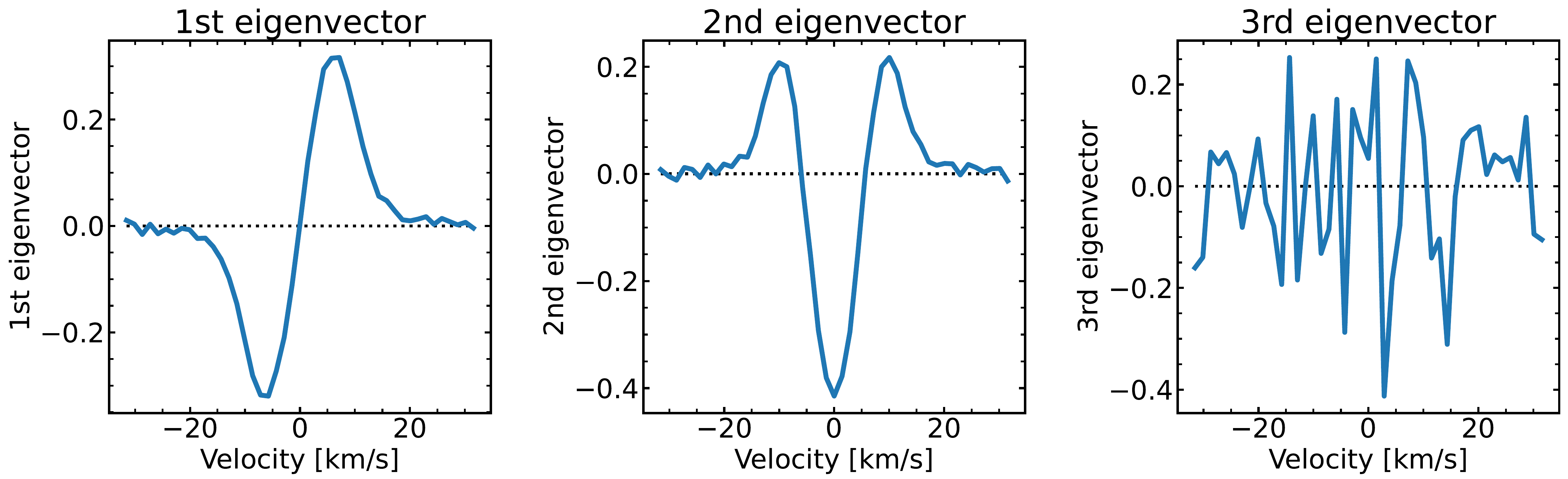} \\
    \includegraphics[width=\columnwidth, trim={0 400 0 0}, clip]{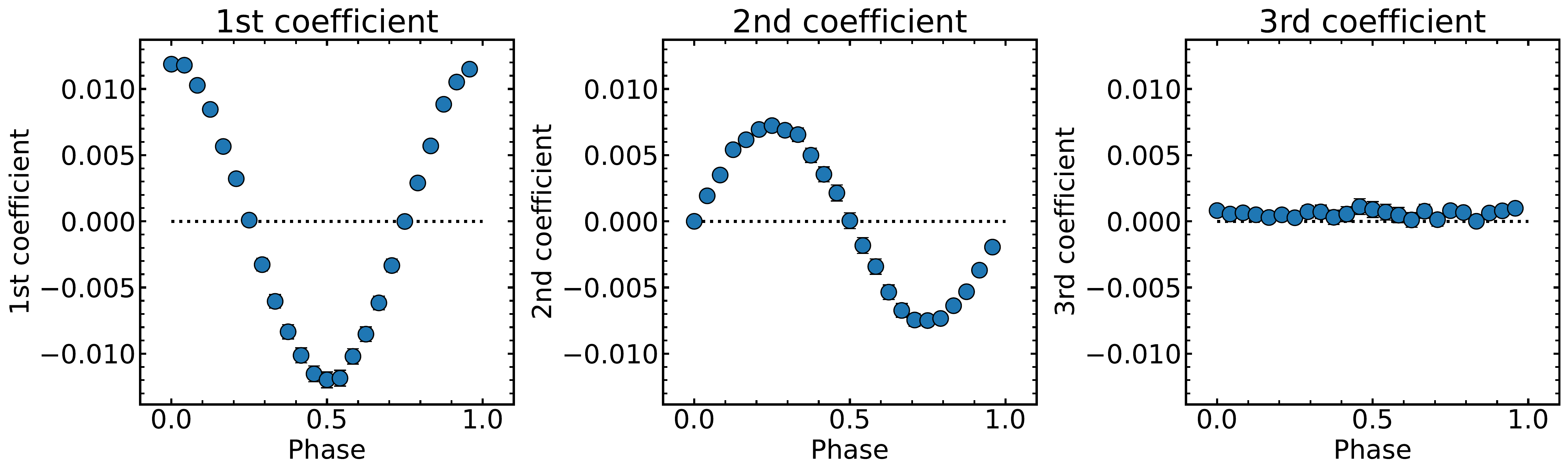} 
    	\end{minipage}
    \caption{The PCA analysis for simulated topology shown in Fig.~\ref{fig:Map_tiltDipol} after restricting the topology to the radial field only. The same format is used as in Fig.~\ref{fig:Map_tiltDipol}.}
    \label{fig:Map_radDipol}
\end{figure*}

Fig.~\ref{fig:Map_radDipol} shows the PCA analysis of the example field topology for Fig.~\ref{fig:Map_tiltDipol} restricted to the radial field component only. The ratio between the coefficient maximum of the symmetric and antisymmetric (second and first) eigenvector is 1.2 for the poloidal dipole shown in Fig.~\ref{fig:Map_tiltDipol} and 0.6 for the radial field restricted map in Fig.~\ref{fig:Map_radDipol}, indicating also that the radial field creates a symmetric but weaker signal.

\begin{figure*} 
	\begin{flushleft}
	\textbf{a.} \hspace{2.5cm} \textbf{b.} \hspace{3.6cm} \textbf{c.}
	\end{flushleft}
    \begin{minipage}{0.11\textwidth}
    \centering
    \includegraphics[height=0.85\columnwidth, angle=270, trim={140 0 0 29}, clip]{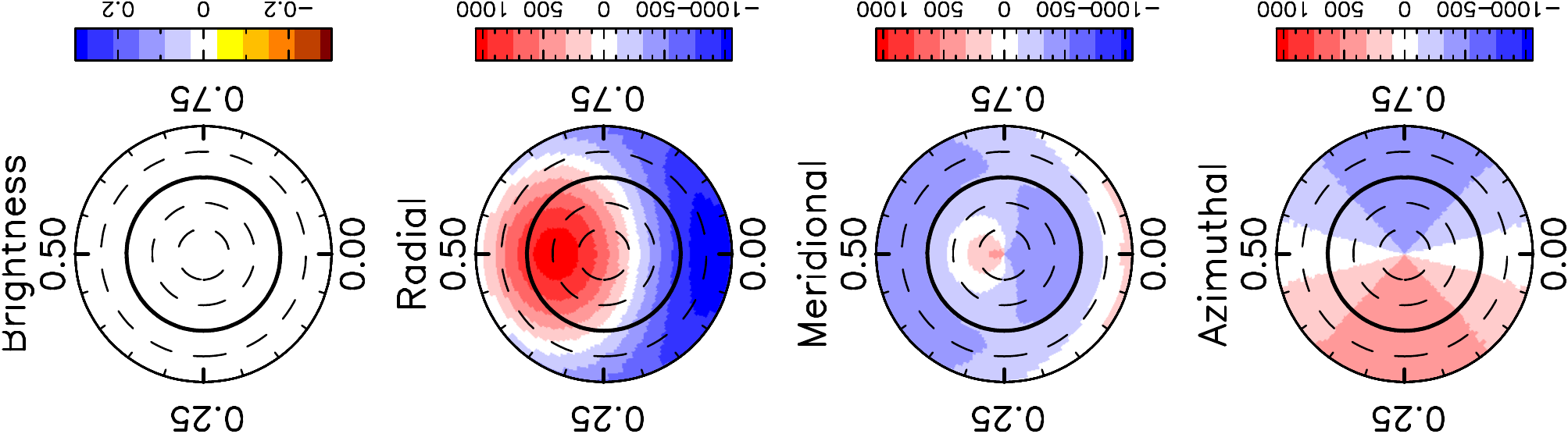} 
	\includegraphics[width=\columnwidth, angle=180, trim={460 130 5 0}, clip]{Figures/MapN5.pdf}
	\end{minipage}
    \begin{minipage}{0.25\textwidth}
    \centering
    \includegraphics[width=\columnwidth, angle=0, trim={0 0 0 0}, clip]{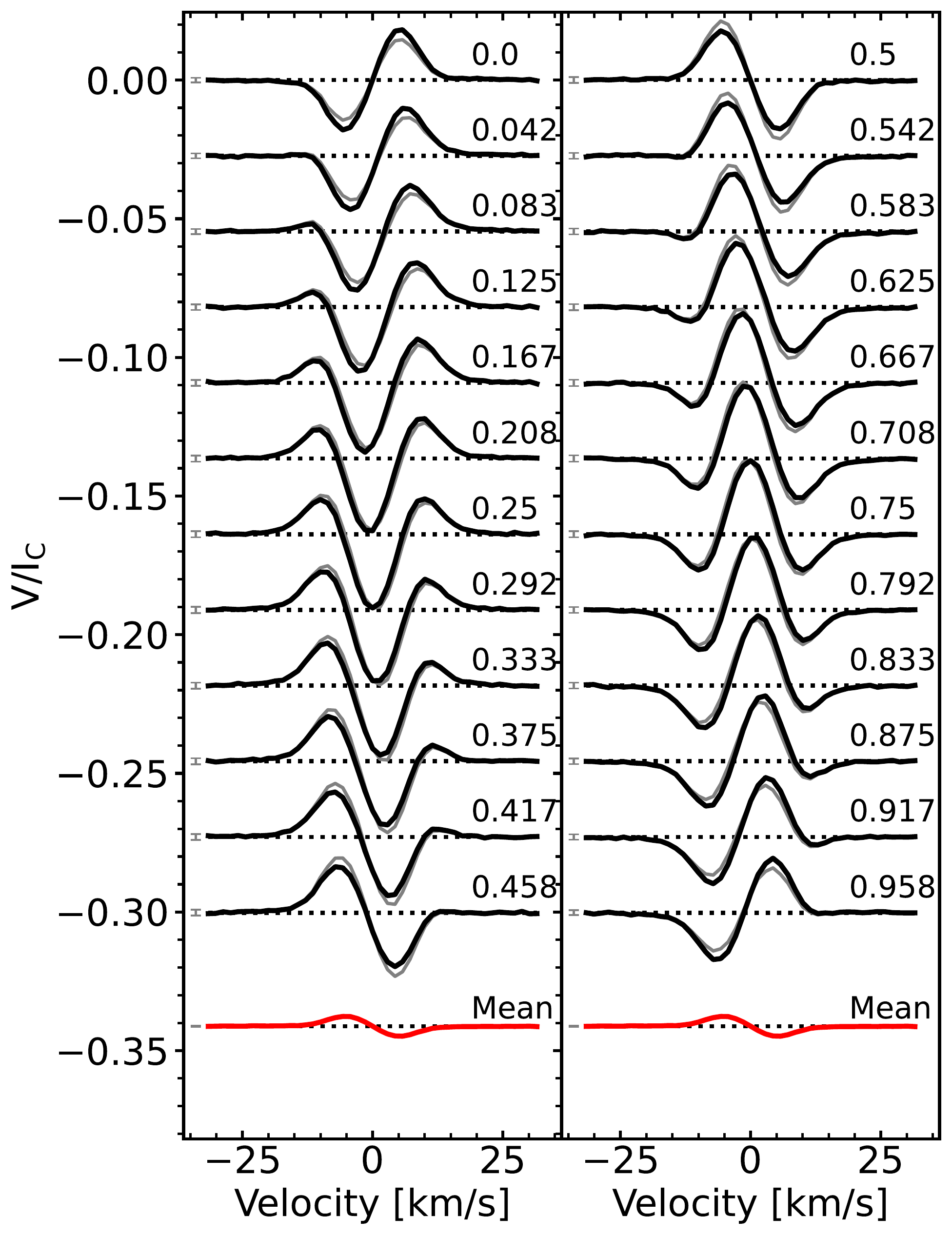} 
    	\end{minipage}
    \begin{minipage}{0.6\textwidth}
    \centering
    \includegraphics[width=\columnwidth, trim={0 400 0 0}, clip]{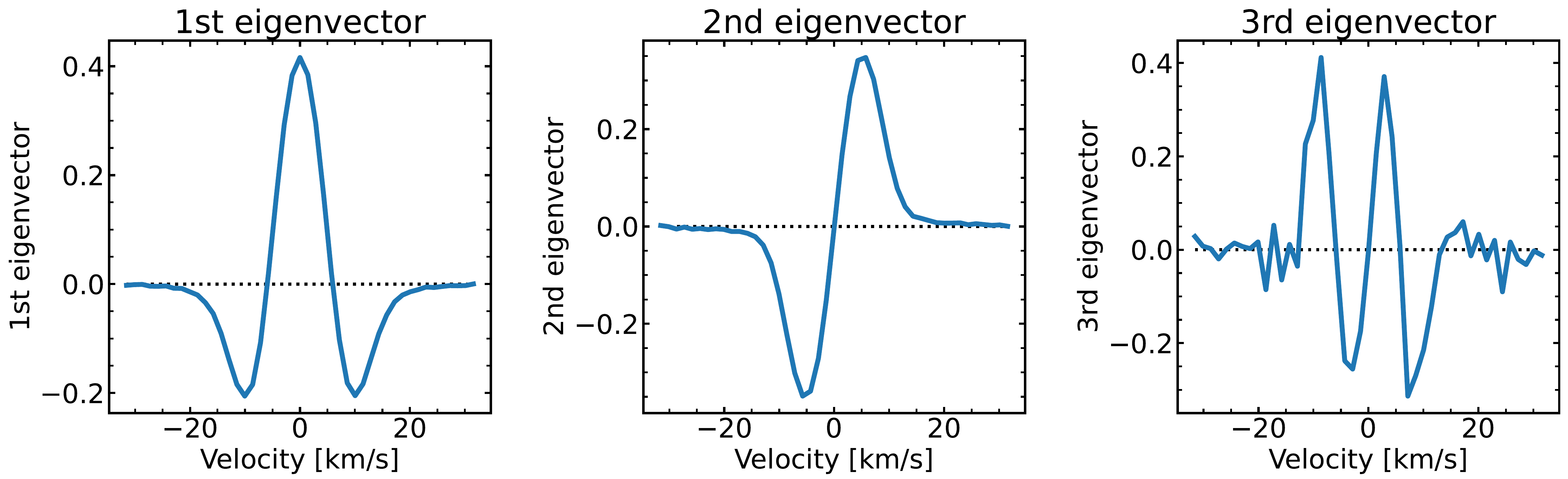} \\
    \includegraphics[width=\columnwidth, trim={0 400 0 0}, clip]{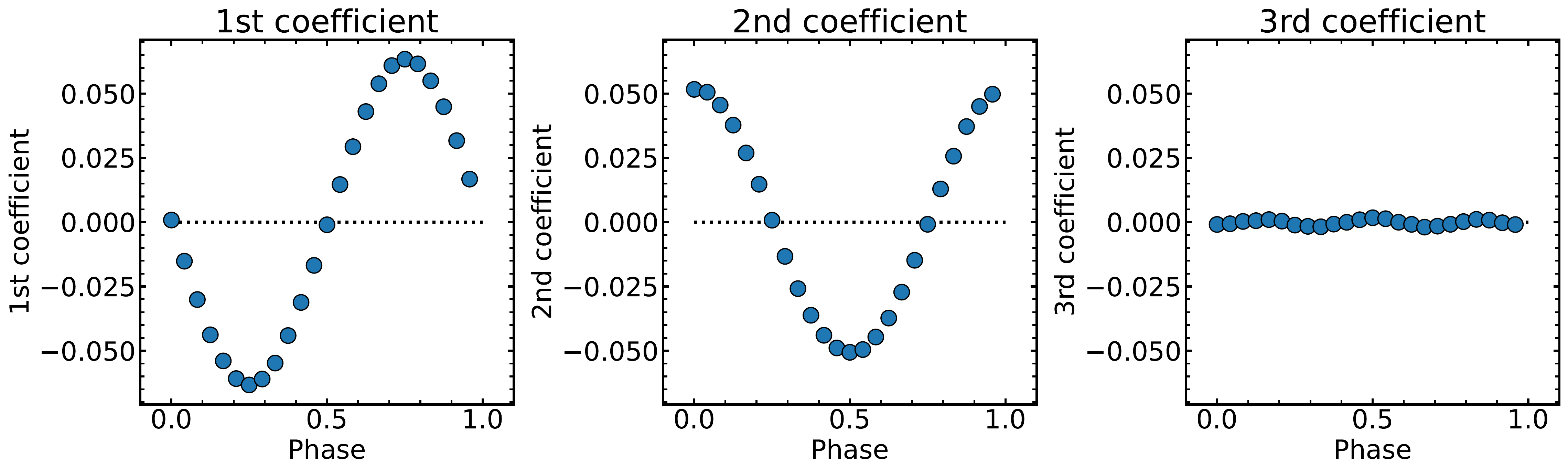} 
    	\end{minipage}
    \caption{The simulated example of a strongly tilted poloidal dipole with tilt angle $\psi = 55^\circ$ presented in the same format as in Fig.~\ref{fig:Map_tiltDipol}.}
    \label{fig:Map_VerytiltDipol}
\end{figure*}

\begin{figure*} 
	\begin{flushleft}
	\textbf{a.} \hspace{2.5cm} \textbf{b.} \hspace{3.6cm} \textbf{c.}
	\end{flushleft}
    \begin{minipage}{0.11\textwidth}
    \centering
    \includegraphics[height=0.85\columnwidth, angle=270, trim={140 0 0 29}, clip]{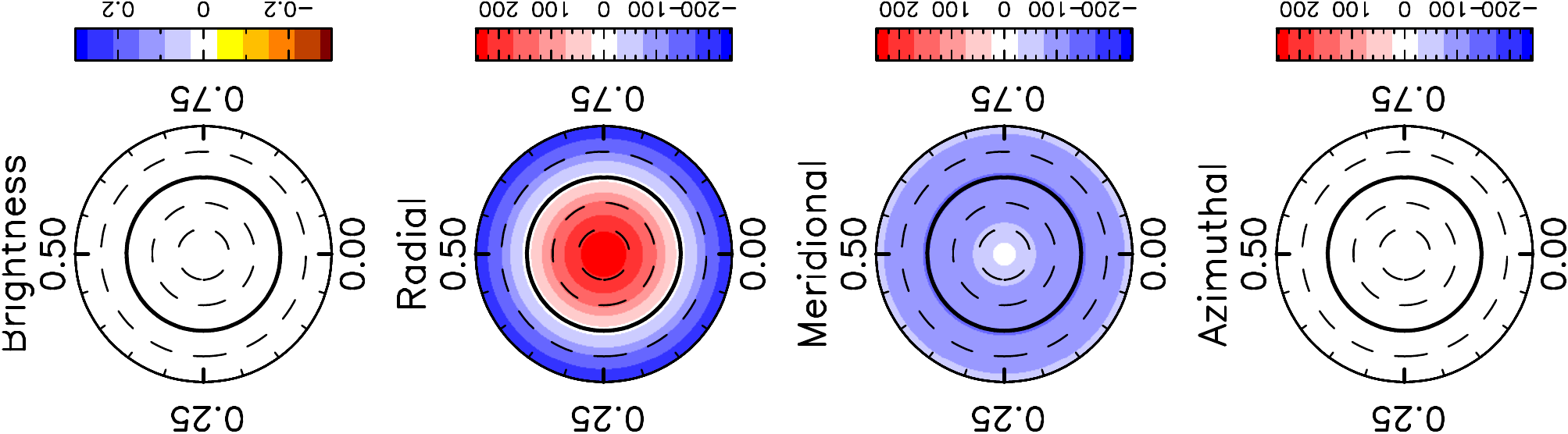} 
	\includegraphics[width=\columnwidth, angle=180, trim={470 130 12 0}, clip]{Figures/MapN4.pdf} 
	\end{minipage}
    \begin{minipage}{0.25\textwidth}
    \centering
    \includegraphics[width=\columnwidth, angle=0, trim={0 0 0 0}, clip]{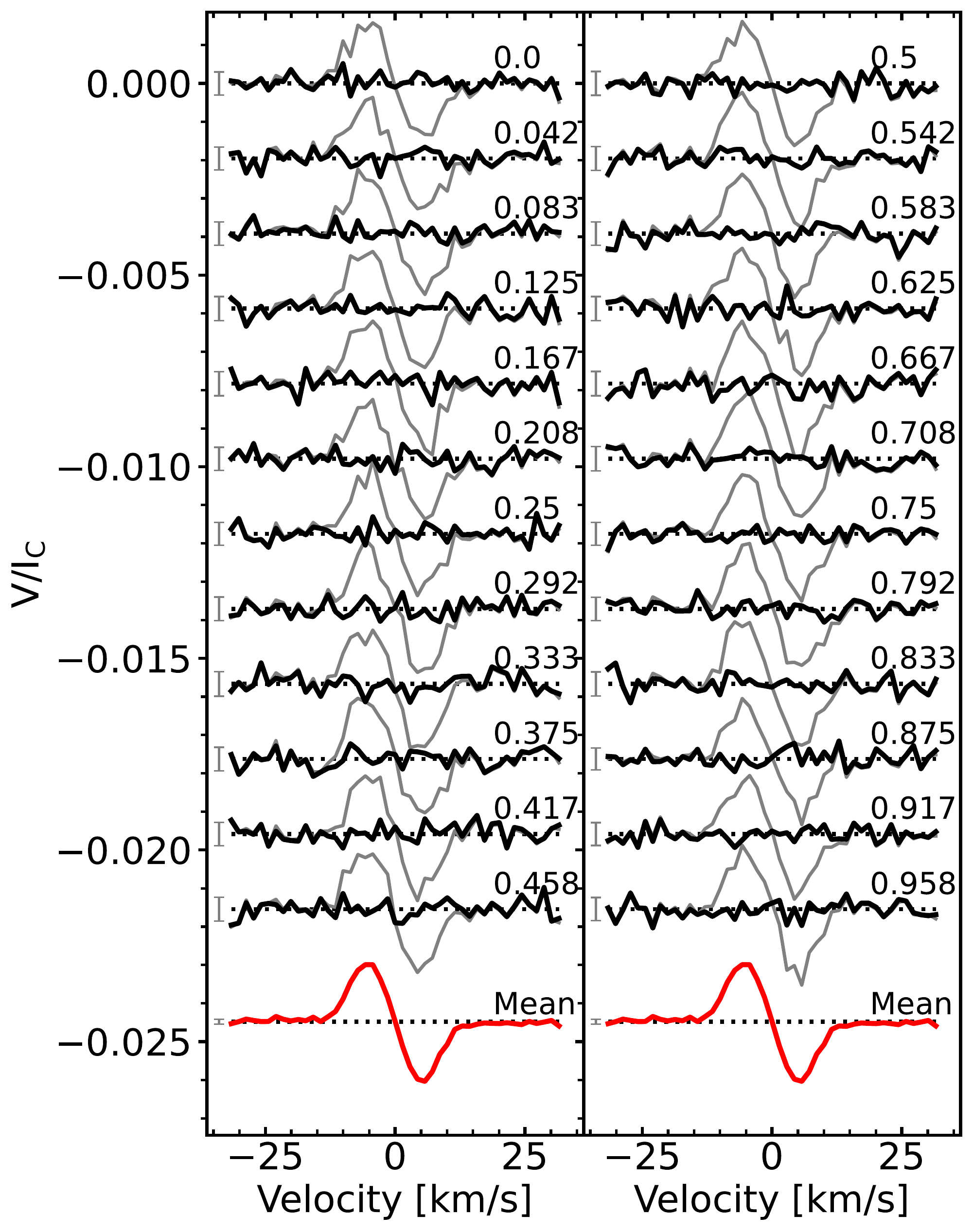} 
    	\end{minipage}
    \begin{minipage}{0.6\textwidth}
    \centering
    \includegraphics[width=\columnwidth, trim={0 400 0 0}, clip]{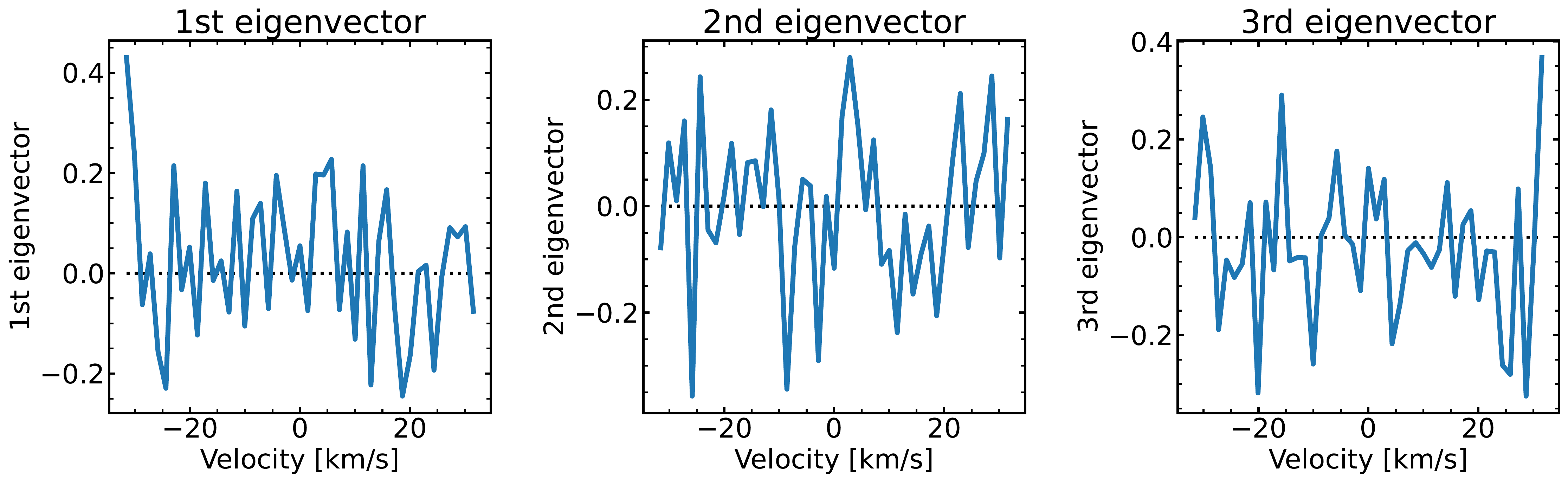} \\
    \includegraphics[width=\columnwidth, trim={0 400 0 0}, clip]{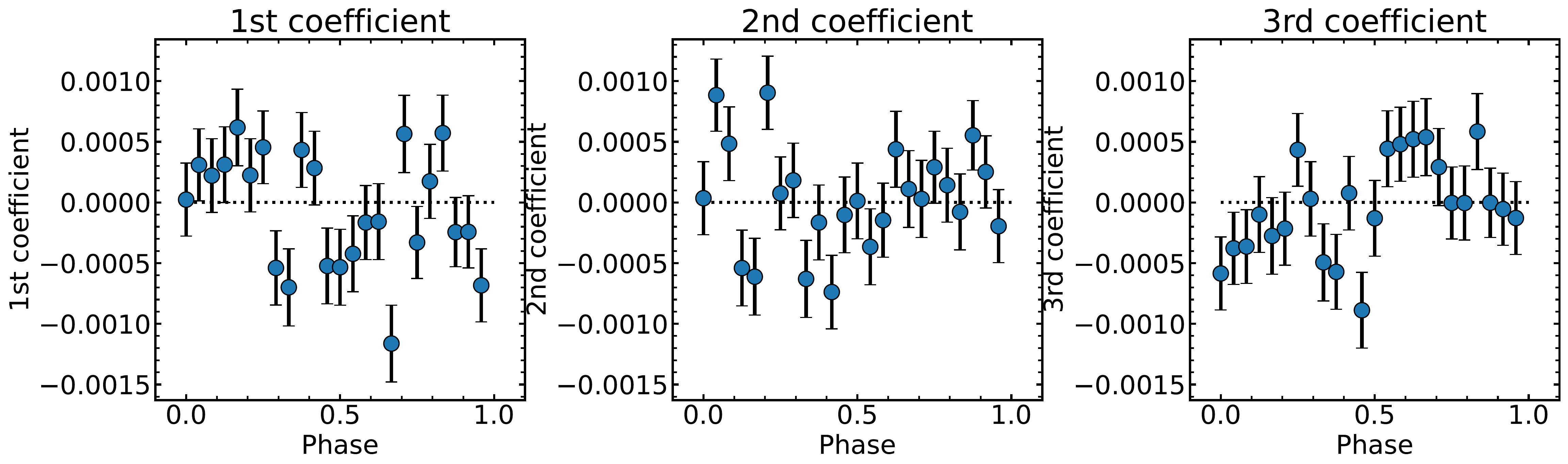} 
    	\end{minipage}
    \caption{The simulated example of a truly axisymmetric poloidal dipole with tilt angle $ \psi = 0^\circ$ presented in the same format as in Fig.~\ref{fig:Map_tiltDipol}.}
    \label{fig:Map_axisymDipol}
\end{figure*}

Further, we provide the simulated example of a highly tilted dipole with a tilt angle $\psi =55^\circ$, see Fig.~\ref{fig:Map_VerytiltDipol} and for a truly axisymmetric configuration with tilt angle $\psi =0^\circ$, see Fig.~\ref{fig:Map_axisymDipol}. For the tilted dipole in Fig.~\ref{fig:Map_VerytiltDipol}, we see a relatively weak mean profile and clear non-noise signals in all three eigenvectors with well-defined trends in the corresponding coefficients. For more tilted configurations further eigenvectors will have a non-noise signal and they will have the shapes of further derivatives of the unpolarised Stokes~$I$. 
For the truly axisymmetric dipole, Fig.~\ref{fig:Map_axisymDipol}, we see, that the Stokes~$V$ signals are constant for all observing phases so that the mean-subtracted Stokes~$V$ profile only contains noise. PCA confirms that by detecting only noise containing eigenvectors and coefficients.

\begin{figure*} 
	\begin{flushleft}
	\textbf{a.} \hspace{2.5cm} \textbf{b.} \hspace{3.6cm} \textbf{c.}
	\end{flushleft}
    \begin{minipage}{0.11\textwidth}
    \centering
    \includegraphics[height=0.85\columnwidth, angle=270, trim={140 0 0 29}, clip]{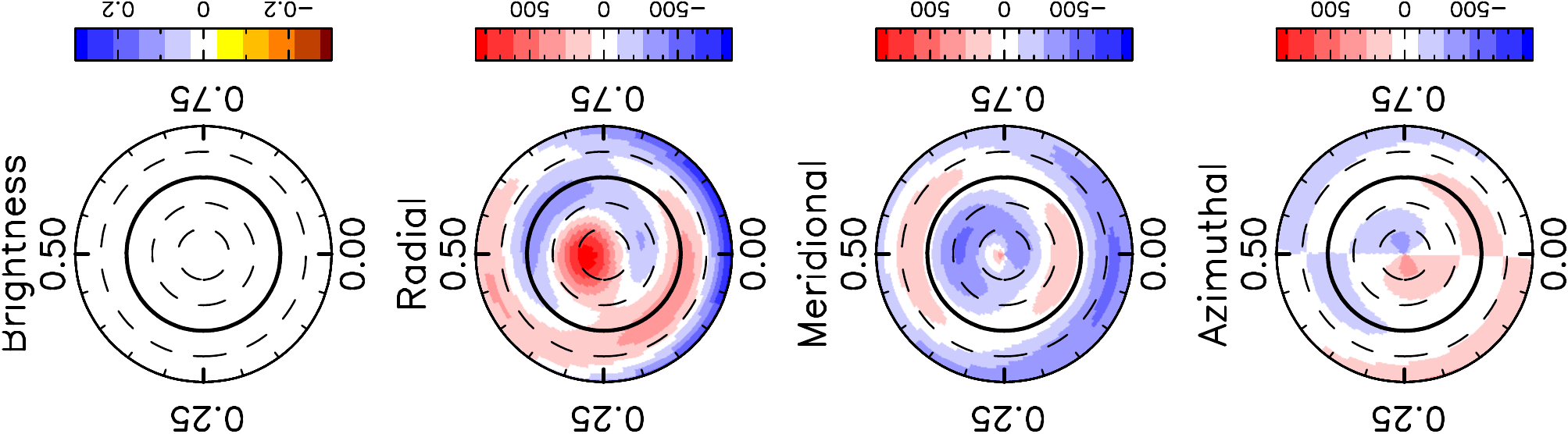} 
	\includegraphics[width=\columnwidth, angle=180, trim={470 130 12 0}, clip]{Figures/MapN13.pdf}
	\end{minipage}
    \begin{minipage}{0.25\textwidth}
    \centering
    \includegraphics[width=\columnwidth, angle=0, trim={0 0 0 0}, clip]{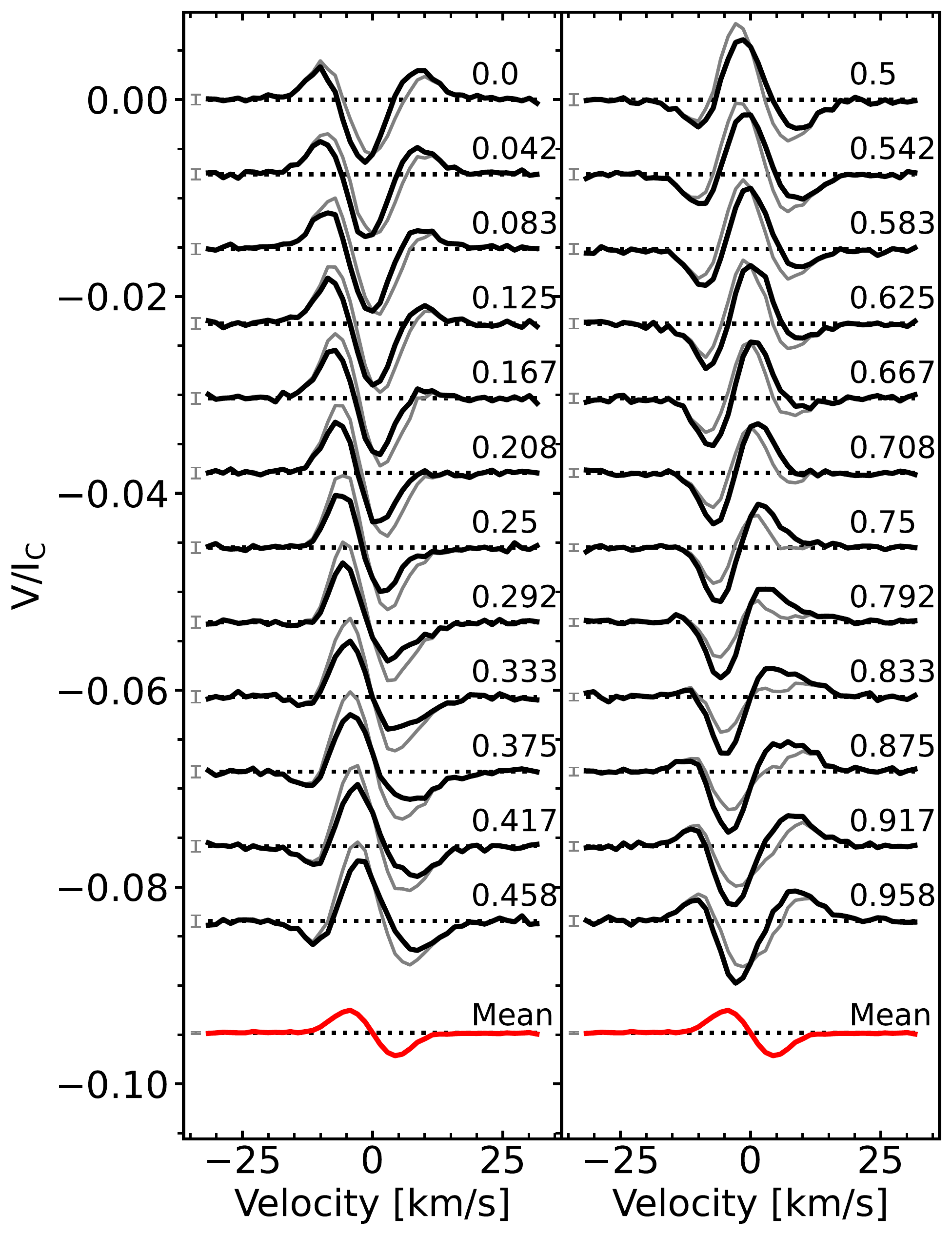} 
    	\end{minipage}
    \begin{minipage}{0.6\textwidth}
    \centering
    \includegraphics[width=\columnwidth, trim={0 400 0 0}, clip]{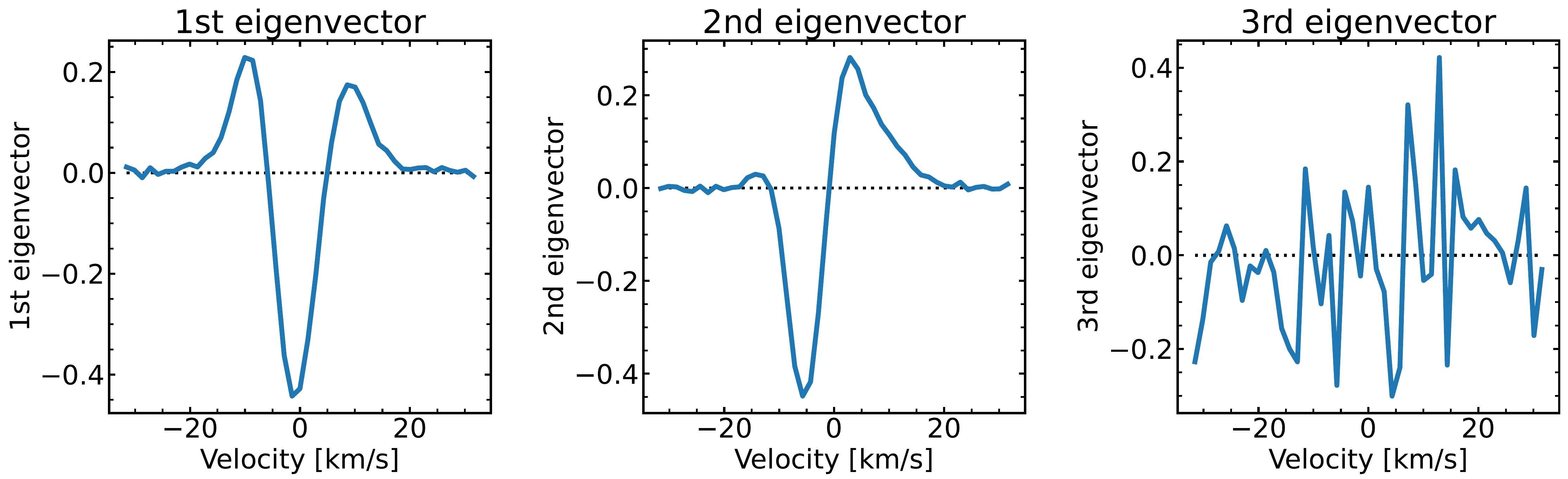} \\
    \includegraphics[width=\columnwidth, trim={0 400 0 0}, clip]{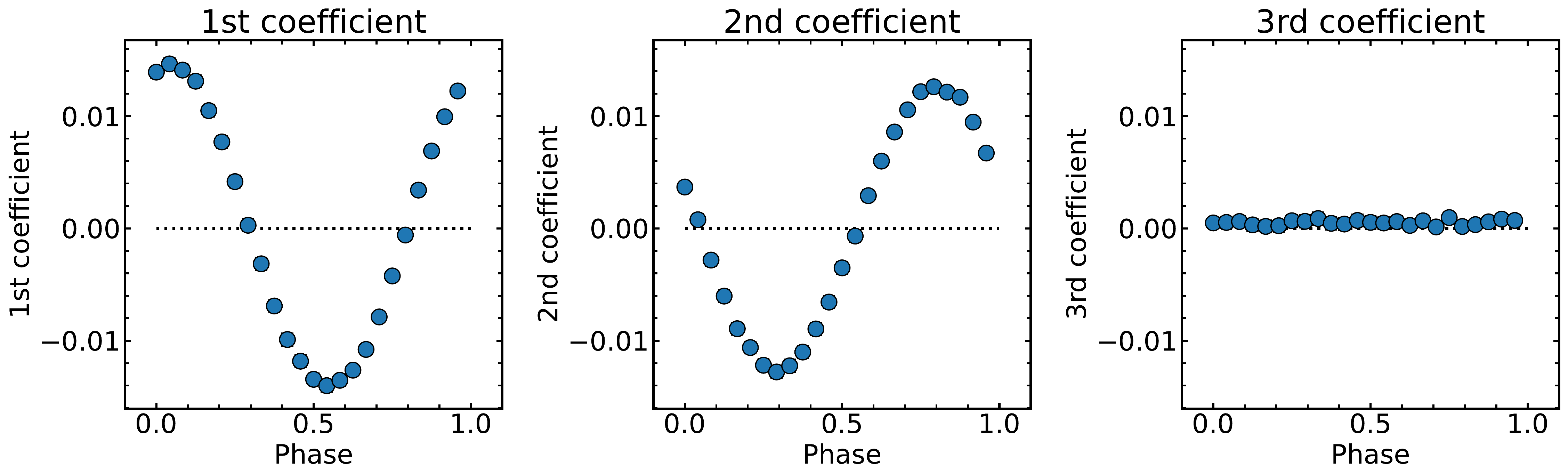} 
    	\end{minipage}
    \caption{The simulated example of a purely poloidal, complex field consisting of a tilted dipole and octopole. The same format as in Fig.~\ref{fig:Map_tiltDipol} is used.}
    \label{fig:Map_ComplexPol}
\end{figure*}

In Fig.~\ref{fig:Map_ComplexPol} we display an example of a poloidal complex field expressed by $\alpha$ coefficients only ($\beta = \gamma = 0$), that shows asymmetric eigenvectors in the PCA analysis. This field configuration contains a poloidal dipole ($\alpha_{\ell = 1}$) and octopole component ($\alpha_{\ell = 3}$), that have different tilt angles and pointing phases. We notice, that asymmetric eigenvectors appear more often if multipoles are shifted to each other in phase.

\section{Magnetic field description}
\label{App:MagField}

We describe the magnetic field using the spherical harmonic description by \cite{Elsasser1946} and \citet[Appendix III]{Chandrasekhar1961}. A right-handed spherical coordinate system $(r, \phi, \theta)$ is applied, where the  radial field component $B_r$ points towards outwards, the azimuthal component $B_\phi$ runs in the anti-clockwise direction as viewed from North pole and the meridional field $B_\theta$ runs with colatitude (from north to south). The poloidal and toroidal component can than be described as follows:
\begin{align}
B_{r,\mathrm{pol}}(\phi, \theta)  &= \sum_{\ell m} \alpha_{\ell m} P_{\ell m} e^{im\phi}, \nonumber \\
B_{\phi,\mathrm{pol}}(\phi, \theta)  &= \sum_{\ell m} (\alpha_{\ell m}+\beta_{\ell m}) \frac{im P_{\ell m} e^{im\phi}}{(\ell + 1) \sin \theta}, \nonumber \\ 
B_{\theta,\mathrm{pol}}(\phi, \theta)  &=\sum_{\ell m} (\alpha_{\ell m}+\beta_{\ell m}) \frac{1}{\ell+1} \frac{\mathrm{d}P_{\ell m}}{\mathrm{d}\theta} e^{im\phi} . \label{Eq:B_pol}
\end{align}
\begin{align}
B_{r,\mathrm{tor}}(\phi, \theta) &= 0, \nonumber \\
B_{\phi,\mathrm{tor}}(\phi, \theta) &= - \sum_{\ell m} \gamma_{\ell m} \frac{1}{\ell+1} \frac{\mathrm{d}P_{\ell m}}{\mathrm{d}\theta} e^{im\phi}, \nonumber \\
B_{\theta,\mathrm{tor}}(\phi, \theta) &= \sum_{\ell m} \gamma_{\ell m} \frac{im P_{\ell m} e^{im\phi}}{(\ell + 1) \sin \theta}, \label{Eq:B_tor}
\end{align}
so that $(B_{r,\rm{pol}}, B_{\phi,\rm{pol}}, B_{\theta,\rm{pol}}) = \vect{B_{\rm{pol}}}$, $(B_{r,\rm{tor}}, B_{\phi,\rm{tor}}, B_{\theta,\rm{tor}}) = \vect{B_{\rm{tor}}}$ and $ \vect{B} = \vect{B_{\mrm{pol}}} + \vect{B_{\mrm{tor}}}$. The associated Legendre polynomial of mode $\ell$ and order $m$ are denoted by $P_{\ell m} \equiv c_{\ell m}P_{\ell m}(\cos \theta)$, where $c_{\ell m}$ is a normalization constant:
\begin{equation}
c_{\ell m} = \sqrt{\frac{2\ell+1}{4\pi}\frac{(\ell - m)!}{(\ell + m)!}}.
\end{equation}
The poloidal field is described by the two coefficients $\alpha_{\ell m}$ and $\beta_{\ell m}$ and a toroidal field by $\gamma_{\ell m}$.

For a pure axisymmetric dipole ($\ell = 1, m=0$) the equations simplify to: 
\begin{align}
B_{r,\mathrm{pol}}(\phi, \theta)  &= \sqrt{\frac{3}{4\pi}}  \alpha_{10} \cos \theta, \nonumber \\
B_{\phi,\mathrm{pol}}(\phi, \theta)  &=  0, \nonumber \\ 
B_{\theta,\mathrm{pol}}(\phi, \theta)  &= - \frac{1}{2}\sqrt{\frac{3}{4\pi}} (\alpha_{10}+\beta_{10}) \sin \theta , \nonumber
\end{align}
\begin{align}
B_{r,\mathrm{tor}}(\phi, \theta) &= 0, \nonumber \\
B_{\phi,\mathrm{tor}}(\phi, \theta) &= \frac{1}{2}\sqrt{\frac{3}{4\pi}} \gamma_{10} \sin \theta, \nonumber \\
B_{\theta,\mathrm{tor}}(\phi, \theta) &= 0.
\label{Eq:B_axi}
\end{align}


\bsp	
\label{lastpage}
\end{document}